\documentclass[12pt,preprint]{aastex} 

\usepackage{natbib}
\usepackage{graphicx}
\bibpunct{(}{)}{;}{a}{}{,}

\newcommand\modified[1]{{\bf #1}}

\begin{document}

   \title{Systematic search for extremely metal poor galaxies in 
        the Sloan Digital Sky Survey}
%
   \author{
        A.~B.~Morales-Luis\altaffilmark{1,2},
          	J.~S\'anchez~Almeida\altaffilmark{1,2},
         	J.~A.~L.~Aguerri\altaffilmark{1,2},  
         	and 
         	C.~Mu\~noz-Tu\~n\'on\altaffilmark{1,2} 
          }
\altaffiltext{1}{Instituto de Astrof\'\i sica de Canarias, E-38205 La Laguna,
Tenerife, Spain}
\altaffiltext{2}{Departamento de Astrof\'\i sica, Universidad de La Laguna,
Tenerife, Spain}
\email{abml@iac.es, jos@iac.es, cmt@iac.es, jalfonso@iac.es}
\begin{abstract}
We carry out a systematic search for extremely metal poor (XMP)
galaxies in the spectroscopic sample of 
Sloan Digital Sky Survey (SDSS) data release 7 (DR7). The XMP 
candidates are found by classifying all the galaxies according 
to the form of their spectra in a region 80\,\AA\ wide 
around H$\alpha$. Due to the data size, the method requires an 
automatic classification algorithm.  We use k-means.
Our systematic search renders 32 galaxies having negligible [NII]~lines,
as expected in XMP galaxy spectra.
Twenty one
of them have been previously identified as XMP galaxies in the 
literature -- the
remaining eleven
are new.
This was established after a thorough bibliographic search 
that yielded only some 130 galaxies known to have 
an oxygen metallicity ten times smaller than the Sun 
(explicitly, with $12+\log({\rm O/H)}\leq 7.65$).
XMP galaxies are rare; they represent 
0.01\% of the galaxies with emission lines in SDSS/DR7.
Although the final metallicity estimate of all candidates
remains pending, strong-line empirical calibrations indicate 
a metallicity about
one-tenth solar, with the 
oxygen metallicity of the
twenty one known targets 
being $12+\log({\rm O/H)}\simeq
7.61\pm 0.19
$.
Since the SDSS catalog is limited in apparent magnitude,
we have been able to estimate the volume number density of XMP
galaxies in the local universe, which turns out to be 
$(1.32\pm 0.23)\cdot 10^{-4}~{\rm Mpc}^{-3}$. The XMP galaxies
constitute 0.1\%\ of the galaxies in the local volume, 
or $\sim$ 0.2\%\ considering only emission line galaxies.
All but four of our candidates are blue compact dwarf galaxies (BCDs),
and 24 of them have either cometary shape or are formed by 
chained knots. 
%
%
\end{abstract}

   \keywords{
   methods: data analysis --
   galaxies: abundances -- 
   galaxies: formation --
   galaxies: starburst --
   galaxies: statistics
               }



\section{Introduction}\label{intro}

The evolution of galaxies involves the birth and death of stars,
therefore, galaxies unavoidably produce metals as they live.
Thus, galaxies with very low metallic content
are probably unevolved objects, 
and if we find them nearby, they
provide a readily accessible fossil record from the early universe. 
These objects are to be expected
according to
the paradigm of hierarchical galaxy formation,
where large galaxies arise through the assembly of
smaller ones in an inefficient process leaving 
many dwarf galaxies as remnant 
\citep[e.g.,][]{1999ApJ...522...82K,2001MNRAS.326.1228B,die07}. 
They seem to be materialized as
the extremely  metal-poor (XMP) dwarf galaxies observed today
which, consequently, would be
the closest examples we can find of these
elementary primordial units from which larger galaxies assembled
\citep[e.g.,][]{2004ApJ...616..768I,2007ApJ...665.1115I}.
Those units must have been extremely common in the past, 
but they cannot be directly observed at high redshift. Nearby
low metallicity galaxies offer a chance for detailed studies 
otherwise impossible. Studies of their interstellar 
medium (ISM) can shed light on the properties of the primordial ISM at the time 
of galaxy formation
\citep[][]{2007ApJ...665.1115I}. For example, even
the most metal-deficient galaxies in the local universe 
formed from matter already enriched by an early
star formation
episode, and the determination of the minimum galactic
metallicity seems to be the best constraint available on 
these first stars \citep[e.g.,][]{bro04,2005ApJS..161..240T}. Because they have
not undergone much chemical evolution, these galaxies are also 
the best objects for the determination of the primordial 
He abundance that constrains cosmological models
\citep[e.g., ][]{pei74,pag92,2004ApJ...602..200I}.

The class of blue compact dwarf galaxies 
\citep[BCDs; e.g.,][]{cai01,cai01c,amo07,amo09} contains the galaxies with 
the lowest known
gas-phase
metallicity \citep[e.g.,][]{kun00,2008A&A...491..113P}. 
It is so far unclear whether such preference for XMP galaxies
to be BCDs is genuine or if the association results from an 
observational bias that systematically disfavors low surface 
brightness objects. The best XMP candidates in the 
local universe are BCDs, but metal poor galaxies are found
among other types of dwarf galaxies as well  
\citep[see][]{kun00}.  

Unfortunately, XMP galaxies are rare. The review by \citet[][]{kun00}
cites only 31 targets with metallicity below one tenth the 
solar value,
which is the threshold customary used to define XMP galaxies. 
For decades I\,Zw\,18 held the record of lowest metallicity
\citep{sar70}, and although a few other examples
have been recently found \citep[e.g.,][]{izo05,izo09},
there is a minimum metallicity close to that
of I~Zw~18, which corresponds to a few hundredths the 
solar value. The existence of such metallicity threshold is suggestive 
of the pre-galactic origin of metals as it 
happens with halo stars \citep[e.g.,][]{spi82}, but it may also be
due to other effects like the 
early self contamination of the ongoing starburst that rises
any original level to a minimum metallicity \citep[e.g.,][]{kun86},
or even the technical difficulty of metallicity determinations 
below a threshold \citep{2008A&A...491..113P}.
The number of low metallicity galaxies has significantly increased
since the work by \citet[][]{kun00}, but they are still rare objects.
The thorough bibliographic compilation described in 
\S~\ref{biblio_search} shows only 129
such targets. The shortage of low metallicity galaxies is
partly a consequence of their low luminosity as expected from the 
luminosity-metallicity relationship \citep[e.g.,][]{leq79,ski89,tre04}. 
They must be faint and so detectable only within a very local volume.

In order to enlarge the list of this rare yet interesting objects, 
we have carried out an automatic search for 
low metallicity galaxies in the seventh Sloan Digital 
Sky Survey data release (SDSS/DR7). The work is reported here.
So far as we are aware of,  this is the first systematic
search of this kind on SDSS/DR7,
even though extensive searches in earlier SDSS data releases 
have been reported \citep[e.g.,][]{2006A&A...454..137I,2009A&A...505...63G}.
The advantage of an orderly search rather than the more traditional 
serendipitous discovery is twofold. First, it maximizes the number 
of potential candidates. Second, the bias of the selection 
is quantifiable, allowing us to estimate for the
first time the volume number density of XMP 
galaxies in the local universe.

Low metallicity galaxies are characterized by having a 
[NII]$\lambda$6583 line negligibly small as compared 
to H$\alpha$. Thus, the ratio between [NII]$\lambda$6583 
and H$\alpha$ is used to measure metallicities through 
the appropriate calibration \citep[e.g.,][]{den02,pet04}.
A low value of this ratio has been imposed as 
a necessary condition in classical works seeking for XMP galaxies 
\citep[e.g.,][]{2006A&A...454..137I}. Building on this 
classical approach, we address
the problem in an original way by automatically
classifying the galaxies according to the 
shape of their spectra in a region around H$\alpha$.
We expected the classification to automatically
separate classes of galaxies whose spectra present this property,
and those targets would be regarded as metal poor candidates.
Note that our approach does not require the detailed knowledge of 
the spectral properties of the XMP galaxies, e.g., we do not have 
to specify a particular ratio [NII]$\lambda$6583  over H$\alpha$ for a 
galaxy to be included.
They are determined by the classification algorithm in view
of the existing spectra. This minimum need of prior knowledge
makes the search novel and 
robust against uncertainties in the selection criteria. The above 
conjecture turned out to work, and the result of the study 
is presented here. We employ a robust classification 
algorithm called {\em k-means}, commonly used in data-mining
\citep[e.g.,][]{eve95}, and which we have already
successfully applied to sort out different types of astronomical 
spectra spanning from polarization profiles in the Sun 
\citep[][]{san00,2011A&A...530A..14V} to galaxy spectra \citep[][]{san09,san10}.

The paper is organized as follows. First, we describe the systematic 
search for low metallicity galaxies (\S~\ref{kmeans}).
SDSS/DR7 renders 32 XMP candidates.  The physical
properties of the galaxies thus selected are analyzed in 
\S~\ref{properties}. These candidates, together with the rest of 
SDSS/DR7 galaxies,  allow us to compute the volume number density 
of low metallicity galaxies in the local universe
(\S~\ref{number_density}). In order to contextualize our work, 
we carried out a comprehensive search for XMP 
in the literature. The results are given in \S~\ref{biblio_search}.
A summary with conclusions and follow up work is
provided in \S~\ref{conclusions}.

\section{K-means based search for galaxies of low metallicity}\label{kmeans}

As we mention in the introduction, the spectrum
around H$\alpha$ is very sensitive to metallicity.
The ratio between the equivalent widths 
of [NII]$\lambda$6583 and H$\alpha$ goes
to zero with decreasing metallicity, as calibrated by, e.g.,
\citet{den02} or \citet{pet04}. We make use of this sensitivity
to select low metallicity galaxies classifying all the galaxies
with spectra in SDSS/DR7 according to their shape around
H$\alpha$. Those classes containing spectra where 
[NII]$\lambda$6583 turns out to be negligible small 
with respect to  H$\alpha$ will be saved as candidates.
For detailed information on the SDSS spectral catalog, see
\citet{sto02}, \citet{aba09}, as well as the comprehensive 
SDSS/DR7 web site\footnote{\tt http://www.sdss.org/dr7}.
The main properties of the catalog affecting our analysis are
the spectral resolution of the spectra, some $2000$ at H$\alpha$, 
and the fact that it includes all galaxies
up to an integrated $r$ magnitude of $17.8$.

\begin{figure}
\includegraphics[width=0.45\textwidth]{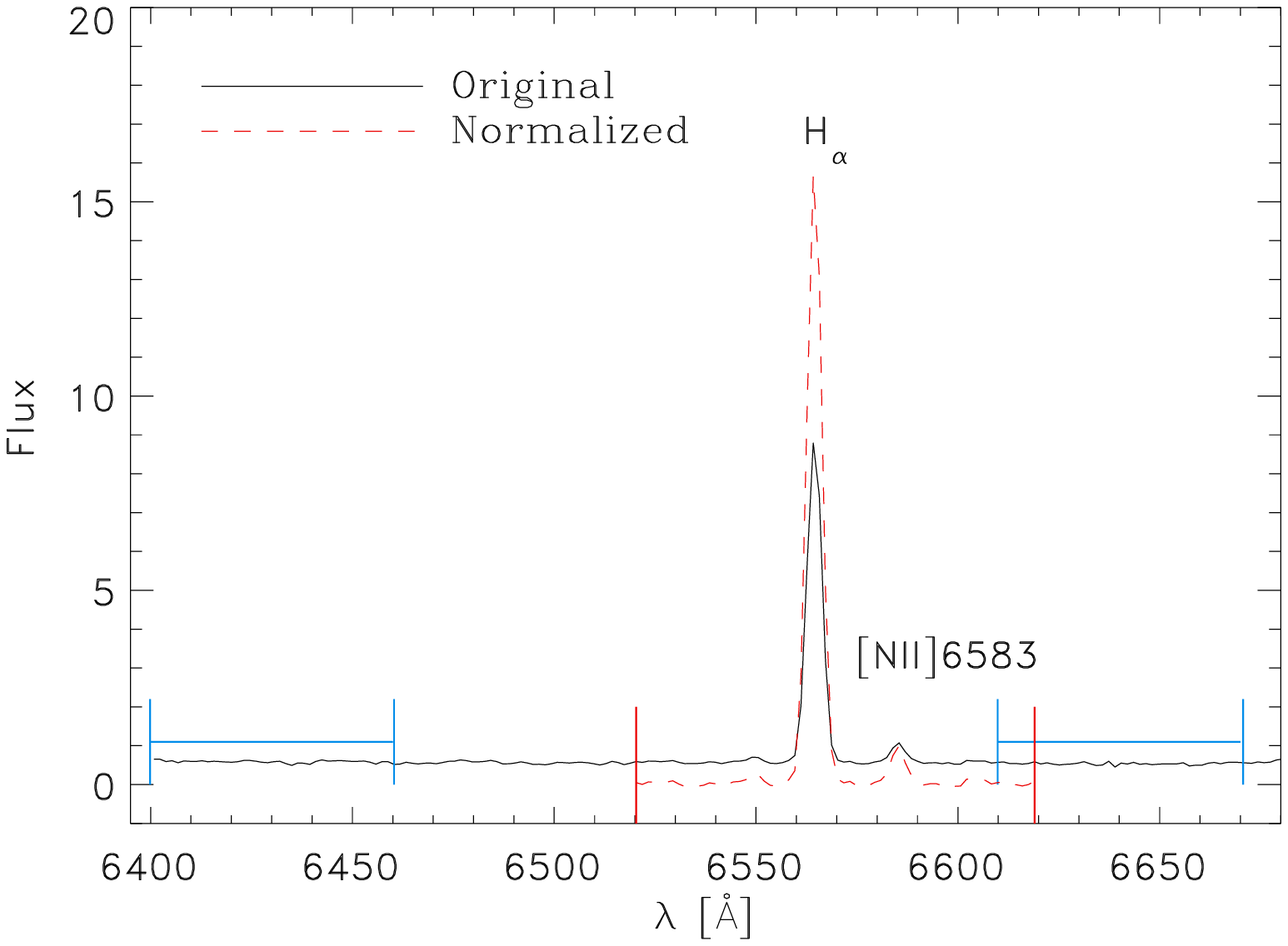}
\caption{
Spectral region classified to search for metal poor galaxies, which
should have [NII]$\lambda$6583 small compared to 
the nearby H$\alpha$ line. The original spectrum (the black solid line) 
is normalized before classification rendering the (red) dashed line.
The normalization includes removing the continuum, and dividing
the residual by the maximum intensity of  [NII]$\lambda$6583.
The blue solid bars at the extremes mark the intervals used to 
determine the continuum. The central 80\,\AA\ wide bar embraces 
the spectral region of interest. Wavelengths $\lambda$ are in \AA , 
and the spectra are given in arbitrary units.
}
\label{normalization}
\end{figure}

Since the classification must be based on the shape 
of the spectrum rather that on other incidental property 
(e.g., the galaxy luminosity), the spectra must go through
a previous normalization procedure.
The original spectra are shifted to restframe wavelength
using SDSS redshifts, and then normalized to the flux
in the $g$ color filter. In addition, the continuum around
H$\alpha$ is subtracted, and this spectrum is re-normalized
to the peak intensity of [NII]$\lambda$6583. Specifically,
if $I(\lambda)$ stands for the spectrum in restframe wavelengths
and normalized to $g$, then the spectrum to be classified
is $S(\lambda)$,
\begin{equation}
S(\lambda)={{I(\lambda)-I_c(\lambda)}\over{|I({\rm 6583\,\AA})-I_c({\rm 6583\,\AA})}|},
\label{def1}
\end{equation} 
where $\lambda$ stands for the wavelength, and $I({\rm 6583\,\AA})$ is the
intensity at the extreme of [NII]$\lambda$6583, i.e.,
the maximum if the line is in emission and the minimum if it is in absorption. 
The continuum intensity $I_c(\lambda)$ is derived by linear interpolation of 
the spectrum in two clean windows at both sides of the spectral region
of interest, namely, from 6400\,\AA\ to 6460\,\AA, and from  
6610\,\AA\ to 6670\,\AA.
Note that the denominator of equation~(\ref{def1}) is always 
positive so that the lines in absorption remain in absorption
after normalization.
We came across this normalization after some alternative trials. 
The normalization to [NII]$\lambda$6583 turns out to
maximize the contrast since it goes to zero for the low metallicity
galaxies we are trying to identify.
Figure~\ref{normalization} summarizes the normalization procedure.
It includes the original spectrum (the black solid line) as well as 
the two continuum windows used to determine the continuum intensity.
After normalization (i.e., after continuum removal plus division 
by the maximum intensity at [NII]$\lambda$6583)
it becomes the dashed line, which is the 80\,\AA\ wide spectral range 
undergoing  classification.

We employ the algorithm k-means for classification.
It treats the spectra as vectors in a $n$-dimensional
space, with $n$ the number of wavelengths.  
It is a rather standard technique in data-mining, machine learning, 
and artificial intelligence \citep[e.g.,][]{eve95,bis06},
and we have already successfully employed it for massive classification
of galaxy spectra \citep{san08,san10}. It works iteratively, 
starting from randomly chosen cluster centers.
Each galaxy spectrum is assigned to the cluster center
that is closest in a least-squares sense. Then the cluster
centers are recomputed as the average spectrum of all the
spectra in the class, and the assignation begins again.
The algorithm ends when no spectra are re-classified in two 
successive iterations. 
The main advantages are: 
(1) it is fast, simple, and robust, as requited to classify large data sets, 
(2) it guarantees that similar spectra end up in the same class, 
(3) it automatically yields the number of classes,
and (4) it provides spectra that are characteristic of the classes.
These template spectra are just the average of all the spectra in
the class, and they can be analyzed and interpreted as individual galaxies 
under the same assumptions followed when applying the popular 
{\em stacking} technique \citep[e.g.,][]{ell00}.
The drawbacks are: 
(1) the starting random seeds influence 
the classification, and such effect must be followed up and, eventually,
corrected for, and (2) the classes do not necessarily represent individual 
clusters in the classification space, but may be parts of clusters.  
The latter downside is not a problem in our particular application
since we are not interested 
in clustering but in separating spectra with different shapes. 
As for the former, one have to carry out several independent
trials to test the robustness of the classes on the 
initialization.

\begin{figure}
\includegraphics[width=0.45\textwidth]{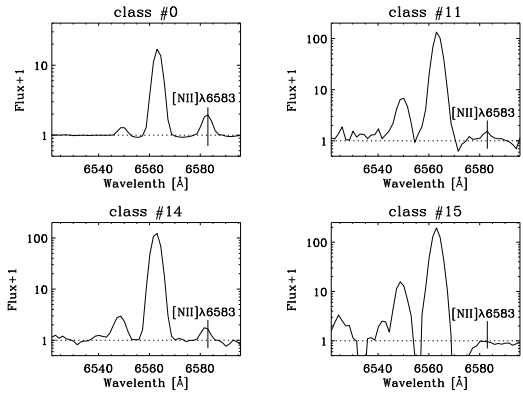}
\caption{
Several templates of the k-means classification
characterized by having H$\alpha\gg {\rm[NII]}\lambda$6583,
which may correspond to classes of
low metallicity galaxies.  The individual
spectra have their continua removed, and have been 
normalized to the peak intensity of [NII]$\lambda$6583.
In order to be plotted in a logarithm scale, the spectra have been 
artificially uplift by one unit.
These classes include spectra where the SDSS pipeline has 
artificially removed  [NII]$\lambda$6583 creating a mock low 
metallicity galaxy (e.g., class\,\#\,15 is mostly formed by these 
spectra). 
Wavelengths are given in \AA .
}
\label{second_class}
\end{figure}

\begin{figure}
\includegraphics[width=0.5\textwidth]{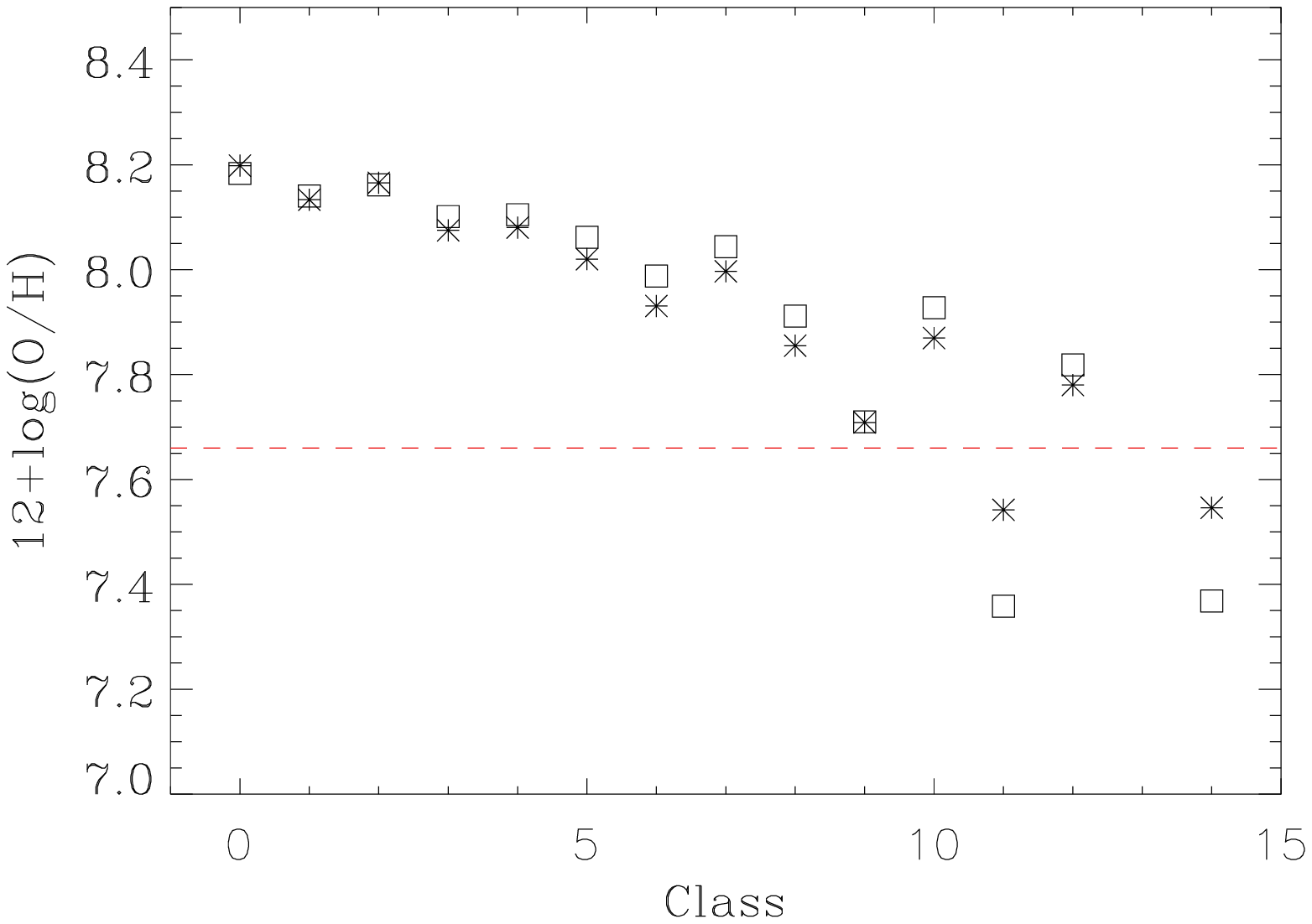}
\caption{
Oxygen abundance of the various classes as inferred using the N2~method 
on the template spectra corresponding to each one the classes. 
The two types of symbols correspond to  the two slightly different calibrations 
in \citet{pet04} -- asterisks for equation~(\ref{pet1}) and boxes
for equation~(\ref{pet2}). The horizontal dashed line represents
one-tenth of the solar metallicity \citep{asp05b}.
}
\label{oxy_abu}
\end{figure}

We apply k-means to the galaxies with spectra 
having redshift $\le 0.25$, which is equivalent to 
classifying all the $\sim9\cdot 10^5$ galaxies with
spectra in SDSS/DR7 since the low 
metallicity targets are expected to be faint
(\S~\ref{intro}) and, consequently they cannot be observed at 
high redshift given the apparent magnitude threshold imposed by SDSS.
(For a galaxy at redshift $\ge$ 0.25 to be observed
spectroscopically,
the absolute $r$ magnitude must be smaller than 
$-22.3$.)
The k-means procedure is applied in two successive steps, because a single
application renders classes too coarse to separate low metallicity
galaxies. In the first application, we remove classes having H$\alpha$ 
in absorption, as well as those whose [NII]$\lambda$6583 is much too 
large. Actually the limit is not imposed directly on [NII]$\lambda$6583,
but on the equivalent oxygen abundance deduced via the so-called
N2-method \citep[e.g.,][]{den02,2005A&A...437..849S} which
depends, exclusively, on the ratio between the equivalent 
widths of [NII]$\lambda$6583 and H$\alpha$,
\begin{equation}
{\rm N2}=\log\big(W_{\rm[NII]\lambda6583}\big/W_{\rm H\alpha}\big).
\label{n2def}
\end{equation}
In the calibration by \citet{pet04}, the oxygen abundance is, 
\begin{equation}
12+\log{\rm (O/H)}\simeq 8.90+0.57\,{\rm N2},
\label{pet1}
\end{equation} 
or alternatively,
\begin{equation}
12+\log{\rm (O/H)}\simeq 9.37+2.03\,{\rm N2}+1.26\,{\rm N2}^2+0.32\,{\rm N2}^3.
\label{pet2}
\end{equation}
Galaxies belonging to classes whose template spectrum has
$12+\log{\rm (O/H)}\le 8.2$ were used for a second k-means run. 
This arbitrary threshold was chosen as a trade off that 
removes enough high metallicity objects, yet allows the second
classification to choose from a broad enough pool. The subset 
undergoing  the second classification contains 
some 5000 galaxies, which correspond to only 0.6\% of the original 
set. 
This second classification renders classes with 
a spectrum characteristic of low metallicity, i.e.,
with H$\alpha\gg {\rm[NII]}\lambda$6583.
Several templates of the classes thus obtained are included 
in Fig.~\ref{second_class}. It shows genuine low metallicity
classes (e.g., class~\#\,14) as well as some others where
the SDSS skyline removal pipeline has artificially cut out 
[NII]$\lambda$6583 (e.g., class~\#\,15). 
The finding of these fake low metallicity galaxies shows a
weakness of the searching technique -- some SDSS/DR7 galaxies 
with negligible [NII]$\lambda$6583 may not be low metallicity
after all,
and the selection must be refined even further. However, this downside
incidentally proves the procedure to be working properly, 
since we know it managed to identify spectra without 
[NII]$\lambda$6583. The problem arises as to how to separate 
genuine from fake low metallicity galaxies. 
Fortunately, it can be sorted out easily as we explain below. 
%
Figure~\ref{oxy_abu} shows the oxygen abundance corresponding
to the templates of the classes inferred from this second step,
discarding those classes where the template shows obvious 
malfunctioning of the SDSS pipeline (i.e., like class~\#\,15 in 
Fig.~\ref{second_class}). 
They are based on the two slightly different calibrations by
\citet{pet04} given in equations~(\ref{pet1}) and (\ref{pet2}).
We use it to select the galaxies in classes \#~11 and \#~14  as
XMP galaxy candidates. The selection 
criteria are somewhat arbitrary but we choose these two classes 
because (1) the metallicity estimated for the template using the
N2-method are smaller than one-tenth 
solar\footnote{Or $W_{\rm H\alpha}/W_{\rm[NII]\lambda6583}\ge 140$, 
using the calibration in equation~(\ref{pet2}) and assuming 
$12+\log({\rm O/H})_\odot=8.65$ \citep{asp05b}, as invoked by \citet{pet04}.}, 
which is the reference value for XMP galaxies (\S~\ref{intro}), (2) there seems to be a gap 
in metallicity between these two classes and the rest 
(see Fig.~\ref{oxy_abu}), and (3) the number of galaxies
the classes contain is small enough to allow a detailed
inspection of the individual spectra. The original 49 spectra included in 
these two classes were visually inspected to discover that 20 of them 
were faults produced by the SDSS pipeline.
These spectra are easily recognized because the pipeline 
linearly interpolates the spectrum around 
${\rm[NII]\lambda6583}$,
and the presence of such a straight line boldly contrasts with the 
rest of the spectrum shape.
They were
excluded. Since faulty spectra ended up in low-metal classes, we
checked the faulty classes for truly XMP spectra sneaked in. 
Most of the spectra were indeed failures from the pipeline,
but we rescued three particularly low
[NII]$\lambda$6583 spectra included in these classes. 
All in all,
our selection rendered 32 galaxies. Their coordinates and the
unique identifier of the galaxy spectrum in SDSS/DR7 are listed in 
Table~\ref{list1}.
The table also includes a column with the oxygen metallicity
inferred applying equation~(\ref{pet2}) to the 
equivalent widths measured on the individual SDSS spectra. 
These values are only tentative since the prescription
has an intrinsic scatter as large as 0.3~dex for individual
galaxies \citep[e.g.,][]{pet04,2005A&A...437..849S,2009MNRAS.398..949P}.
Keeping this caveat in mind, the mean $12+\log({\rm O/H})$
of the candidates is 7.60, with a standard deviation of 0.22.
As we will see in \S~\ref{biblio_search}, this estimate is in
agreement with the metallicity of the subset of candidates whose 
abundance has been derived with more precise means.
\begin{deluxetable}{llccccccc}
\tablewidth{0pc}
\tabletypesize{\scriptsize}
\tablecaption{XMP candidates found by classifying 
all SDSS/DR7 galaxies according to their spectra 
around H$\alpha$.}
\tablehead{
\colhead{Index\tablenotemark{a}}&
\colhead{Name}&
\colhead{$g$}&	
\colhead{Redshift}&
\multicolumn{2}{c}{$12+\log({\rm O/H})$}&	
\colhead{SpecObjID\tablenotemark{b}}&	
\colhead{Comment\tablenotemark{c}}\\
&&&&d&e&
}
\startdata
  1 & SDSS J003630.40+005234.6 & 18.8 & 0.028 & 7.64 &--& 194442194734546944 & single knot\\
  2$^\star$ & SDSS J012534.19+075924.4 & 18.4 & 0.010 & 7.60 & 7.60 & 655785970125242368 & knotted
cometary\\
  3 & SDSS J015809.39+000637.2 & 18.1 & 0.012 & 7.75 &--& 302812946306695168 & cometary\\
  4$^\star$ & SDSS J030331.27-010947.1 & 19.7 & 0.030 & 7.48 & 7.22 & 225967511814799360 &
cometary\\
  5 & SDSS J031300.05+000612.1 & 19.2 & 0.029 & 7.82 &--& 299996706848636928 & cometary\\
  6$^\star$ & SDSS J080840.85+172856.4 & 19.2 & 0.044 & 7.36 & 7.48 & 585978593608204288 & single
knot \\
  7 & SDSS J082540.45+184617.2 & 19.0 & 0.038 & 7.75 &--& 640023302407979008 & single knot  \\
  8 & SDSS J084236.58+103313.9 & 17.7 & 0.011 & 7.58 &--& 724467307371298816 & cometary  \\
  9$^\star$ & SDSS J093402.03+551427.7 & 16.4 & 0.002 & 6.88 & 7.17 & 156443095175528448 &
2-knot cometary\\
  10 & SDSS J094254.27+340411.8 & 19.1 & 0.023 & 7.67 &--& 547698126572486656 & cometary?  \\
  11 & SDSS J100348.66+450457.7 & 17.5 & 0.009 & 7.65 &--& 265375788289753088 & single knot  \\
  12$^\star$ & SDSS J101624.52+375445.9 & 15.9 & 0.004 & 7.61 & 7.58 & 401892408779866112 &
cometary  \\
  13$^\star$ & SDSS J103137.28+043422.0 & 16.2 & 0.004 & 7.52 & 7.70 & 162635977557278720 &
cometary  \\
  14$^\star$ & SDSS J104457.80+035313.1 & 17.5 & 0.013 & 7.01 & 7.44 & 162917331083722752 &
cometary  \\
  15$^\star$ & SDSS J111934.34+513012.1 & 16.8 & 0.004 & 7.75 & 7.51 & 247359886311030784 & 2-knot
cometary  \\
  16 & SDSS J114506.25+501802.4 & 17.8 & 0.006 & 7.71 & -- & 272412374642720768 & cometary  \\
  17$^\star$ & SDSS J115132.94-022222.0 & 16.7 & 0.004 & 7.58 & 7.78 & 93111671593107456 & 2-knot
cometary  \\
  18 & SDSS J115754.18+563816.7 & 16.9 & 0.001 & 7.83 &--& 369803376499621888 & cometary  \\
  19$^\star$ & SDSS J120122.31+021108.3 & 17.6 & 0.003 & 7.69 & 7.49 & 145464500043120640 &
cometary  \\
  20$^\star$ & SDSS J121402.48+534517.4 & 17.3 & 0.003 & 7.54 & 7.64 & 287049377199947776 &
cometary  \\
  21$^\star$ & SDSS J123048.60+120242.8 & 16.7 & 0.004 & 7.78 & 7.73 & 454810434122285056 &
cometary  \\
  22$^\star$ & SDSS J125526.07-021334.0 & 19.1 & 0.052 & 7.73 & 7.83 & 95080395053203456 & single
knot  \\
  23$^\star$ & SDSS J132347.46-013252.0 & 18.1 & 0.023 & 7.21 & 7.78 & 96204976459612160 & single
knot  \\
  24 & SDSS J132723.29+402204.1 & 19.0 & 0.012 & 7.67 &--& 412307391283986432 & single knot  \\
  25$^\star$ & SDSS J133126.91+415148.3 & 17.1 & 0.012 & 7.64 & 7.75 & 412307391565004800 &
cometary?  \\
  26$^\star$ & SDSS J135525.66+465151.3 & 19.2 & 0.028 & 7.86 & 7.63 & 361921789577658368 &
cometary?  \\
  27 & SDSS J141851.13+210239.7 & 17.6 & 0.009 & 7.64 &--& 784423532984532992 & cometary  \\
  28$^\star$ & SDSS J142342.88+225728.7 & 17.8 & 0.033 & 7.67 & 7.72 & 600334402043510784 & single
knot  \\
  29$^\star$ & SDSS J150934.17+373146.1 & 17.3 & 0.033 & 7.74 & 7.85 & 394011865673367552 &
cometary?  \\
  30$^\star$ & SDSS J164710.66+210514.5 & 17.2 & 0.009 & 7.74 & 7.75 & 442143986740625408 &
knotted cometary  \\
  31$^\star$ & SDSS J223831.12+140029.7 & 18.9 & 0.021 & 7.64 & 7.45 & 533060792673632256 &
2-knot cometary  \\
  32$^\star$ & SDSS J230210.00+004938.8 & 18.7 & 0.033 & 7.57 & 7.71 & 190784502518251520 &
2-knot cometary  \\
\enddata
\tablenotetext{a}{Those indexes marked with an asterisk correspond to 
known XMP galaxies. The rest are new.}
\tablenotetext{b}{Unique identifier of the galaxy spectrum in SDSS/DR7.}
\tablenotetext{c}{Sketch of galaxy shape.}
\tablenotetext{d}{
Using the calibration by \citet{pet04}, given in our equation~(\ref{pet2}).
}
\tablenotetext{e}{
From the literature as listed in Table~\ref{revision}.
}
\label{list1}
\end{deluxetable}

%

K-means renders final classes that depend on the random initialization.
This problem is not particularly severe in our application, since
we are not trying to find clusters. We just try to separate 
spectra with a particular shape, independently on whether they
appear on a single class or in several classes. Therefore,
we did not expected the random initialization to represent
a serious problem, yet we study the impact on the 
selected spectra by repeating the k-means classification 
5~times. Classes with one-tenth solar metallicity are always present.
They contain most of the 49 spectra in the
original
classification purged later on. 
Specifically, the two classes of lowest metallicity always 
include between 79\% and 93\% of the spectra in the  
classification.

Our search is systematic. If SDSS/DR7 has spectra
with H$\alpha\gg {\rm[NII]}\lambda$6583, they will 
appear in one of the classes. The assumption that we detect all
the spectra with this property is used 
in \S~\ref{number_density} to estimate the number density of XMP
galaxies in the local universe. Deviations from this assumption are 
also analyzed in \S~\ref{number_density}.

\section{Properties of the XMP candidates}\label{properties}

As pointed out in \S~\ref{intro}, the known XMP galaxies 
tend to be BCD galaxies, which are characterized by blue colors, 
high surface brightness, and low luminosity. Obviously, 
the metal-poor galaxies represent only a small fraction of the BCDs. 
In order to explore the physical properties of our candidates,
we use the diagnostics developed to identify BCDs in large galaxy
samples, to see whether our candidates conform to the properties
of known XMP galaxies. Following \citet{san08}, which condense
the criteria by \citet{gil03} and \citet{mal05},  BCDs
can be observationally characterized by,
\begin{center}
\begin{tabular}{ccc}
(a)& $\mu_g-\mu_r \leq 0.43$ mag arcsec$^{-2}$,\\
(b)&$\mu_g < 21.83 - 0.47 (\mu_g-\mu_r)$ mag arcsec$^{-2}$,\\
(c)& $M_g > -19.12+1.72 (M_g-M_r)$ mag,\\
(d)& $W_{{\rm H}\alpha} > 50~$\AA,\\
(e)& $12+\log({\rm O/H}) < 8.2$ ($\equiv$  1/3 $\odot$),\\
(f)& no AGN,\\
(g)& no bright galaxy within 10 $R_{50}$.
\end{tabular}
\end{center}
The various new symbols have their usual meaning --
$\mu_g$ and $\mu_r$ stand for the mean surface brightness in the 
SDSS filters $g$ and $r$, $M_g$ and $M_r$ represent
the absolute 
magnitudes in these two filters, and $R_{50}$ is
the radius containing 50\% of the galaxy light.
The criteria above assure that (a) BCDs are blue, (b) they
have high surface brightness, (c) they are dwarf, (d) they
have a large star-formation rate,  (e) they are
metal poor, (f) they are not confused 
with active galactic nuclei (AGN), and (g) they have no close companions.
%
\begin{figure*}
\includegraphics[width=0.7\textwidth,angle=90]{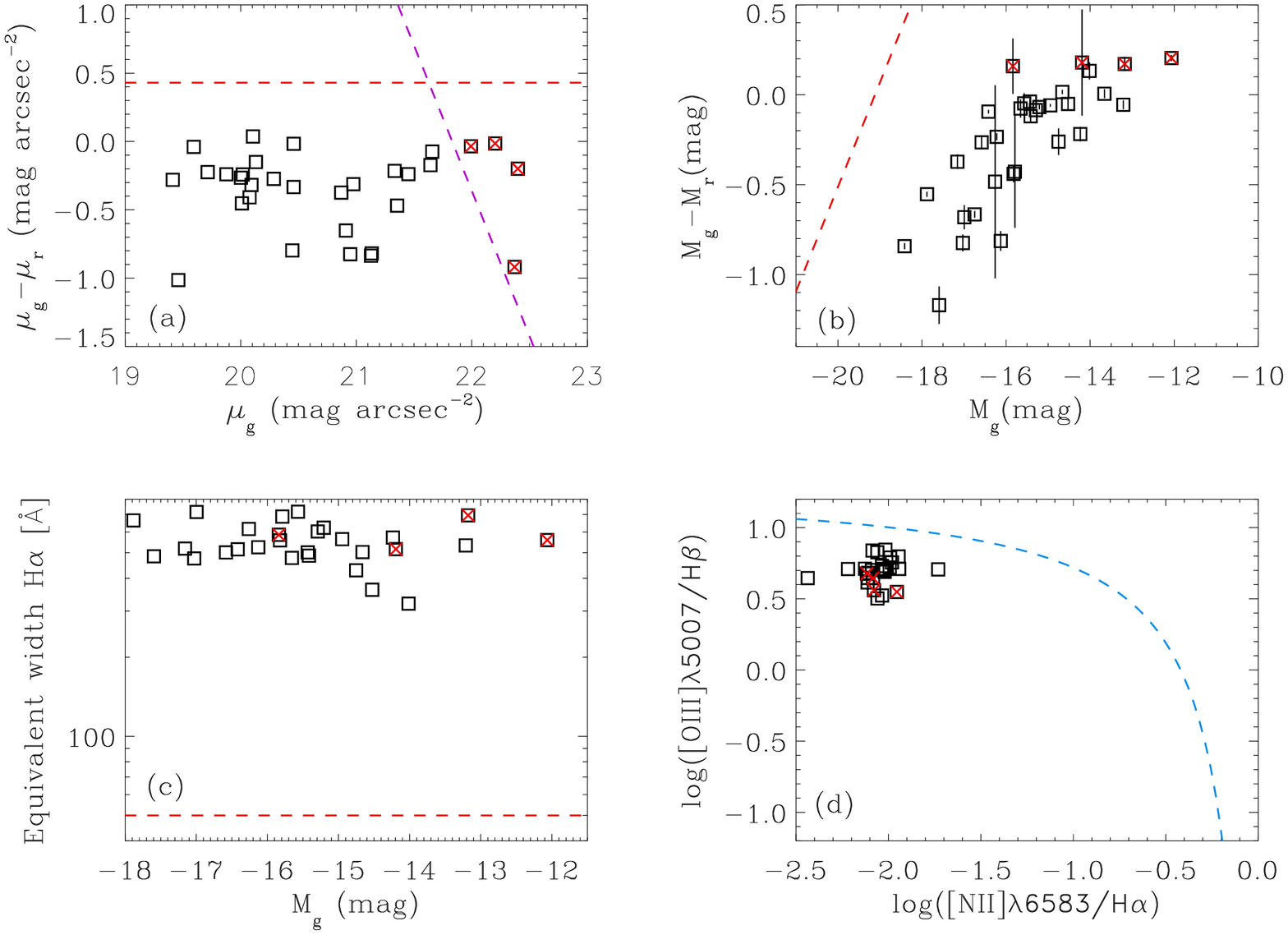}
\caption{
Diagnostic plots to characterize the XMP galaxies resulting from the search.
(a) Color $\mu_g-\mu_r$ vs surface brightness in the $g$ filter
$\mu_g$. Only 4 out of the 32 galaxies have 
a surface brightness beyond the slanted line that bounds BCD galaxies. 
These four outliers are
marked with times symbols (in red) in all the plots.
(b) Color $M_g-M_r$ vs absolute magnitude $M_g$ in the 
$g$ filter. All the galaxies are consistent with being BCDs.
The error bars have been taken directly from SDSS.
(c)  H$\alpha$ equivalent width vs absolute magnitudes.
The star-formation rate inferred from these equivalent widths
exceeds in all cases the BCD threshold 
(the dashed line).
(d) BPT diagram showing all the targets to be starforming 
galaxies as opposed to AGNs 
\citep[separated by the dashed curve worked out by][]{kau03}.
}
\label{diag_bcd}
\end{figure*}
%
Figure~\ref{diag_bcd} shows various scatter plots relating the physical 
parameters involved in BCD characterization. The required
colors and sizes have been taken directly from the SDSS/DR7 
database using Petrosian magnitudes \citep[][]{sto02,aba09}.
The color-magnitude plot, Fig.~\ref{diag_bcd}b, shows
that all our XMP candidates are dwarf -- the dashed line represents 
constraint (c). Their H$\alpha$ equivalent width 
exceeds the required threshold (constraint d), and the galaxies 
reside in a region of the BPT diagram \citep[after][]{bal81}
discarding the AGN nature -- the 
dashed line in Fig.~\ref{diag_bcd}d divide AGN
activity and star-forming activity as
worked out by \citet{kau03}. The XMP candidates obviously fulfill
criterion (e) (see Fig.~\ref{oxy_abu}), and they are
also blue-enough, staying below the horizontal dashed line in 
Fig.~\ref{diag_bcd}a corresponding to criterion (a). 
As far as the compactness criterion
(b) is concerned, most galaxies comply with it, being to the
left of the slanted dashed line in Fig.~\ref{diag_bcd}a.
Four of them do not -- they are represented as asterisks in 
Fig.~\ref{diag_bcd}a, as well as in the remaining
panels of the figure. 
The low surface brightness of these galaxies is to some extent
apparent. Visual inspection of their SDSS composite-color images
show two of them to present a marked cometary shape or double nucleus, 
and the SDSS reduction pipeline has probably overestimated the apparent 
size of those galaxies.

Spurred by the unusual shape of these non-BCD
galaxies, we carry out an eyeball inspection of all
the 32 targets. Many of them turned out to have distorted morphologies, 
often with cometary  shape, knots and/or chains of knots. 
Only 8 appear as a single knot without obvious substructure. 
The results of this crude morphological
classification are given in Table~\ref{list1}.
Figure~\ref{mugshot} includes SDSS mugshots of 
all XMP candidates not marked as {\em single knot}
in Table.~\ref{list1}. Those galaxies tagged with a question
mark in Table~\ref{list1} represent ambiguous
cases where the cometary shape is less obvious, 
however, even in theses cases the images show elongated 
asymmetrical substructure (e.g., \#~29).
\begin{figure*}
\includegraphics[width=0.24\textwidth]{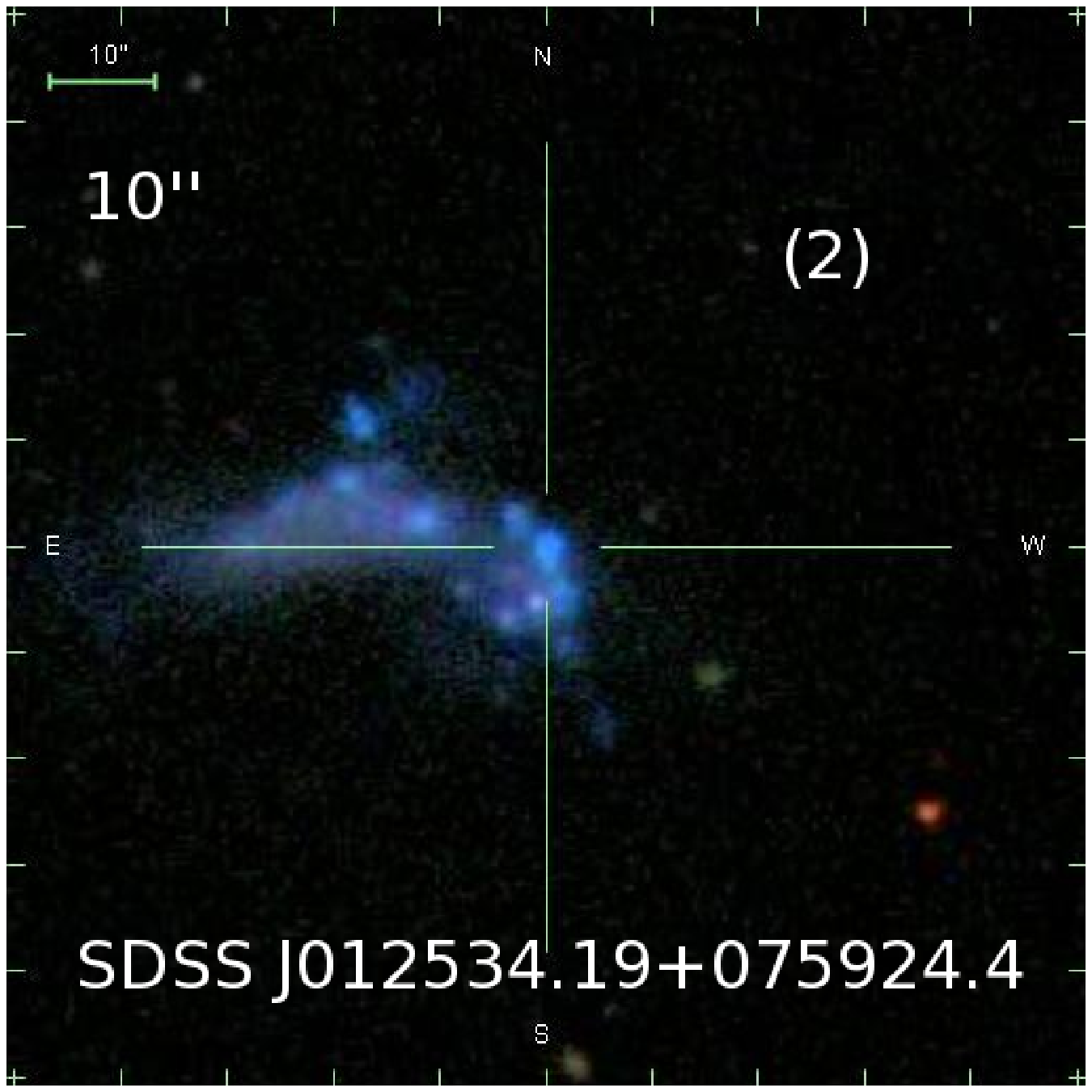}
\includegraphics[width=0.24\textwidth]{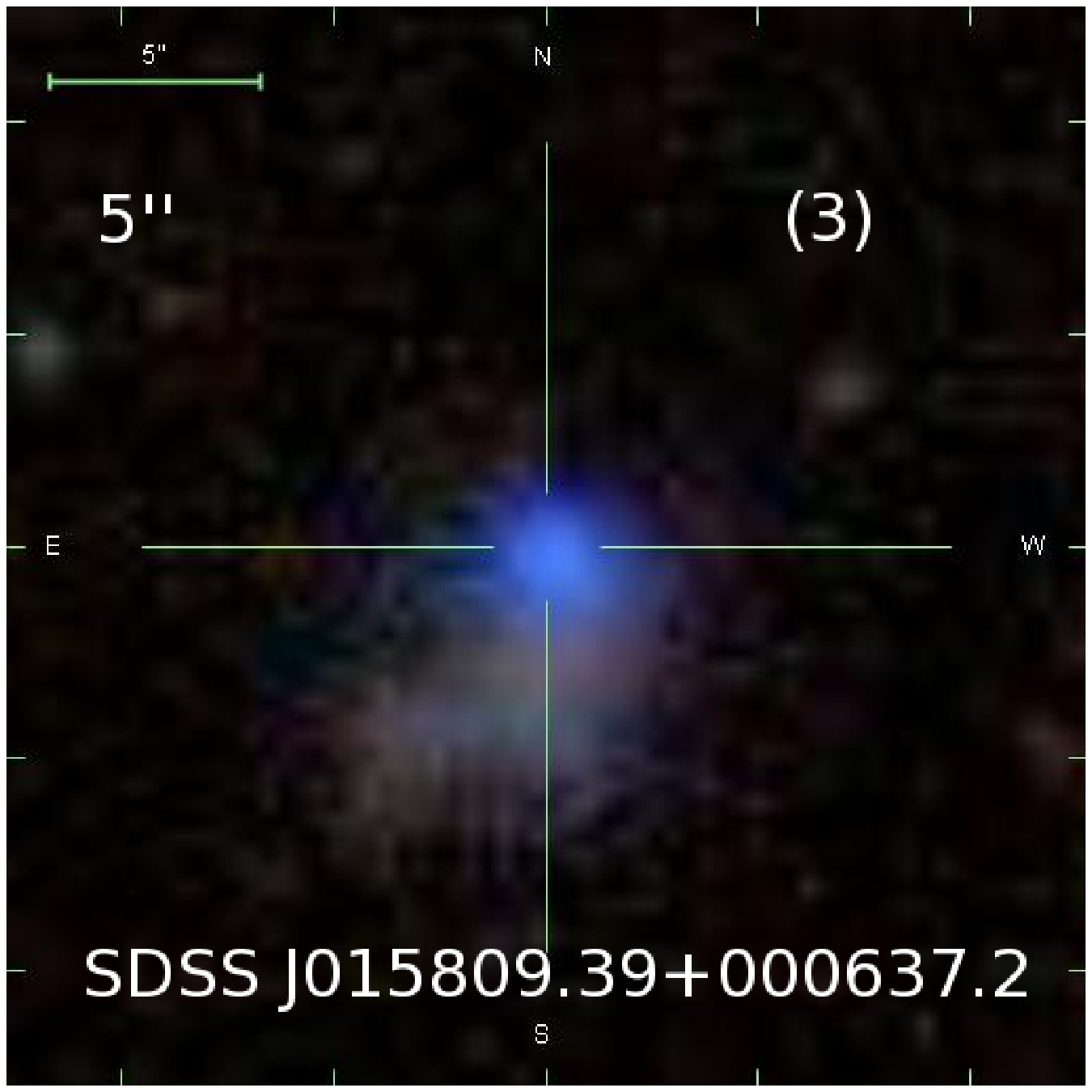}
\includegraphics[width=0.24\textwidth]{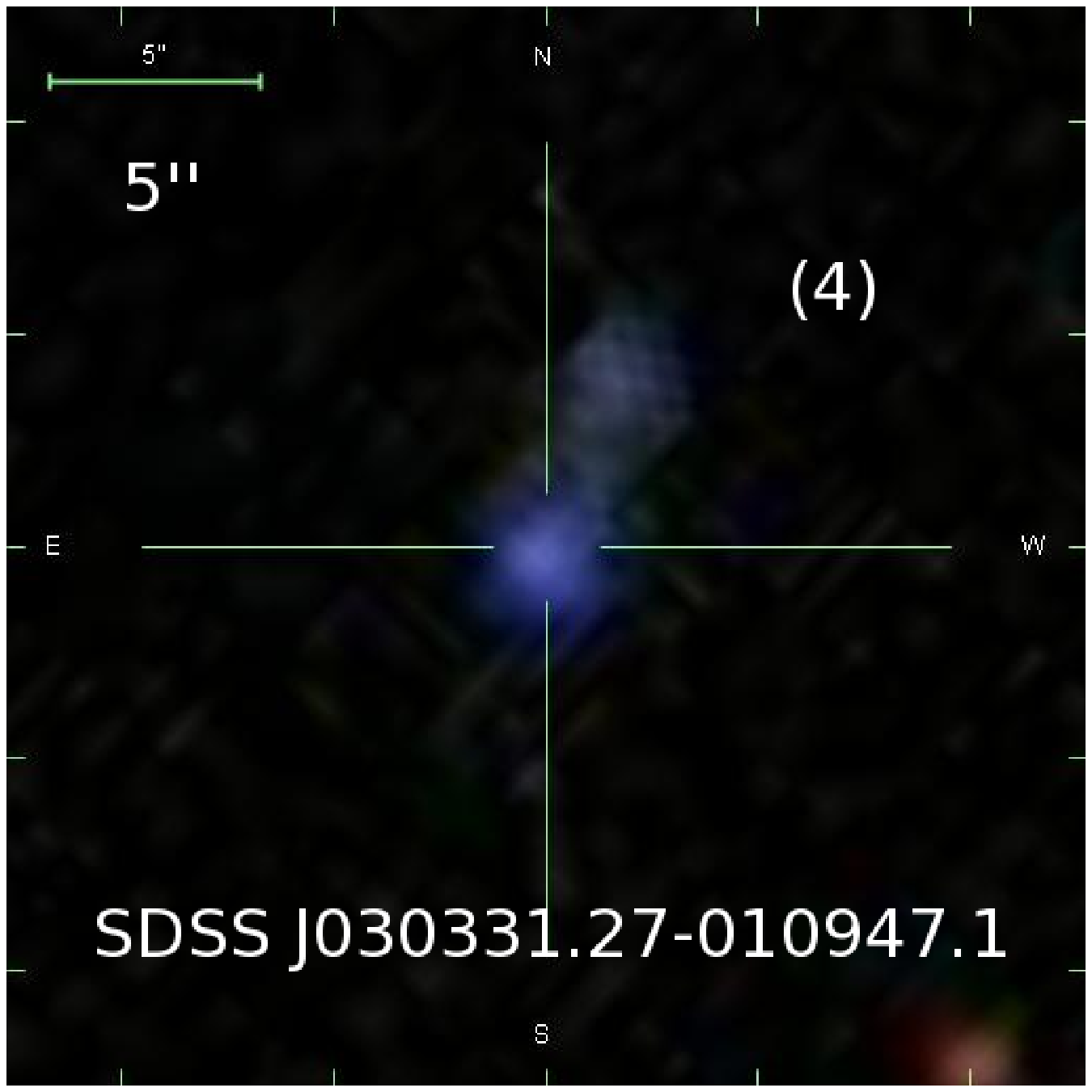}
\includegraphics[width=0.24\textwidth]{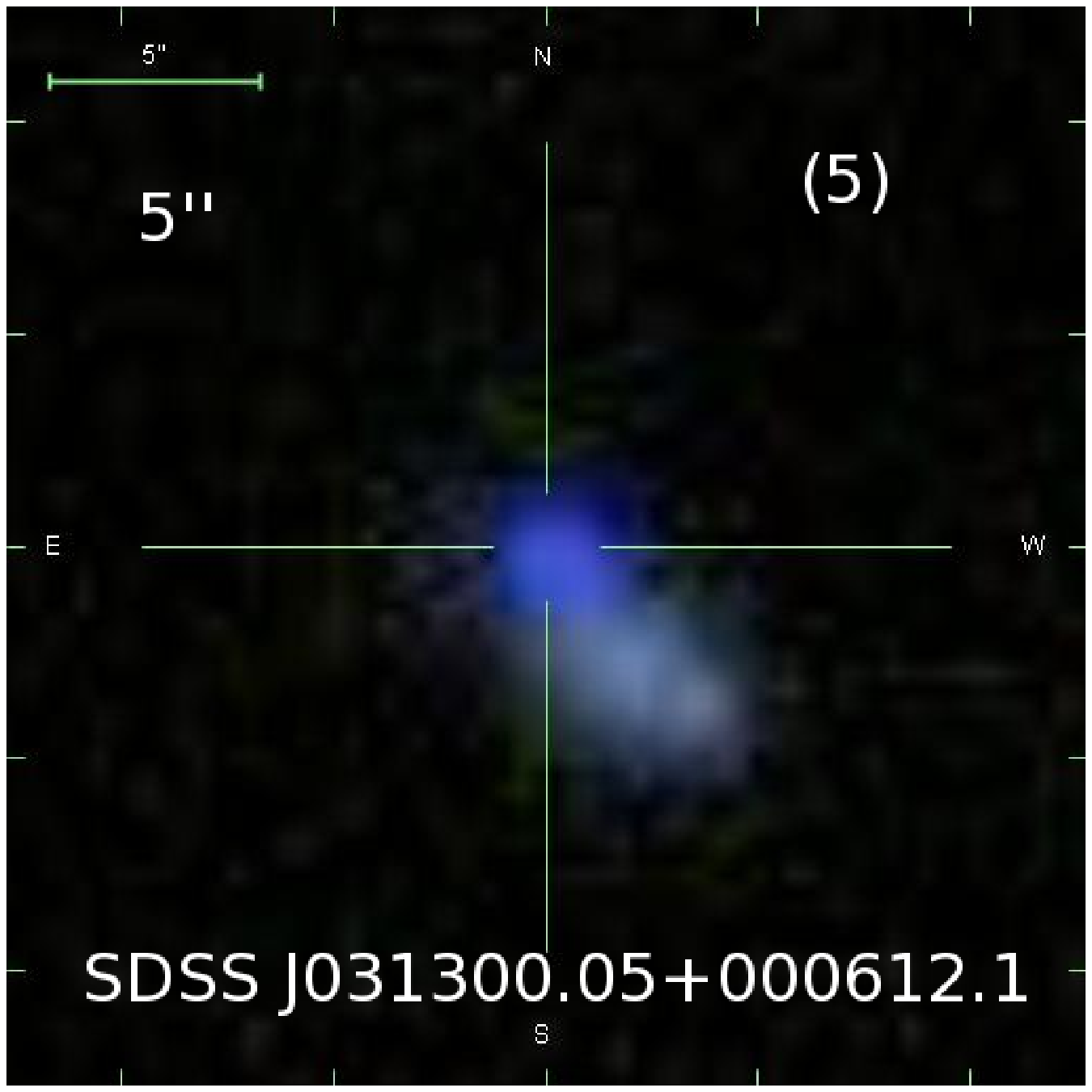}\\
\includegraphics[width=0.24\textwidth]{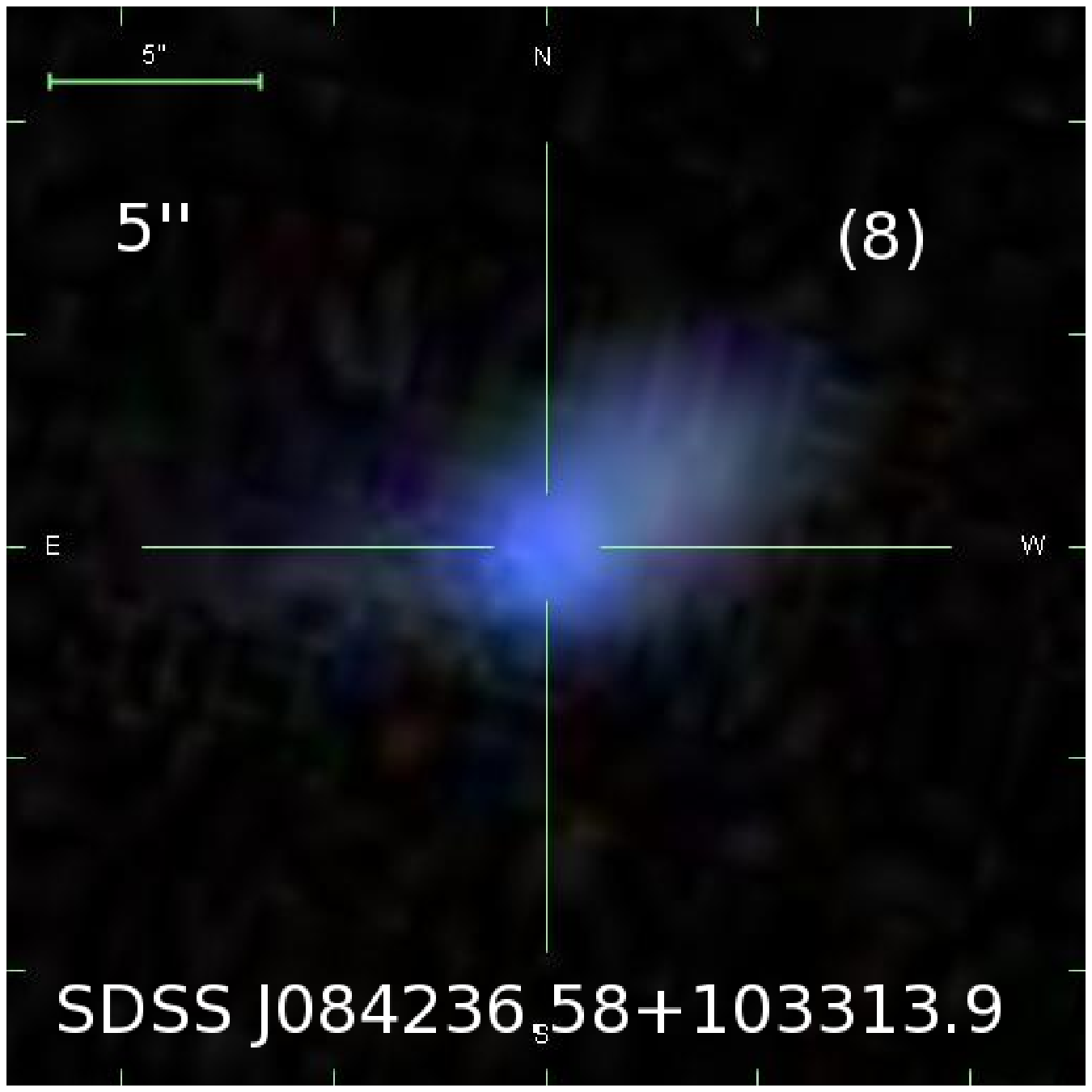}
\includegraphics[width=0.24\textwidth]{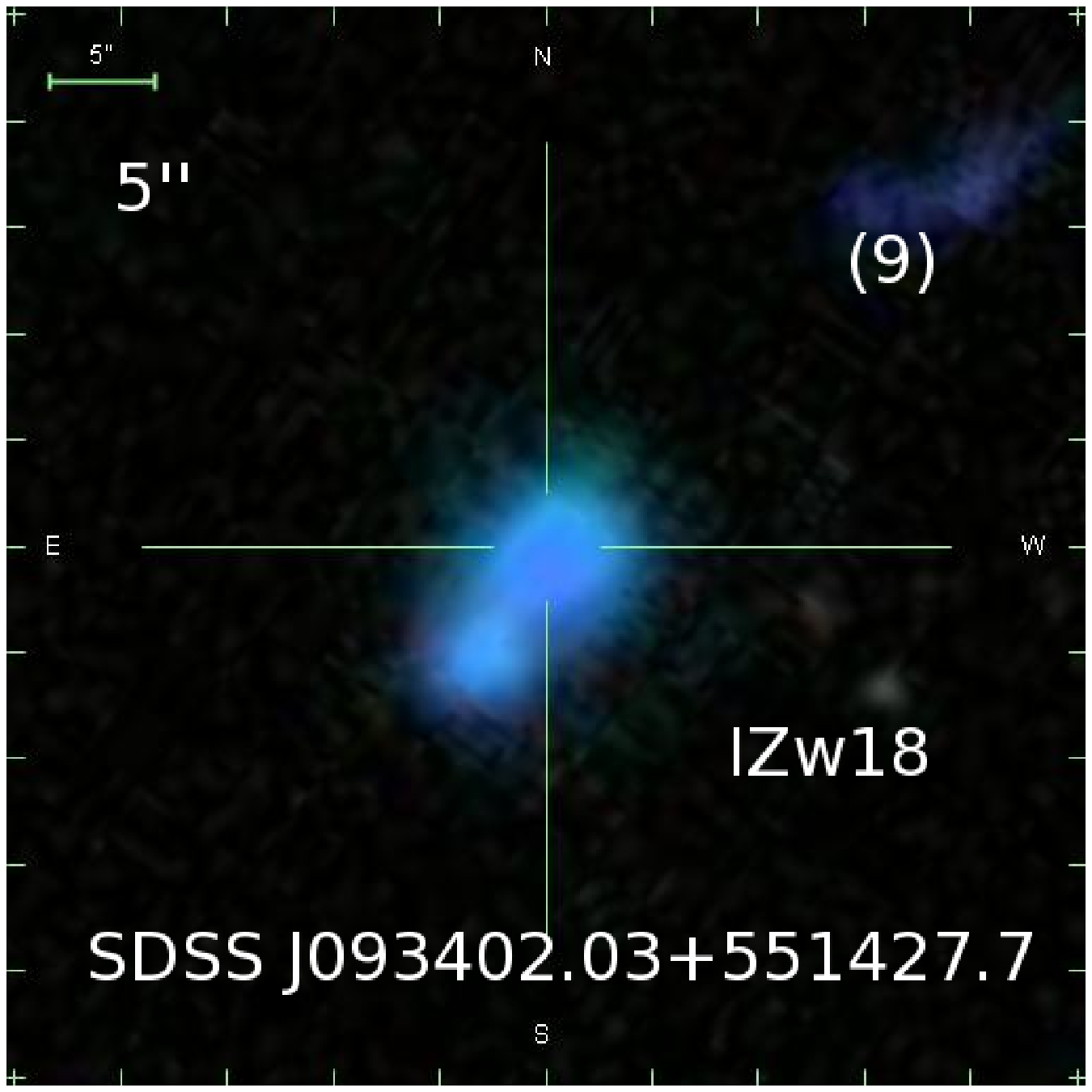}
\includegraphics[width=0.24\textwidth]{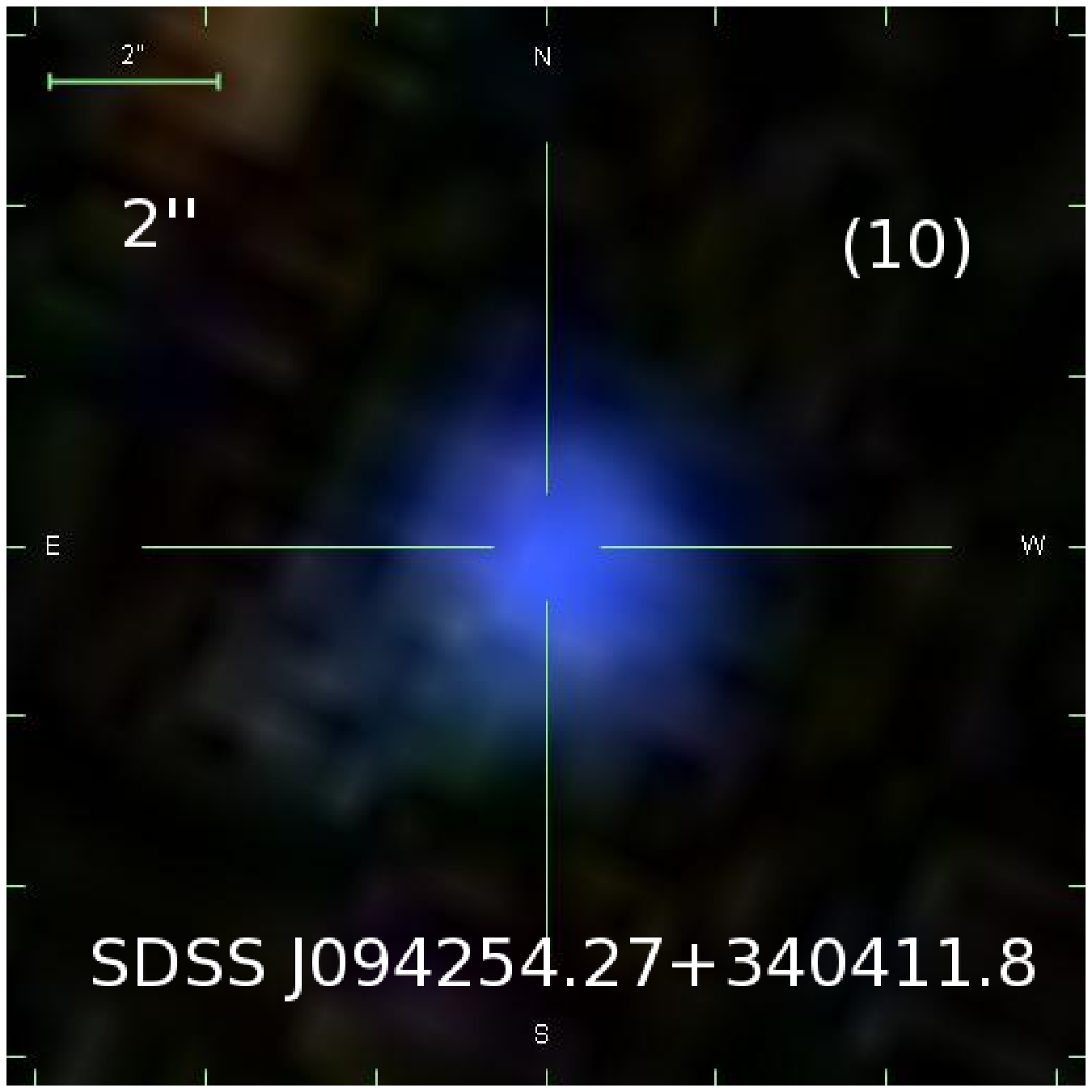}
\includegraphics[width=0.24\textwidth]{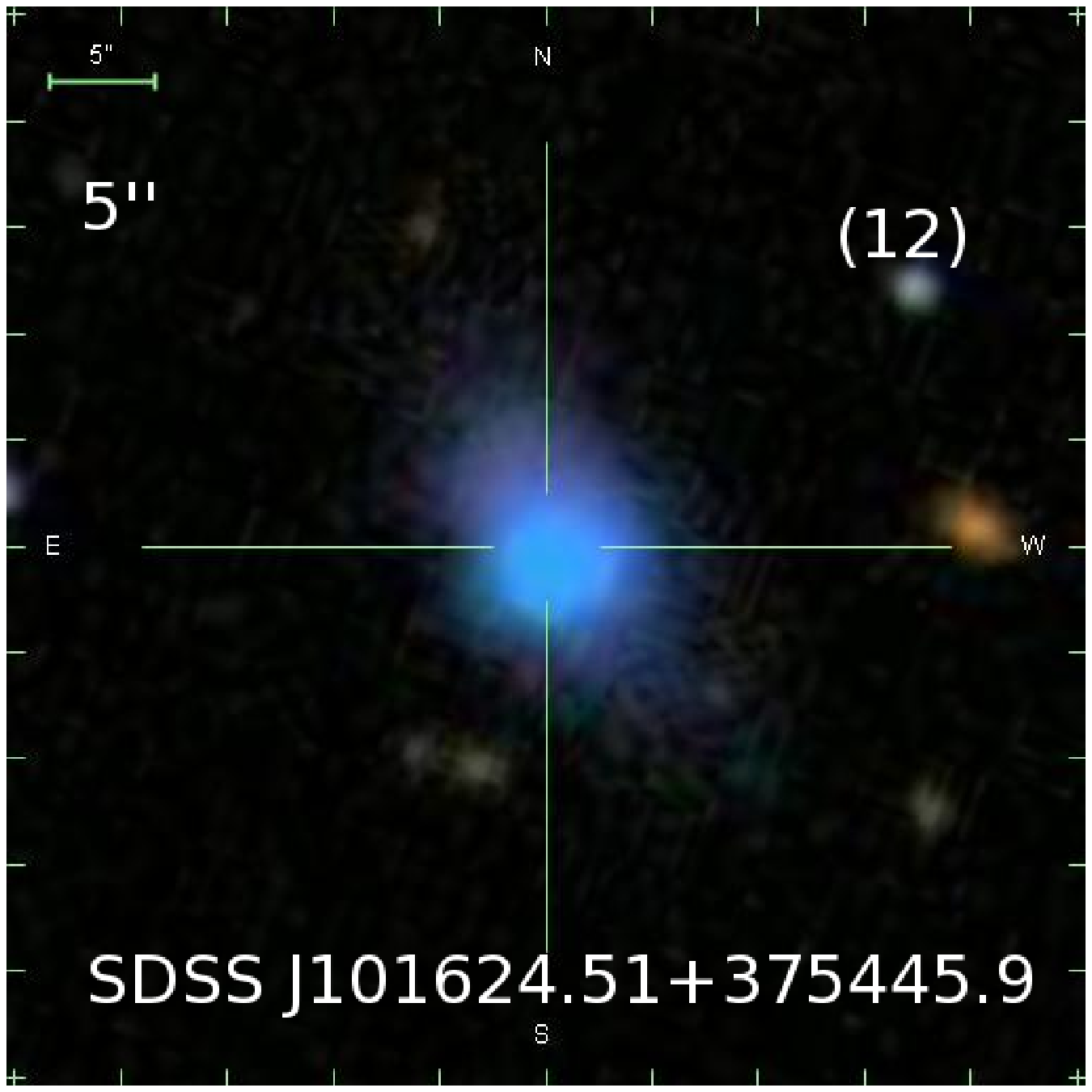}\\
\includegraphics[width=0.24\textwidth]{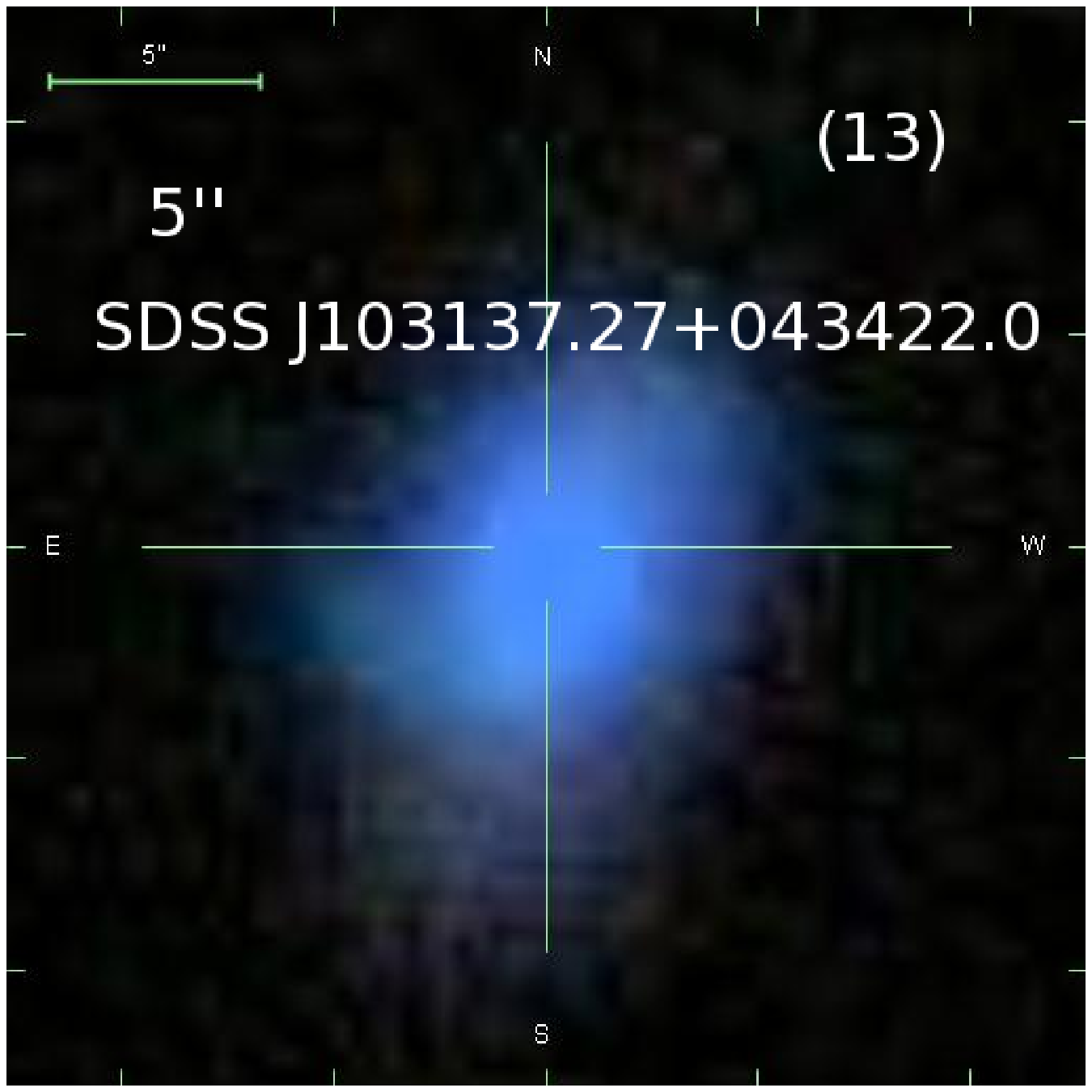}
\includegraphics[width=0.24\textwidth]{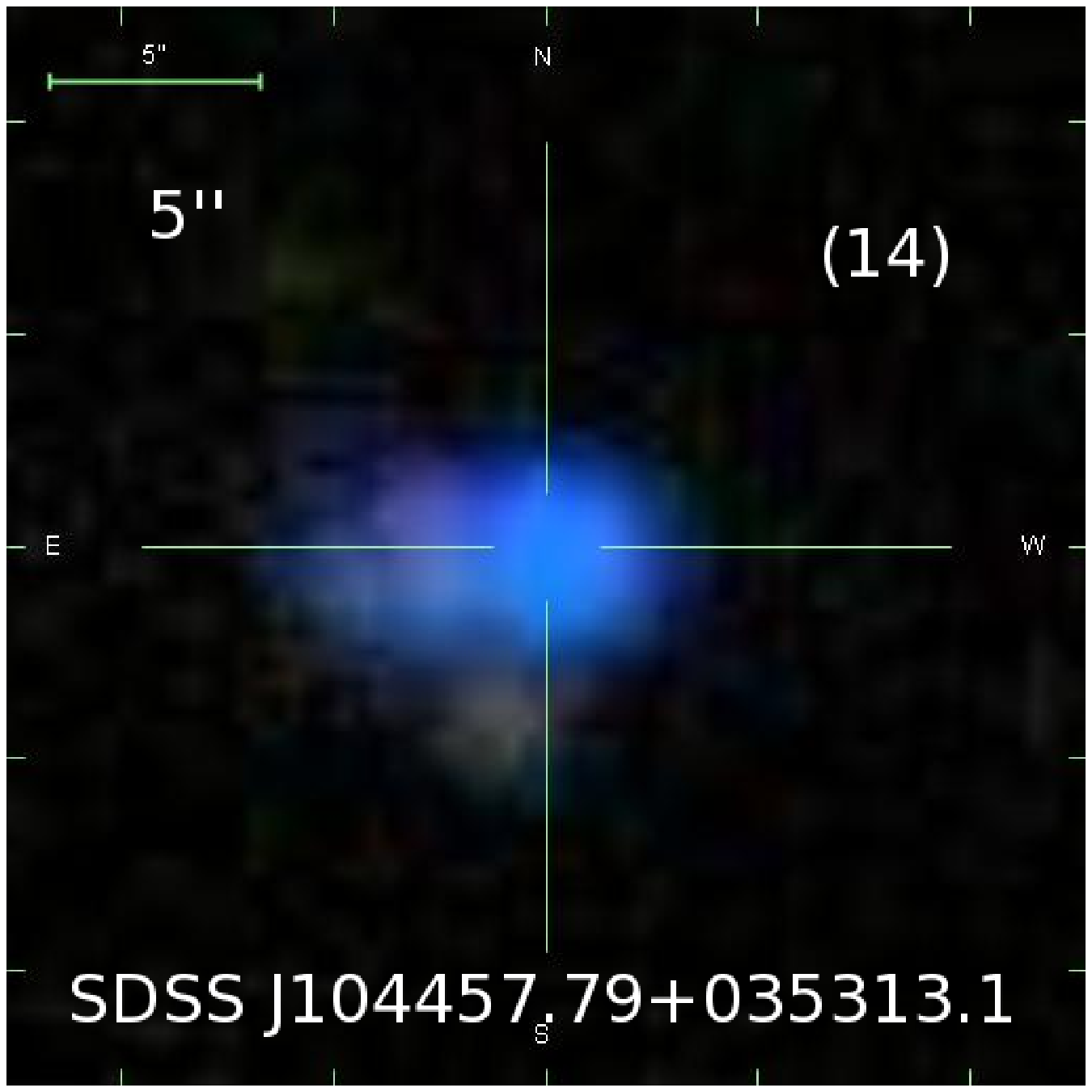}
\includegraphics[width=0.24\textwidth]{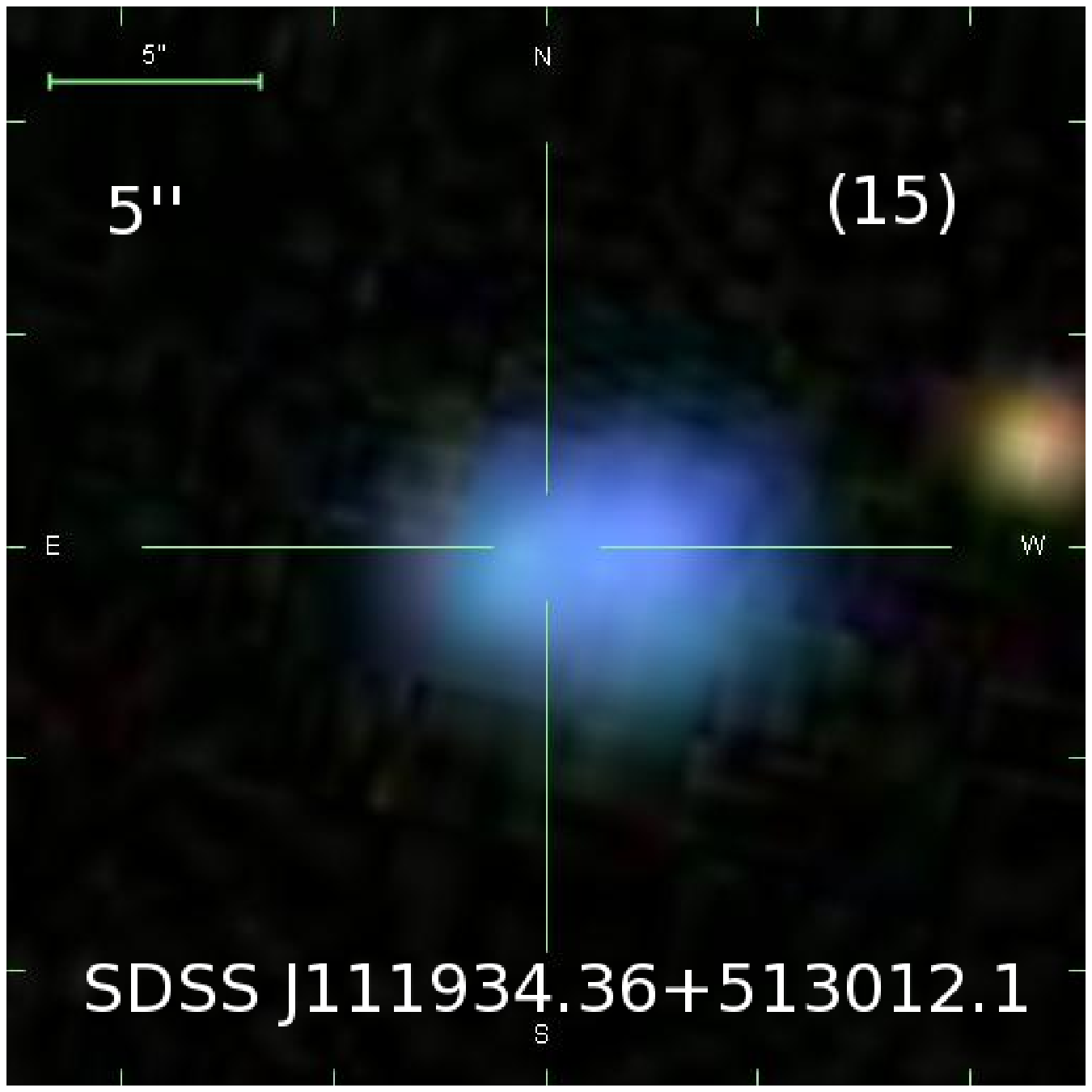}
\includegraphics[width=0.24\textwidth]{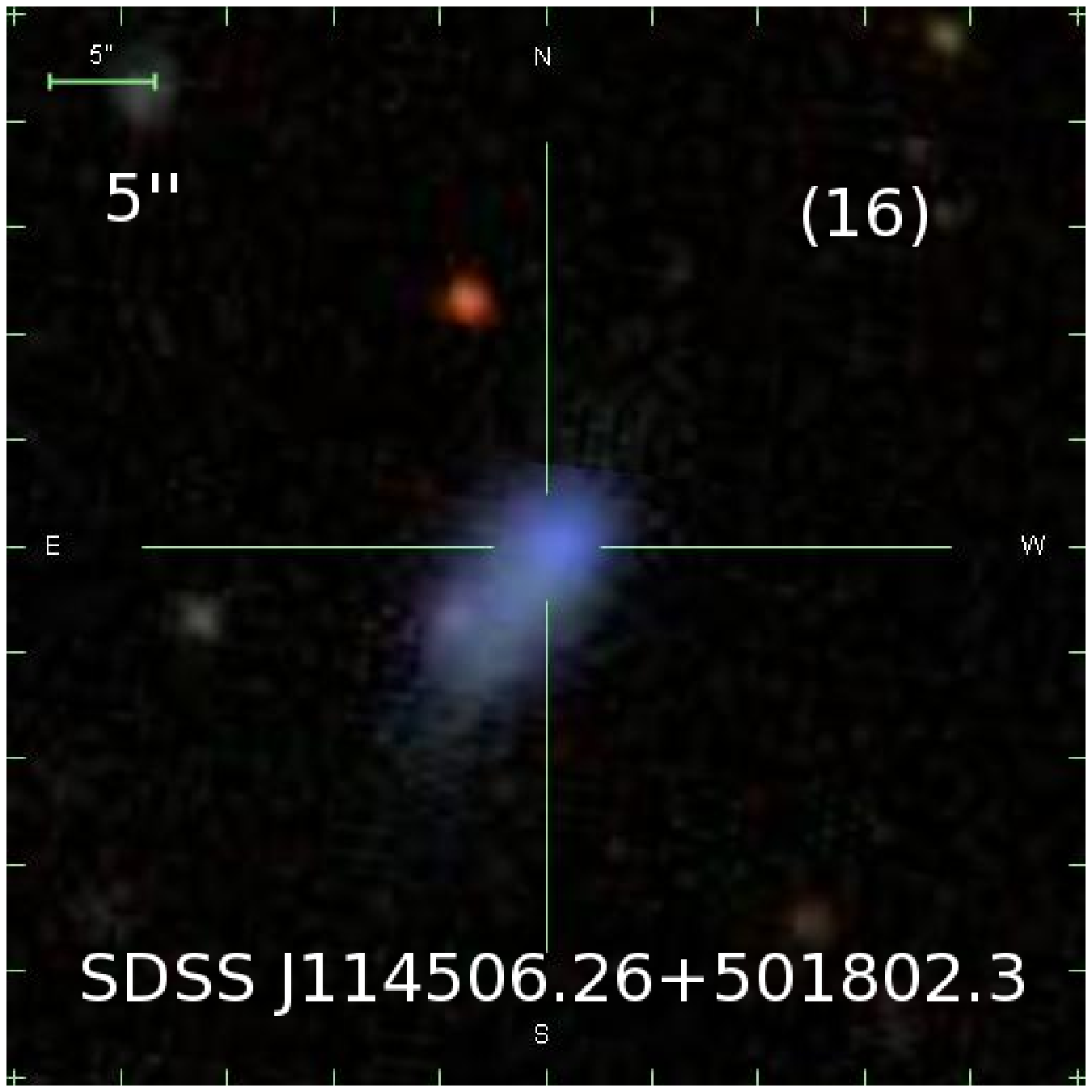}
\caption{
SDSS mugshots of all XMP candidates not marked as {\em single knot}
in Table.~\ref{list1}. They are cometary (e.g., \#~5), 
knotted-cometary (e.g., \#~2 ), or doubtful (e.g., \#~13).
The numbers correspond to the index 
in Table.~\ref{list1}, whereas the scales on the top left
corner of the panels represent 2, 5 or 10~arcsec, as indicated
by the insets. The figure continues in Fig.~\ref{mugshot2}.
}
\label{mugshot}
\end{figure*}
\setcounter{figure}{4} 
\begin{figure*}
\includegraphics[width=0.24\textwidth]{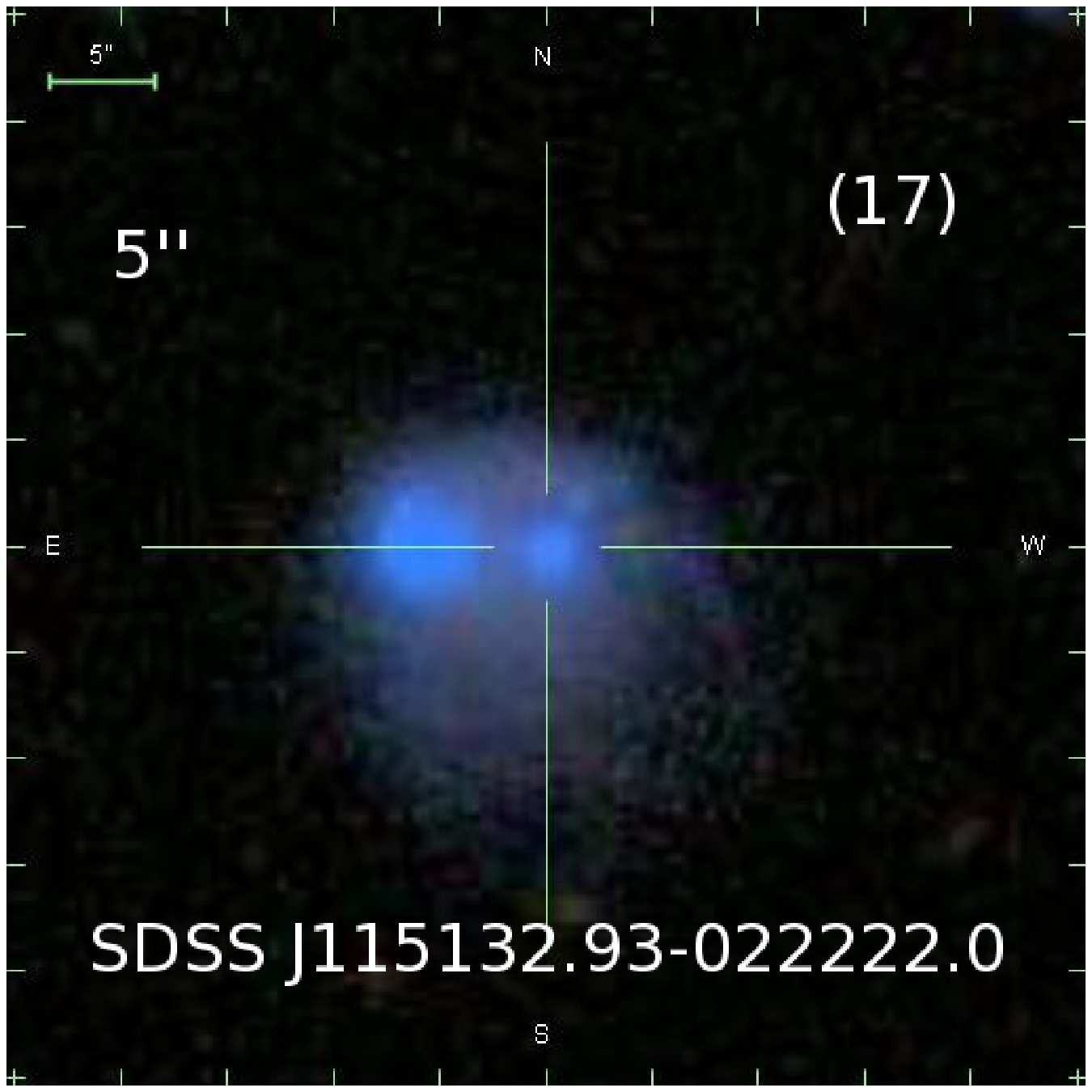}
\includegraphics[width=0.24\textwidth]{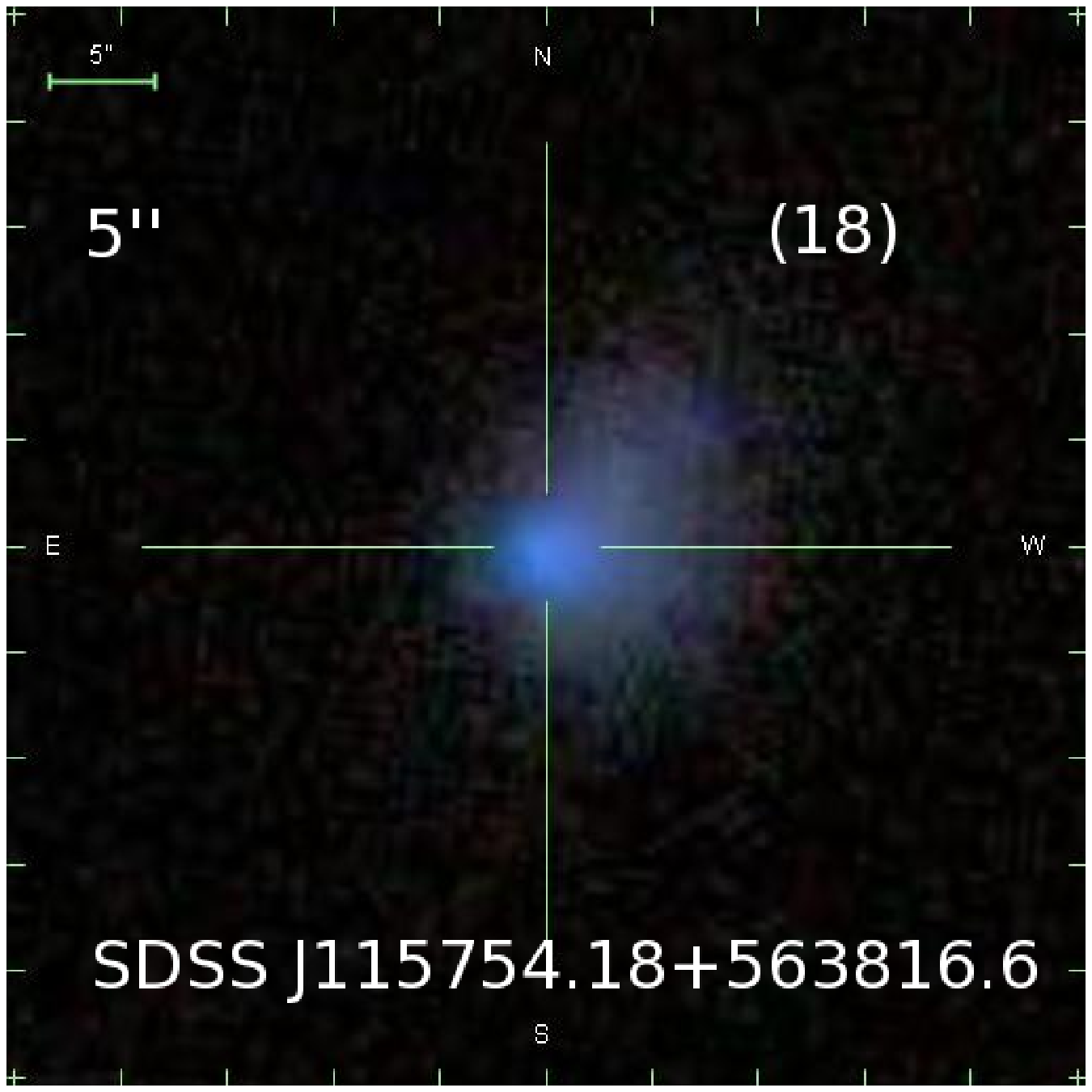}
\includegraphics[width=0.24\textwidth]{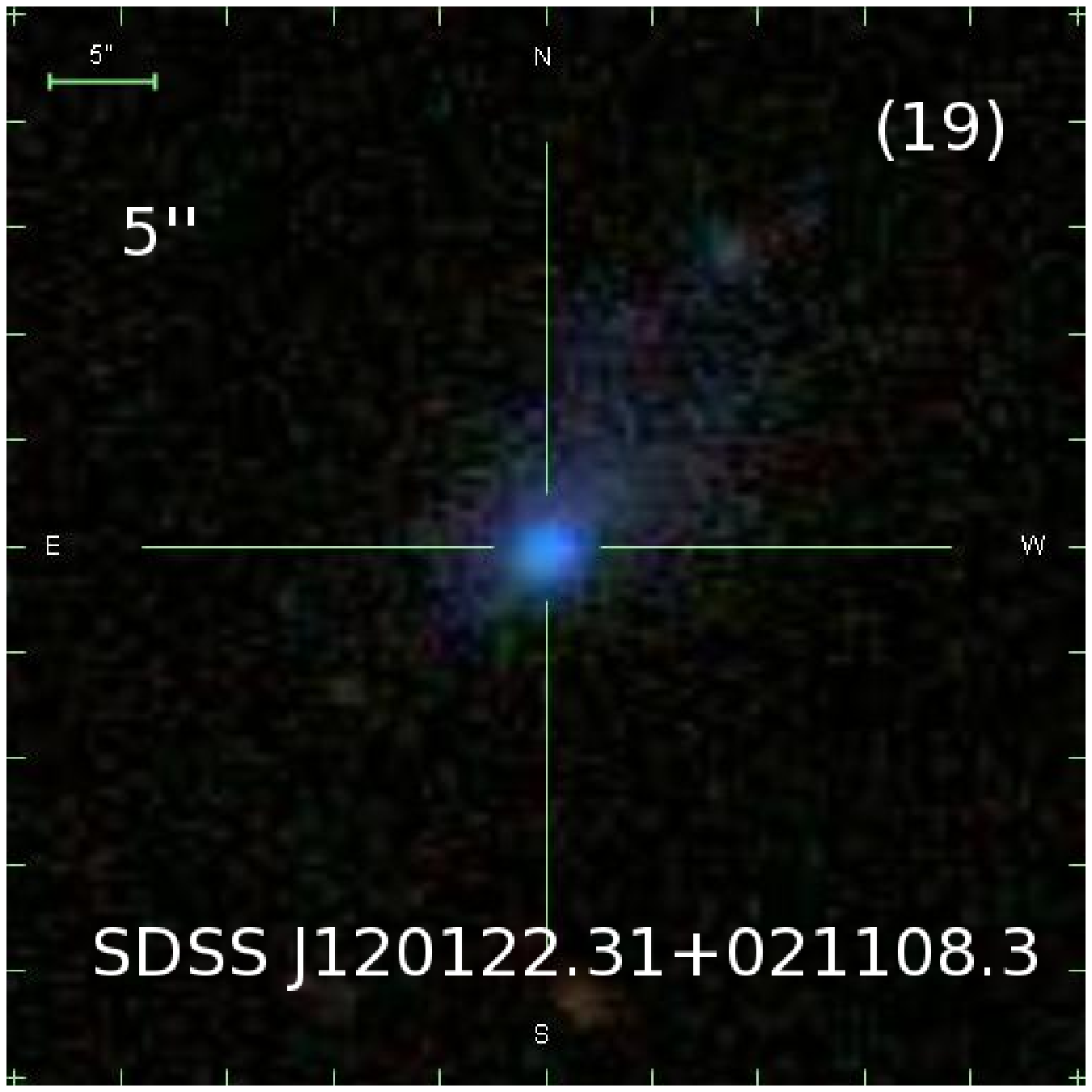}
\includegraphics[width=0.24\textwidth]{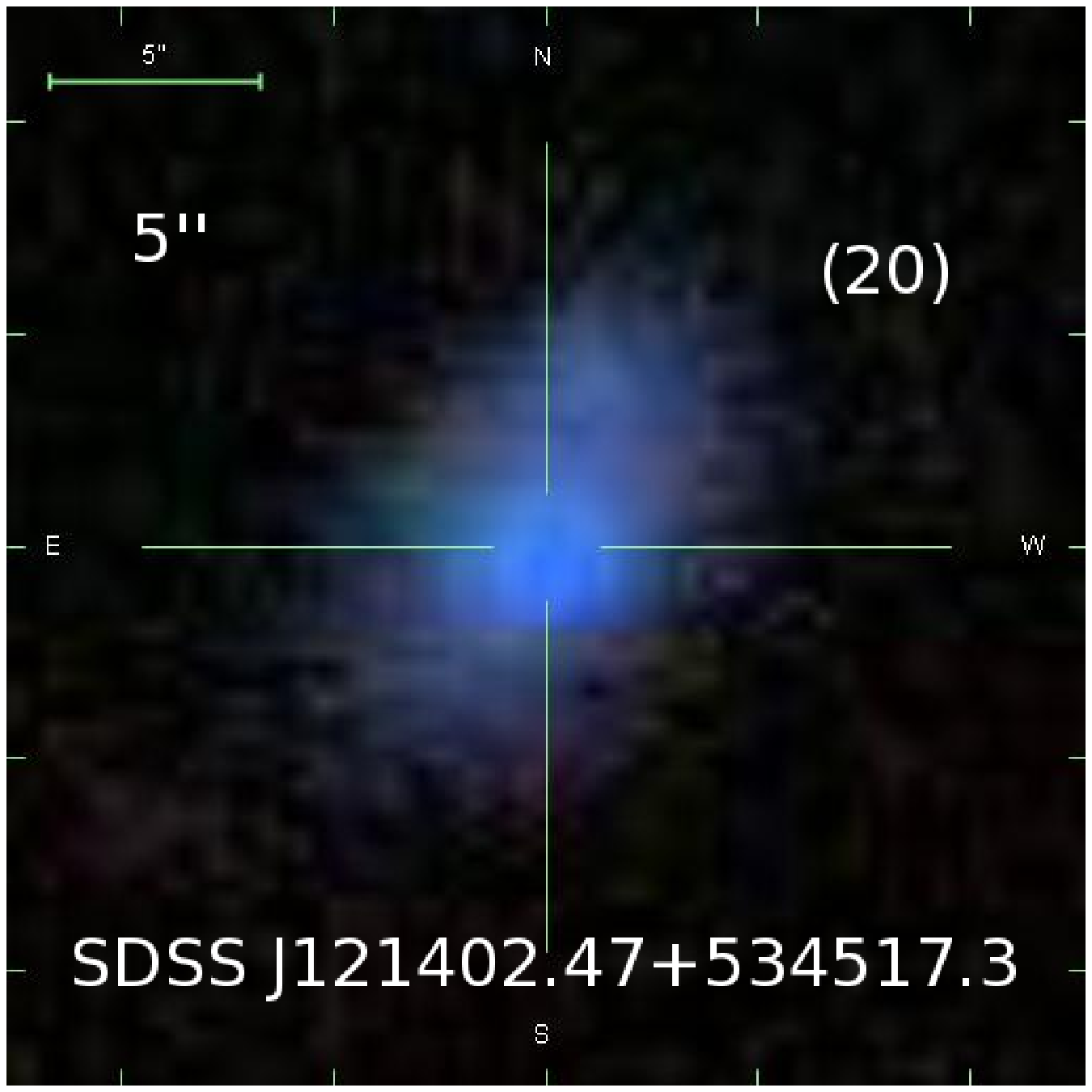}\\
\includegraphics[width=0.24\textwidth]{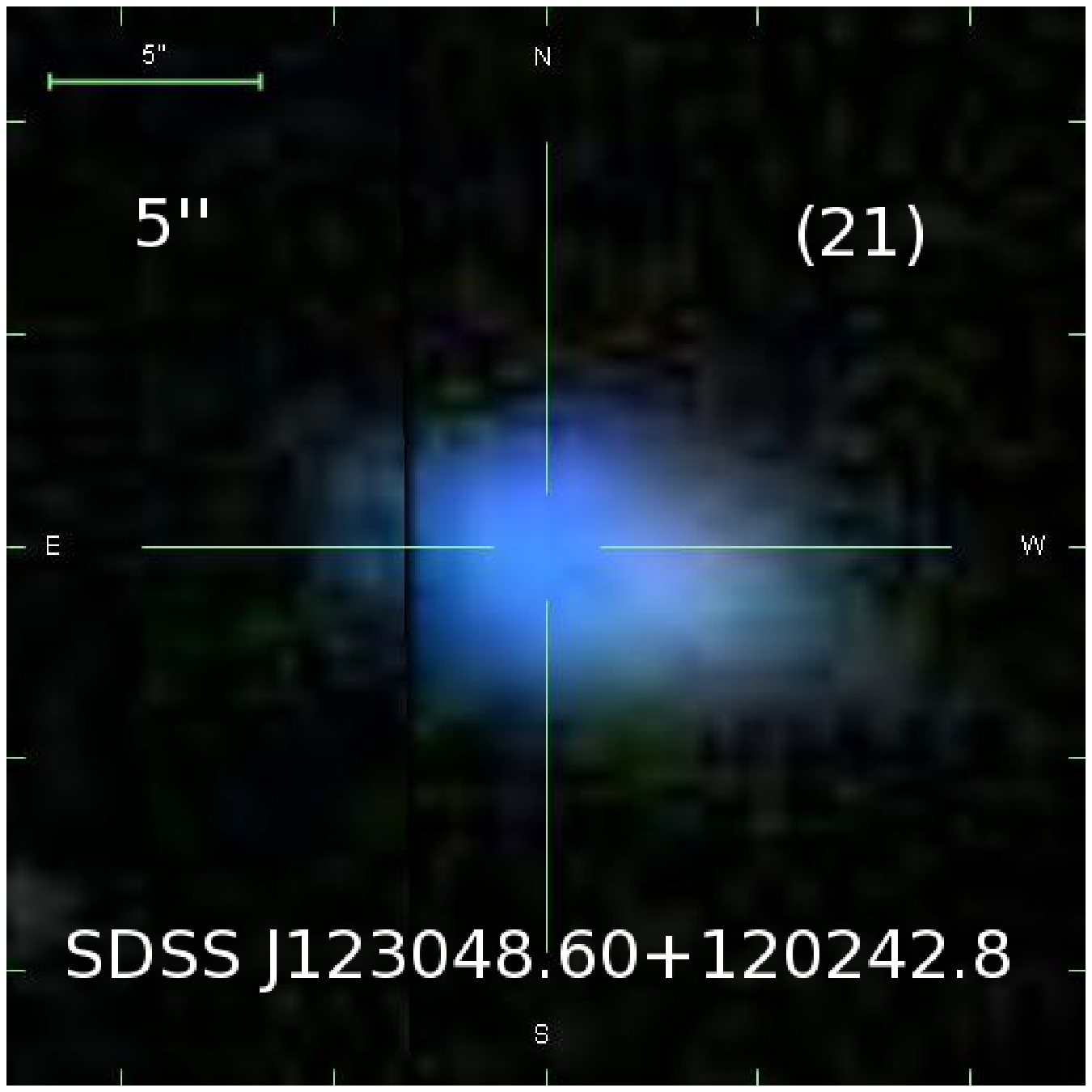}
\includegraphics[width=0.24\textwidth]{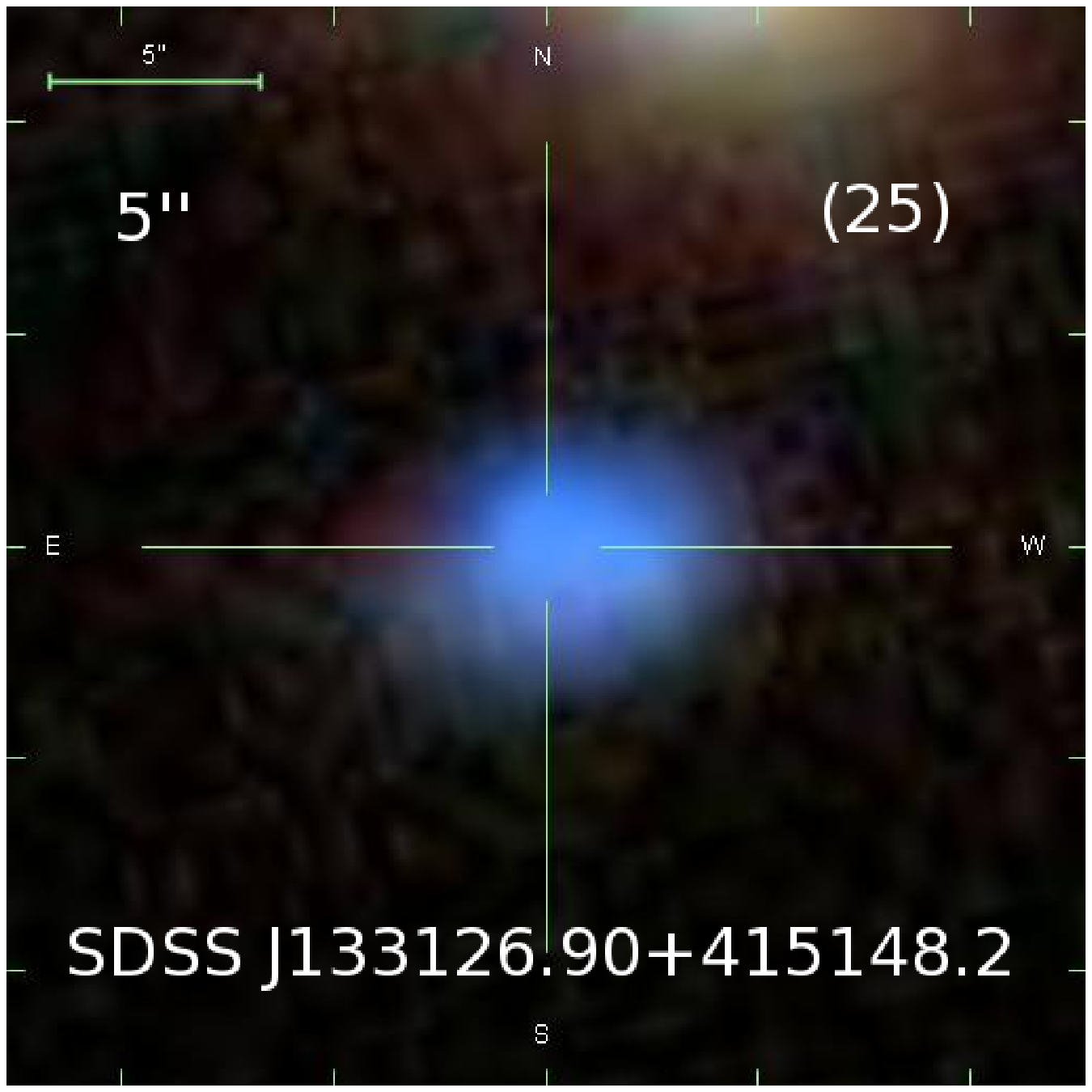}
\includegraphics[width=0.24\textwidth]{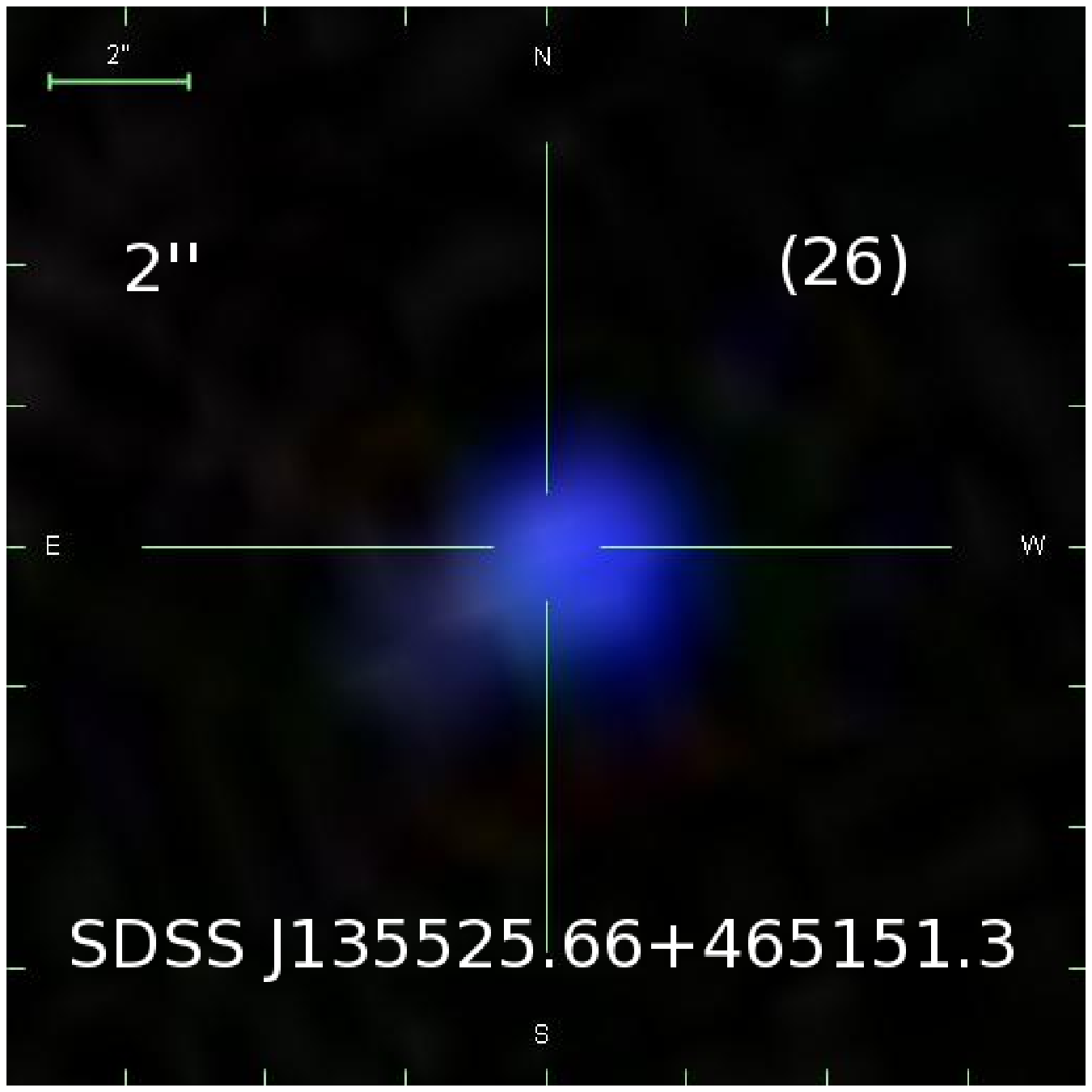}
\includegraphics[width=0.24\textwidth]{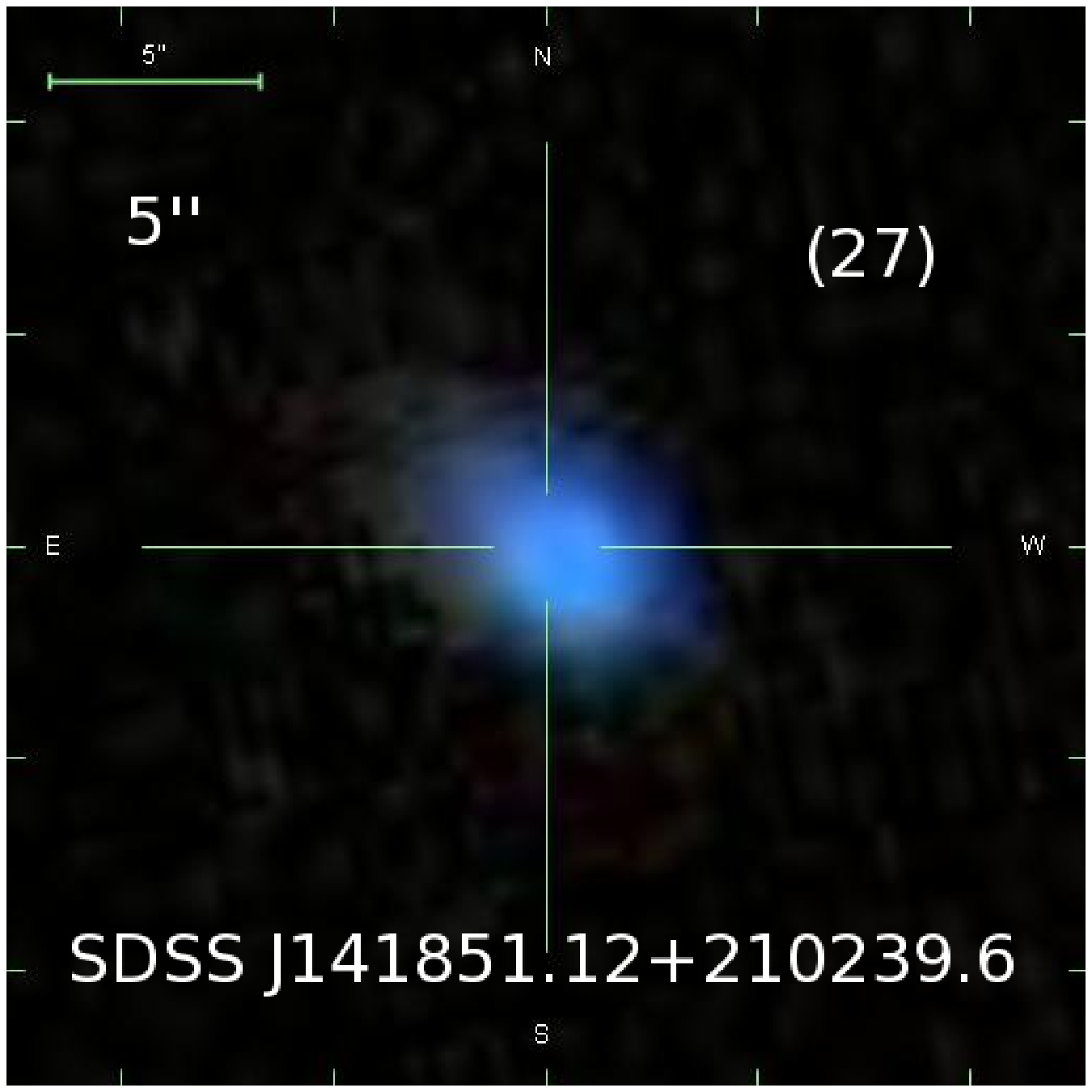}\\
\includegraphics[width=0.24\textwidth]{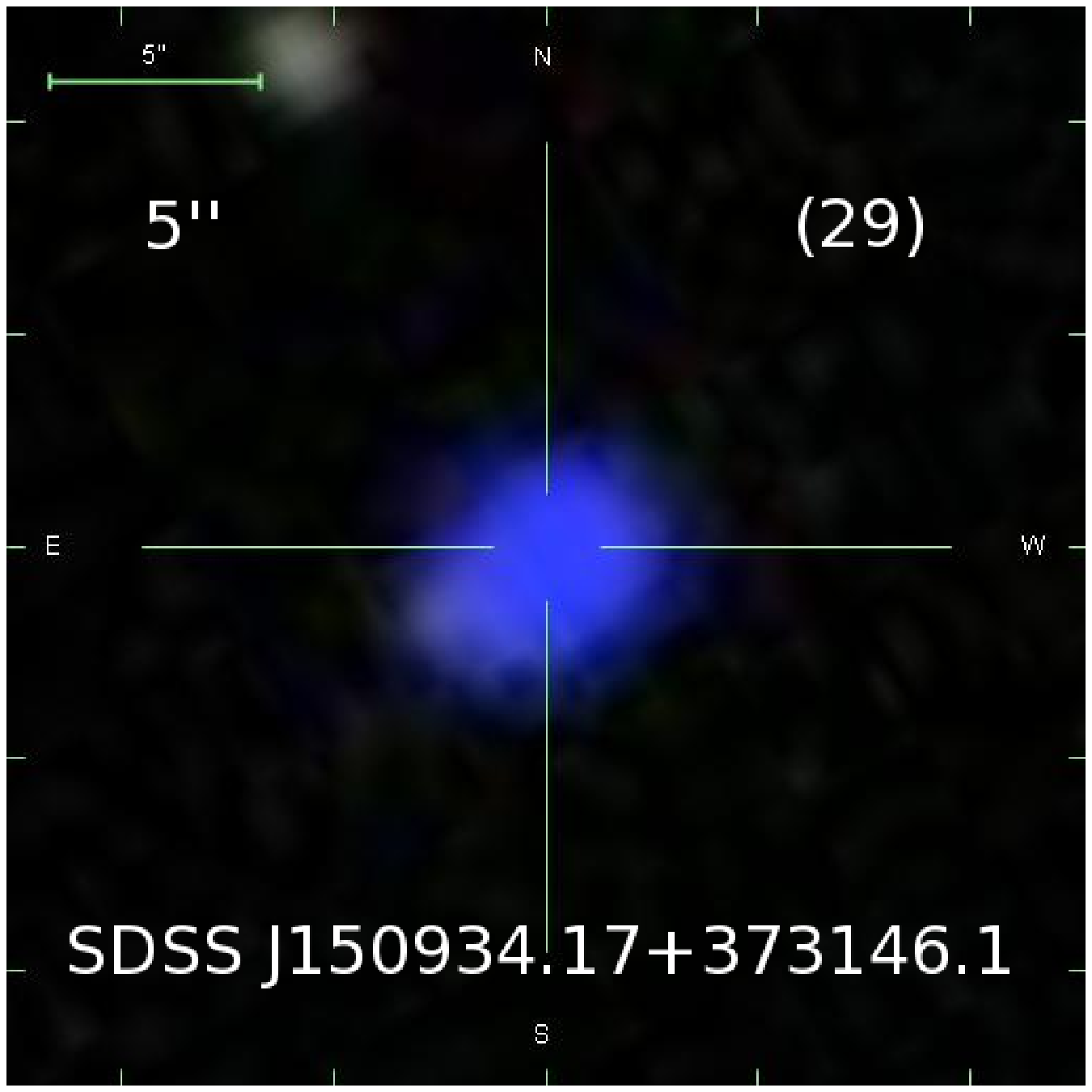}
\includegraphics[width=0.24\textwidth]{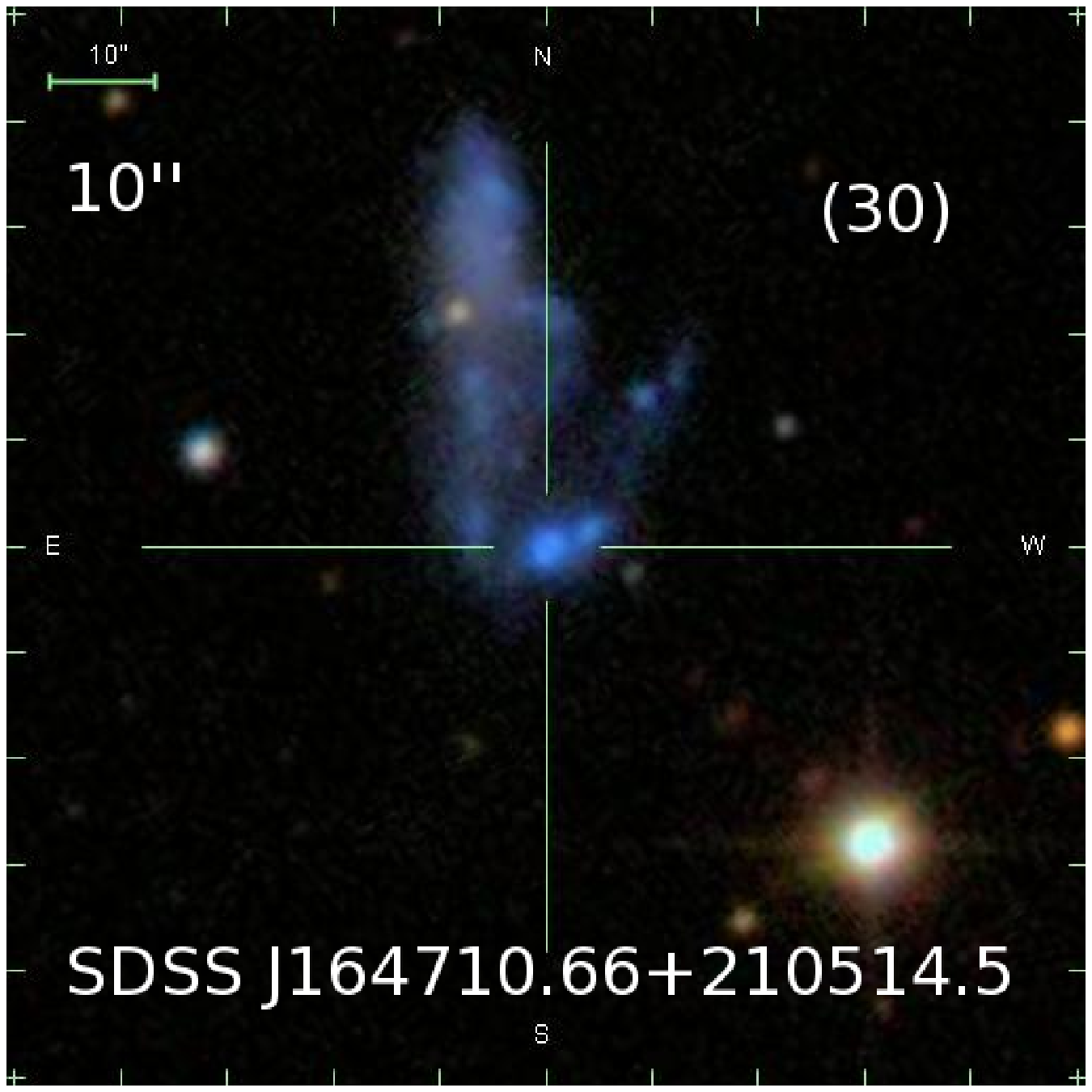}
\includegraphics[width=0.24\textwidth]{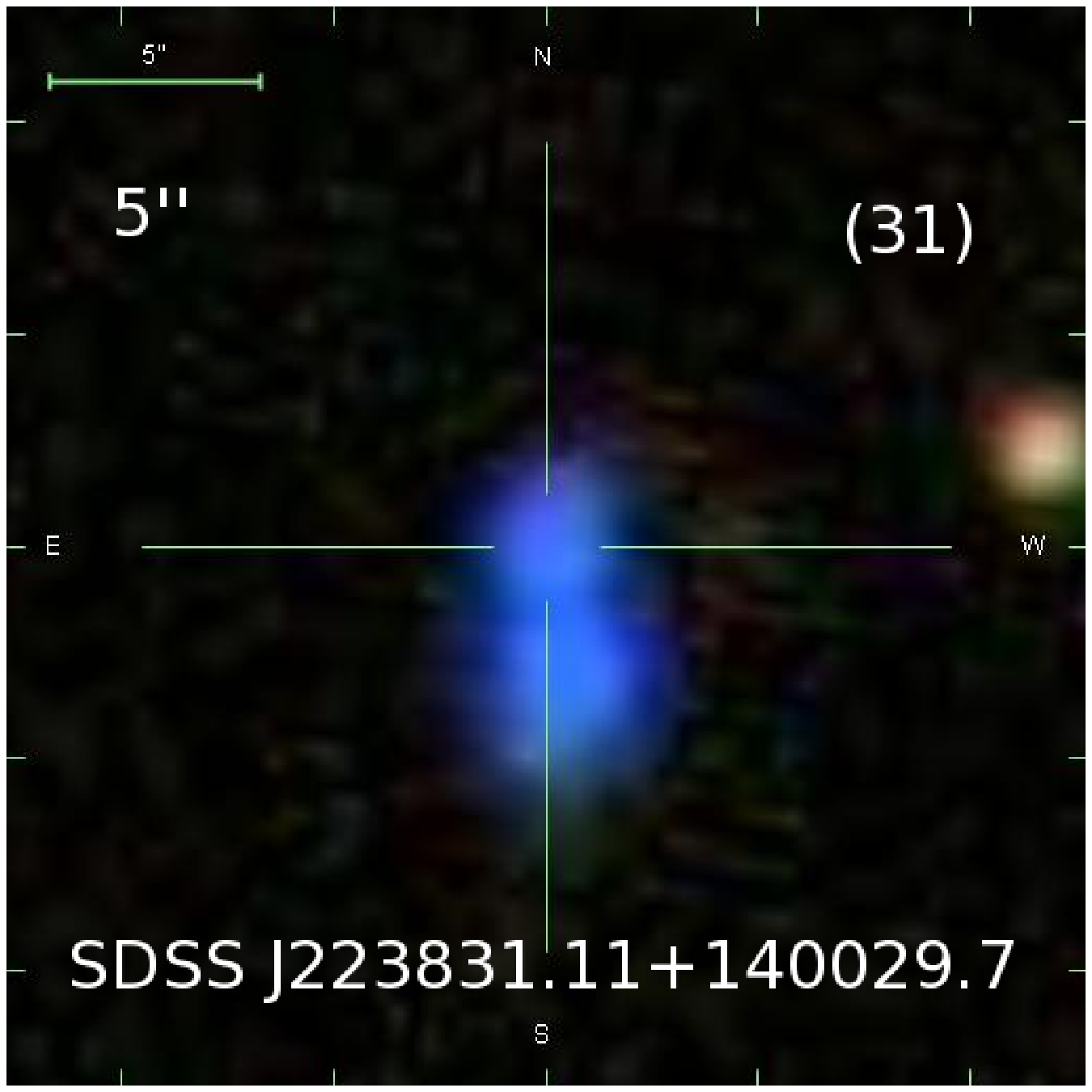}
\includegraphics[width=0.24\textwidth]{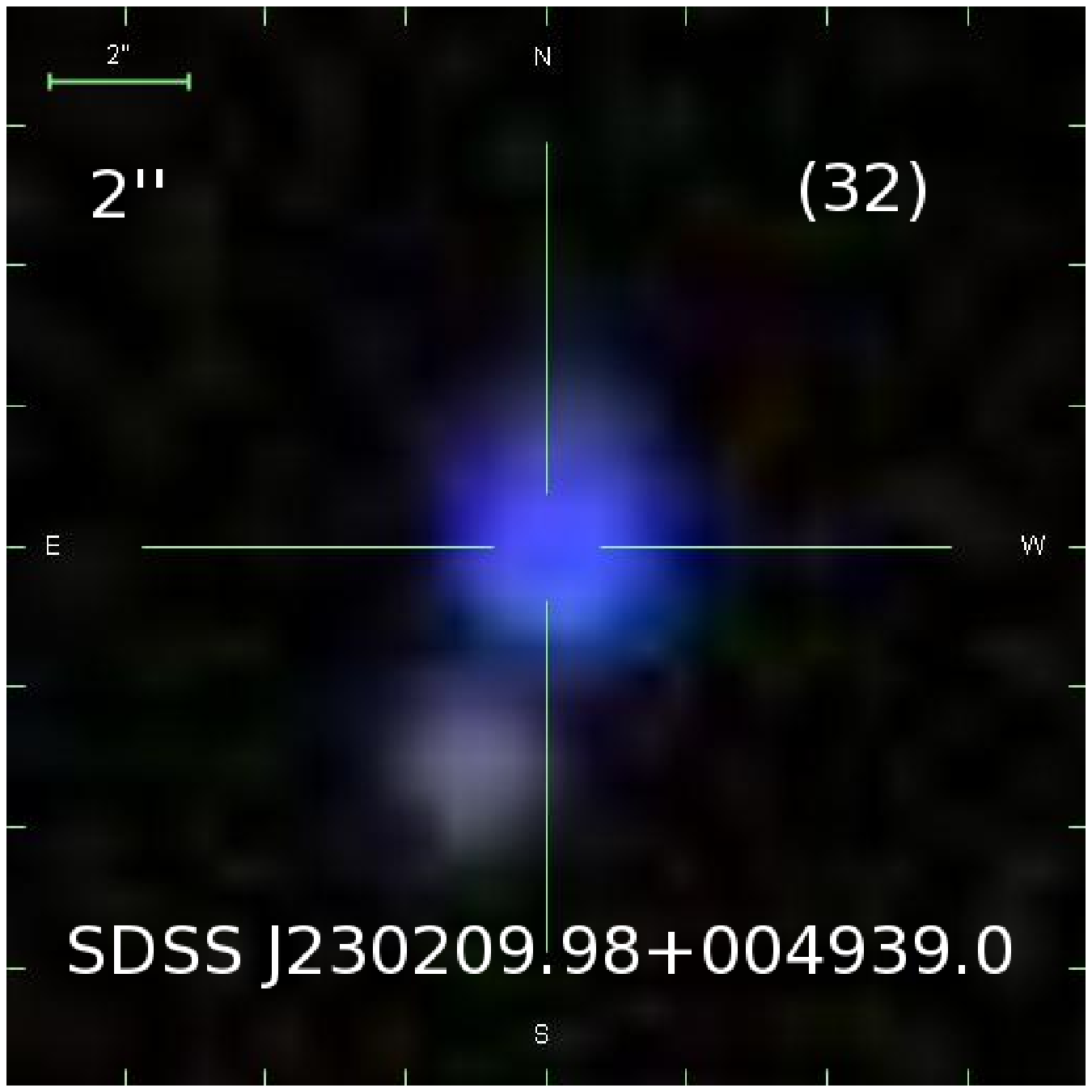}
\caption{
Continuation of Fig.~\ref{mugshot}.
}
\label{mugshot2}
\end{figure*}

The color magnitude diagram in Fig.~\ref{diag_bcd}b 
deserves a separate comment. First, it shows how several XMP 
candidates are extremely blue, reaching up to $g-r\lesssim-1$.
These extreme colors are rare, but they have been quoted
in connection with luminous star-forming galaxies 
\citep[e.g.,][]{2011ApJ...728..161I}, and with 
blue compact dwarfs \citep[e.g.,][]{san08}. 
Second, the figure suggests a trend that contrasts with 
usual behavior where brighter galaxies tend to be redder 
\citep[e.g.,][]{2009ARA&A..47..159B} --
the brighter the XMP candidate the bluer.
This unusual trend cannot be ascribed to photometric 
errors (the error bars provided by SDSS are included in 
Fig.~\ref{diag_bcd}b). The behavior remains 
independently of the type of magnitude 
\citep[Fig.~\ref{diag_bcd}b uses Petrosian magnitudes, but we also tried
the other magnitudes provided by SDSS; see][]{2002AJ....123..485S}.
Moreover, the trend disappears when colors other than 
$g-r$ are used. This unusual trend calls for an 
explanation but, so far, we can only offer conjectures. 
It may be a random fluctuation, since only a handful of 
galaxies define the trend. Alternatively, it may be due to a
subtle effect on the integrated colors of the large emission 
lines that dominate the spectrum of these galaxies.

In short, all but four of the 32 candidates fulfill the 
criteria to be BCDs, and 24 of them show cometary or knotted
shape.  The fact that the XMP candidates have these properties is
by no means trivial. We have 
selected our sample according to the form of their spectra in an 
narrow spectral window around H$\alpha$, and this narrow bit
of spectrum turns out to determine many global properties of the 
galaxy such as color, compactness, star formation rate, and 
even morphology.

%
%
\section{Search for XMP galaxies in the literature}\label{biblio_search}

%
%
%
\begin{deluxetable}{ccccccc}
\tabletypesize{\scriptsize}
\tablewidth{0pc}
\tablecaption{XMP targets found in the literature.}
\tablehead{
\colhead{Name\tablenotemark{a}}&
\colhead{RA}&	
\colhead{DEC}&	
\colhead{$g$}&	
\colhead{SpecObjID\tablenotemark{b}}&	
\colhead{$12+\log({\rm O/H})$}&
\colhead{Comment\tablenotemark{c}}	
}
\startdata
%
  UGC 12894 & 00 00 22 & +39 29 44 &  &  & 7.64 & {\citet{2006ApJ...636..214V}}\\
  J0004+0025 & 00 04 22 & +00 25 36 & 19.4 & 433417380224303104 & 7.37 &
{\citet{2009A&A...505...63G}}\\
  J0014-0044 & 00 14 29 & -00 44 44 & 18.7 & 306753510077104128 & 7.63 &
{\citet{2009A&A...505...63G}}\\
  J0015+0104 & 00 15 21 & +01 04 37 & 18.3 & 433983718529433600 & 7.07 &
{\citet{2009A&A...505...63G}}\\
  J0016+0108 & 00 16 28 & +01 08 02 & 18.9 & 193598873954942976 & 7.53 &
{\citet{2009A&A...505...63G}}\\
  HS 0017+1055 & 00 20 21 & +11 12 21 &  &  & 7.63 & {\citet{2007A&A...464..859P}}\\
  J0029-0108 & 00 29 05 & -01 08 26 & 19.2 & 434546692474273792 & 7.35 &
{\citet{2009A&A...505...63G}}\\
  J0029-0025 & 00 29 49 & -00 25 40 & 20.4 & 434546692394582016 & 7.29 &
{\citet{2009A&A...505...63G}}\\
  ESO 473- G024 & 00 31 22 & -22 45 57 &  &  & 7.45 & {\citet{2003AJ....125..610S}}\\
  Andromeda IV & 00 42 32 & +40 34 19 &  &  & 7.49 & {\citet{2008AstBu..63..102P}}\\
  J0057-0022 & 00 57 13 & -00 21 58 & 19.1 & 305062972235972608 & 7.60 &
{\citet{2009A&A...505...63G}}\\
  IC 1613 & 01 04 48 & +02 07 04 &  &  & 7.64 & {\citet{2006A&A...459...85N}}\\
  J0107+0001 & 01 07 51 & +00 01 28 & 19.4 & 422158628955357184 & 7.23 &
{\citet{2009A&A...505...63G}}\\
  AM 0106-382 & 01 08 22 & -38 12 34 &  &  & 7.56 & {\citet{2003A&A...401..141L}}\\
  J0113+0052 & 01 13 40 & +00 52 39 & 20.1 & 422158629852938240 & 7.24 &
{\citet{2007ApJ...665.1115I}}\\
  J0119-0935 & 01 19 14 & -09 35 46 & 19.5 & 185997585616470016 & 7.31 &
{\citet{2010MNRAS.406.1238E}}\\
  HS 0122+0743$^\star$ & 01 25 34 & +07 59 24 & 15.7 & 655785970125242368 & 7.60 &
{\citet{2004ApJ...602..200I}}\\
  J0126-0038 & 01 26 46 & -00 38 39 & 18.4 & 422724752210132992 & 7.51 &
{\citet{2009A&A...505...63G}}\\
  J0133+1342 & 01 33 53 & +13 42 09 & 18.1 & 120131172889001984 & 7.56 &
{\citet{2008A&A...491..113P}}\\
  J0135-0023 & 01 35 44 & -00 23 17 & 18.9 & 303656125474013184 & 7.38 &
{\citet{2009A&A...505...63G}}\\
  UGCA20 & 01 43 15 & +19 58 32 & 18.0 &  & 7.60 & K\"O, {\citet{2006ApJ...636..214V}}\\
  UM133 & 01 44 42 & +04 53 42 & 15.4 &  & 7.63 & K\"O, {\citet{2001A&A...371..404K}}\\
  HKK97L14 & 02 00 10 & +28 49 53 &  &  & 7.56 & {\citet{2006ApJ...636..214V}}\\
  J0204-1009 & 02 04 26 & -10 09 35 & 17.1 & 187686313107914752 & 7.36 &
{\citet{2007ApJ...665.1115I}}\\
  J0205-0949 & 02 05 49 & -09 49 18 & 15.3 & 187967849057222656 & 7.61 &
{\citet{2003ApJ...593L..73K}}\\
  J0216+0115 & 02 16 29 & +01 15 21 & 17.4 & 424413702440091648 & 7.44 &
{\citet{2009A&A...505...63G}}\\
  096632 & 02 51 47 & -30 06 32 & 16.3 &  & 7.51 & {\citet{2009A&A...505...63G}}\\
  J0254+0035 & 02 54 29 & +00 35 50 & 19.8 & 425817950054776832 & 7.28 &
{\citet{2007ApJ...665.1115I}}\\
  J0301-0059 & 03 01 26 & -00 59 26 & 21.5 & 300559785177120768 & 7.64 &
{\citet{2009A&A...505...63G}}\\
  J0301-0052 & 03 01 49 & -00 52 57 & 18.8 & 300559785005154304 & 7.52 &
{\citet{2007ApJ...665.1115I}}\\
  J0303-0109$^\star$ & 03 03 31 & -01 09 47 & 19.8 & 225967511814799360 & 7.22 &
{\citet{2009A&A...505...63G}}\\
  J0313+0010 & 03 13 02 & +00 10 40 & 18.9 & 226248880952442880 & 7.44 &
{\citet{2007ApJ...665.1115I}}\\
  J0315-0024 & 03 16 00 & -00 24 26 & 20.2 & 426661932058017792 & 7.41 &
{\citet{2009A&A...505...63G}}\\
  UGC2684 & 03 20 24 & +17 17 45 & 22.8 &  & 7.60 & K\"O, {\citet{2006ApJ...636..214V}}\\
SBS0335-052W & 03 37 38 & -05 02 37 & 19.0 &  & 7.11 & K\"O, {\citet{2005ApJS..161..240T}}\\
SBS0335-052E & 03 37 44 & -05 02 40 & 16.3 &  & 7.31 & K\"O, {\citet{2005ApJS..161..240T}}\\
  J0338+0013 & 03 38 12 & +00 13 13 & 24.4 & 227094714526990336 & 7.64 &
{\citet{2009A&A...505...63G}}\\
  J0341-0026 & 03 41 18 & -00 26 28 & 18.8 & 325892648806121472 & 7.26 &
{\citet{2009A&A...505...63G}}\\
  ESO 358- G 060 & 03 45 12 & -35 34 15 &  &  & 7.26 & {\citet{2003A&A...401..141L}}\\
  G0405-3648 & 04 05 19 & -36 48 49 &  &  & 7.25 & {\citet{2009A&A...505...63G}}\\
  J0519+0007 & 05 19 03 & +00 07 29 & 18.4 &  & 7.44 & {\citet{2009A&A...505...63G}}\\
  To0618-402 & 06 20 02 & -40 18 09 &  &  & 7.56 & K\"O, {\citet{1994ApJ...420..576M}}\\
  ESO489-G56 & 06 26 17 & -26 15 56 & 15.6 &  & 7.49 & K\"O, {\citet{1995A&A...302..353R}}\\
  J0808+1728$^\star$ & 08 08 41 & +17 28 56 & 19.2 & 585978593608204288 & 7.48 &
{\citet{2008AJ....135...92B}}\\
  J0812+4836 & 08 12 39 & +48 36 46 & 16.0 & 124071834843873280 & 7.28 &
{\citet{2007ApJ...665.1115I}}\\
  UGC 4305 & 08 19 05 & +70 43 12 &  &  & 7.65 & {\citet{2006A&A...459...85N}}\\
  HS0822+03542 & 08 25 55 & +35 32 31 & 17.8 &  & 7.35 & K\"O, {\citet{2000A&A...357..101K}}\\
  DDO53 & 08 34 07 & +66 10 54 & 20.3 &  & 7.62 & K\"O, {\citet{2006ApJ...637..269V}}\\
  UGC4483 & 08 37 03 & +69 46 31 & 15.1 &  & 7.58 & K\"O, {\citet{2002ApJ...567..875I}}\\
  HS0837+4717 & 08 40 30 & +47 07 10 & 17.6 &  & 7.64 & {\citet{2004A&A...419..469P}}\\
  HS 0846+3522 & 08 49 40 & +35 11 39 & 18.2 &  & 7.65 & {\citet{2007A&A...464..859P}}\\
  J0859+3923 & 08 59 47 & +39 23 06 & 17.2 &  & 7.57 & {\citet{2007ApJ...665.1115I}}\\
  J0910+0711 & 09 10 29 & +07 11 18 & 16.9 & 336307481519063040 & 7.63 &
{\citet{2010MNRAS.406.1238E}}\\
  J0911+3135 & 09 11 59 & +31 35 36 & 17.8 & 448054218540974080 & 7.51 &
{\citet{2007ApJ...665.1115I}}\\
  J0926+3343 & 09 26 09 & +33 43 04 & 17.8 & 448617233443192832 & 7.12 &
{\citet{2010MNRAS.401..333P}}\\
  IZw18$^\star$ & 09 34 02 & +55 14 25 & 16.4 & 156443095175528448 & 7.17 &
K\"O, {\citet{2005ApJS..161..240T}}\\
  J0940+2935 & 09 40 13 & +29 35 30 & 16.5 & 546853820680896512 & 7.65 &
{\citet{2007ApJ...665.1115I}}\\
  CGCG 007-025 & 09 44 02 & -00 38 32 & 16.0 & 75094093385957376 & 7.65 &
{\citet{2007A&A...464..885G}}\\
  SBS940+544 & 09 44 17 & +54 11 34 & 19.1 &  & 7.46 & K\"O, {\citet{2001A&A...378..756G}}\\
  CS 0953-174 & 09 55 00 & -17 00 00 &  &  & 7.58 & K\"O, {\citet{1994ApJ...420..576M}}\\
  J0956+2849 & 09 56 46 & +28 49 44 & 15.9 & 548261264221011968 & 7.13 &
{\citet{2007ApJ...665.1115I}}\\
  LeoA & 09 59 26 & +30 44 47 & 19.0 &  & 7.30 & K\"O, {\citet{2006ApJ...636..214V}}\\
  Sextans B & 10 00 00 & +05 19 56 & 20.5 &  & 7.53 & {\citet{2006ApJ...647..970L}}\\
  Sextans A & 10 11 00 & -04 41 34 &  &  & 7.54 & K\"O, {\citet{2005AJ....130.1558K}}\\
  KUG 1013+381$^\star$ & 10 16 24 & +37 54 44 & 15.9 & 401892408779866112 & 7.58 &
K\"O, {\citet{1998BSAO...46...23K}}\\
  SDSS J1025+1402 & 10 25 30 & +14 02 07 & 20.4 & 491964741116231680 & 7.36 &
{\citet{2007ApJ...671.1297I}}\\
  UGCA 211 & 10 27 02 & +56 16 14 & 16.2 &  & 7.56 & {\citet{2010A&A...520A..90C}}\\
  J1031+0434$^\star$ & 10 31 37 & +04 34 22 & 16.2 & 162635977557278720 & 7.70 &
{\citet{2007ApJ...671.1297I}}\\
  HS 1033+4757 & 10 36 25 & +47 41 52 & 17.5 & 271004930616066048 & 7.65 &
{\citet{2007A&A...464..859P}}\\
  J1044+0353$^\star$ & 10 44 58 & +03 53 13 & 17.5 & 162917331083722752 & 7.44 &
{\citet{2008A&A...491..113P}}\\
  HS 1059+3934 & 11 02 10 & +39 18 45 & 17.9 &  & 7.62 & {\citet{2007A&A...464..859P}}\\
  J1105+6022 & 11 05 54 & +60 22 29 & 16.4 & 218086199200317440 & 7.64 &
{\citet{2003ApJ...593L..73K}}\\
  J1119+5130$^\star$ & 11 19 34 & +51 30 12 & 16.9 & 247359886311030784 & 7.51 &
K\"O, {\citet{2003ApJ...593L..73K}}\\
  J1121+0324 & 11 21 53 & +03 24 21 & 18.1 & 235538034580258816 & 7.64 &
{\citet{2003ApJ...593L..73K}}\\
  UGC 6456 & 11 28 00 & +78 59 39 &  &  & 7.35 & {\citet{2006A&A...459...85N}}\\
  SBS1129+576 & 11 32 02 & +57 22 46 & 16.7 &  & 7.36 & {\citet{2003A&A...407..105G}}\\
  J1151-0222$^\star$ & 11 51 32 & -02 22 22 & 16.8 & 93111671593107456 & 7.78 &
{\citet{2006A&A...459...85N}}\\
  J1201+0211$^\star$ & 12 01 22 & +02 11 08 & 17.6 & 145464500043120640 & 7.49 &
{\citet{2008A&A...491..113P}}\\
  SBS1159+545 & 12 02 02 & +54 15 50 & 18.7 &  & 7.41 & K\"O, {\citet{2006ApJ...645.1076N}}\\
  SBS 1211+540$^\star$ & 12 14 02 & +53 45 17 & 17.4 & 287049377199947776 & 7.64 &
{\citet{2006A&A...459...85N}}\\
  J1215+5223 & 12 15 47 & +52 23 14 & 15.2 & 248767590061572096 & 7.43 &
{\citet{2003ApJ...593L..73K}}\\
  Tol1214-277 & 12 17 17 & -28 02 33 &  &  & 7.55 & K\"O, {\citet{2004A&A...421..539I}}\\
  VCC 0428 & 12 20 40 & +13 53 22 & 17.0 & 497314452257898496 & 7.64 &
{\citet{2003ApJS..145..225V}}\\
  HS 1222+3741 & 12 24 37 & +37 24 37 & 17.9 & 564023911881113600 & 7.64 &
{\citet{2000A&AS..142..247P}}*\\ 
  Tol65 & 12 25 47 & -36 14 01 & 17.5 &  & 7.54 & K\"O, {\citet{2004A&A...421..539I}}\\
  J1230+1202$^\star$ & 12 30 49 & +12 02 43 & 16.7 & 454810434122285056 & 7.73 &
K\"O, {\citet{2002A&A...389..405P}}\\
  KISSR 85 & 12 37 18 & +29 14 55 & 19.9 &  & 7.61 & {\citet{2004ApJ...616..752L}}\\
  UGCA 292 & 12 38 40 & +32 46 01 & 18.9 &  & 7.28 & {\citet{2001A&A...374..412P}}\\
  HS 1236+3937 & 12 39 20 & +39 21 05 & 18.5 &  & 7.47 & {\citet{2000A&AS..142..247P}}*\\ 
  J1239+1456 & 12 39 45 & +14 56 13 & 19.8 &  & 7.65 & {\citet{2008AJ....135...92B}}\\
  SBS 1249+493 & 12 51 52 & +49 03 28 & 18.0 &  & 7.64 & {\citet{2006ApJ...645.1076N}}\\
  J1255-0213$^\star$ & 12 55 26 & -02 13 34 & 19.1 & 95080395053203456 & 7.83 &
{\citet{2007A&A...462..535Y}}\\
  Gr8 & 12 58 40 & +14 13 03 & 17.9 &  & 7.65 & K\"O, {\citet{2006ApJ...636..214V}}\\
  KISSR 1490 & 13 13 16 & +44 02 30 & 19.0 &  & 7.56 & {\citet{2004ApJ...616..752L}}\\
  DDO 167 & 13 13 23 & +46 19 22 &  &  & 7.20 & {\citet{2002A&A...389..836H}}\\
  HS 1319+3224 & 13 21 20 & +32 08 25 & 18.6 &  & 7.59 & {\citet{2000A&AS..142..247P}}\\
  J1323-0132$^\star$ & 13 23 47 & -01 32 52 & 18.2 & 96204976459612160 & 7.78 &
{\citet{2007ApJ...671.1297I}}\\
  J1331+4151$^\star$ & 13 31 27 & +41 51 48 & 17.1 & 412307391565004800 & 7.75 &
{\citet{2007ApJ...662...15I}}\\
  ESO577-G27 & 13 42 47 & -19 34 54 &  &  & 7.57 & K\"O, {\citet{1995A&A...302..353R}}\\
  J1355+4651$^\star$ & 13 55 26 & +46 51 51 & 19.3 & 361921789577658368 & 7.63 &
{\citet{2005A&A...442..109P}}\\
  J1414-0208 & 14 14 54 & -02 08 23 & 18.0 & 258056040799535104 & 7.28 &
{\citet{2008A&A...491..113P}}\\
  SBS1415+437 & 14 17 01 & +43 30 05 & 17.8 &  & 7.43 & K\"O, {\citet{2006ApJ...645.1076N}}\\
  J1422+5145 & 14 22 51 & +51 45 16 & 20.2 &  & 7.41 & {\citet{2008AJ....135...92B}}\\
  J1423+2257$^\star$ & 14 23 43 & +22 57 29 & 17.9 & 600334402043510784 & 7.72 &
{\citet{2007ApJ...662...15I}}\\
  J1441+2914 & 14 41 58 & +29 14 34 & 20.1 &  & 7.47 & {\citet{2008AJ....135...92B}}\\
  HS1442+4250 & 14 44 13 & +42 37 44 & 15.9 &  & 7.54 & {\citet{2003A&A...407...91G}}\\
  J1509+3731$^\star$ & 15 09 34 & +37 31 46 & 17.3 & 394011865673367552 & 7.85 &
{\citet{2007ApJ...662...15I}}\\
  KISSR 666 & 15 15 42 & +29 01 40 & 19.1 &  & 7.53 & {\citet{2006A&A...459...85N}}\\
  KISSR 1013 & 16 16 39 & +29 03 33 & 18.2 & 396262437692637184 & 7.63 &
{\citet{2006ApJ...645.1076N}}\\
  J1644+2734 & 16 44 03 & +27 34 05 & 17.7 & 475922385527111680 & 7.48 &
{\citet{2007ApJ...671.1297I}}\\
  J1647+2105$^\star$ & 16 47 11 & +21 05 15 & 17.3 & 442143986740625408 & 7.75 &
{\citet{2007ApJ...662...15I}}\\
  W1702+18 & 17 02 33 & +18 03 06 & 18.4 &  & 7.63 & {\citet{2011arXiv1106.4844G}}\\
  HS 1704+4332 & 17 05 45 & +43 28 49 & 18.4 &  & 7.55 & {\citet{2007A&A...464..859P}}\\
  SagDIG & 19 29 59 & -17 40 41 &  &  & 7.44 & K\"O, {\citet{2006ApJ...637..269V}}\\
  J2053+0039 & 20 53 13 & +00 39 15 & 19.4 & 288175754707992576 & 7.33 &
{\citet{2009A&A...505...63G}}\\
  J2104-0035 & 21 04 55 & -00 35 22 & 17.9 & 288457262211530752 & 7.05 &
{\citet{2009A&A...505...63G}}\\
  J2105+0032 & 21 05 09 & +00 32 23 & 19.0 & 288457264749084672 & 7.42 &
{\citet{2009A&A...505...63G}}\\
  J2120-0058 & 21 20 26 & -00 58 27 & 18.8 & 289300532823064576 & 7.65 &
{\citet{2009A&A...505...63G}}\\
  HS2134+0400 & 21 36 59 & +04 14 04 &  &  & 7.44 & {\citet{2006AstL...32..228P}}\\
  J2150+0033 & 21 50 32 & +00 33 05 & 19.3 & 414839880762261504 & 7.60 &
{\citet{2009A&A...505...63G}}\\
  ESO146-G14 & 22 13 00 & -62 04 03 &  &  & 7.59 & K\"O, {\citet{1995A&A...302..353R}}\\
  2dF 171716 & 22 13 26 & -25 26 43 &  &  & 7.54 & {\citet{2006A&A...457...45P}}\\
  PHL293B & 22 30 37 & -00 06 37 & 17.2 &  & 7.62 & {\citet{2009A&A...505...63G}}\\
  2dF 115901 & 22 37 02 & -28 52 41 &  &  & 7.57 & {\citet{2006A&A...457...45P}}\\
  J2238+1400$^\star$ & 22 38 31 & +14 00 30 & 19.0 & 533060792673632256 & 7.45 &
{\citet{2009A&A...505...63G}}\\
  J2250+0000 & 22 50 59 & +00 00 33 & 19.8 & 190501187865280512 & 7.61 &
{\citet{2007ApJ...671.1297I}}\\
  J2259+1413 & 22 59 01 & +14 13 43 & 19.1 &  & 7.37 & {\citet{2008AJ....135...92B}}\\
  J2302+0049$^\star$ & 23 02 10 & +00 49 39 & 18.8 & 190784502518251520 & 7.71 &
{\citet{2009A&A...505...63G}}\\
  J2354-0005 & 23 54 37 & -00 05 02 & 18.7 & 307596757485748224 & 7.35 &
{\citet{2009A&A...505...63G}}\\
%
\enddata
\tablenotetext{a}{Those names marked with an asterisk correspond to galaxies
also identified in this work (Table~\ref{list1}).
}
\tablenotetext{b}{Unique identifier of the galaxy spectrum in SDSS/DR7.}
\tablenotetext{c}{Reference to the work where the oxygen abundance was obtained, 
plus the K\"O  flag when the galaxy belongs to the compilation by \citet{kun00}.}
\label{revision}
\end{deluxetable}
%


In order to put our search into context, we carried out a careful 
revision of the literature to select all the nearby galaxies with 
metallicity reported to be one-tenth solar or less. 
($12+\log({\rm O/H})< 7.65$).
Those are listed in Table~\ref{revision}.
We start off from Table~3 in \citet{kun00}, who compile all the
galaxies found in the literature prior to year 2000.
It comprises 31 targets, from which 
\modified{
5 
}
were discarded because their 
metallicity has been revised upwards.
Then we cover the last ten years by 
checking cross-references existing in recent known papers. We begin with the 
work by \citet{2009A&A...505...63G}, that analyzes 
\modified{
44 
}
targets selected in SDSS/DR6.
The spectra used in the analysis are not spectra from SDSS but obtained
elsewhere.
We choose from \citet{2009A&A...505...63G} all the galaxies with oxygen abundance
below the threshold. Then, using %
ADS\footnote{The digital library NASA Astrophysical Data System
{\tt http://www.adsabs.harvard.edu/}}, we revised all papers 
cited in \citet{2009A&A...505...63G} and published after \citet{kun00}. 
Based on their title and abstract, we examined those 
papers dealing with galaxy metallicity, separating the appropriate 
XMP galaxies. High redshift targets were not included 
\citep[e.g., those in][]{2003ApJ...597..730L}.
The procedure was repeated with all the papers containing
XMP galaxies, until no new reference earlier than \citet{kun00} was found.
In addition, our recursive searching strategy was 
repeated with all the papers citing \citet{2009A&A...505...63G}. 
For example, \citet{2007ApJ...665.1115I} is cited by \citet{2009A&A...505...63G} and 
provides
%
13 targets from SDSS/DR5.
\citet{2008A&A...491..113P} study spectroscopically and morphologically 
7 galaxies identified in SDSS/DR4 and 6dFGS 
\citep[six-degree field galaxy survey,][]{jon04}.
\modified{
Three
}
galaxies are exhaustively analyzed in three
separate papers
by \citet{gus03c,2003A&A...407...91G,2003A&A...407..105G}, and are also included in
Table~\ref{revision}. 
Two additional galaxies are from the study by \citet{2006A&A...454..137I}. Through this
step-by-step search procedure, we revised all the papers that seems to be 
relevant, and a number of them rendered one or several additional 
XMP galaxies for Table~\ref{revision} 
\citep[][]{
2000A&AS..142..247P,
2000A&A...357..101K,
2001A&A...371..404K,
2001A&A...378..756G,
2001A&A...374..412P,
2002A&A...389..836H,
2002ApJ...567..875I,
2003A&A...401..141L,
2003A&A...407..105G,
2003AJ....125..610S,
2003ApJS..145..225V,
2003ApJ...593L..73K,
2004A&A...419..469P,
2004A&A...421..539I,
2004ApJ...602..200I,
2005ApJS..161..240T,
2005AJ....130.1558K,
2006A&A...457...45P,
2006A&A...459...85N,
2006ApJ...636..214V,
2006ApJ...637..269V,
2006ApJ...645.1076N,
2006ApJ...647..970L,
2006AstL...32..228P,
2007A&A...464..885G,
2007A&A...464..859P,
2007ApJ...665.1115I,
2007ApJ...671.1297I,
2008A&A...491..113P,
2008AstBu..63..102P,
2008AJ....135...92B,
2009A&A...505...63G,
2010MNRAS.401..333P,
2010MNRAS.406.1238E,
2010A&A...520A..90C}.
Finally, the galaxies in Table~\ref{list1} were 
inspected individually in NED\footnote{NASA Extragalactic Data base {\tt
http://nedwww.ipac.caltech.edu/}.}, to assure that
they are included in our bibliographic search even when the 
estimated metallicity exceeds the imposed threshold.
Table~\ref{revision} contains all galaxies we found in the 
literature
through this search. It includes only 
\modified{
129 
}
galaxies, proving that the XMP targets are really 
rare objects. 
The table includes name, coordinates, 
SDSS SpecObjID (when available), the reference for the
oxygen metallicity, as well as a tag to identify galaxies
included in the work by \citet{kun00} used as reference. 
The match between Tables~\ref{list1} and \ref{revision}
renders 21 objects, i.e., only 21
of the 32 candidates had been previously identified as 
XMP galaxies. They are marked in the tables with asterisks. 
The remaining 11 objects are new. 
The overlapping between the two tables allows us to confirm the 
low metallicity of the XMP candidates selected using state-of-the-art metallicity
determinations. The oxygen abundance in Table~\ref{revision} are based on 
self-consistently determined electron temperatures. Assuming
the twenty one
known objects to be representative of the full set, then 
mean metallicity turns out to be  
$12+\log({\rm O/H})=7.61\pm 0.19$,
with the error bar accounting for the standard deviation.

\begin{figure*}
\includegraphics[width=.7\textwidth,angle=90]{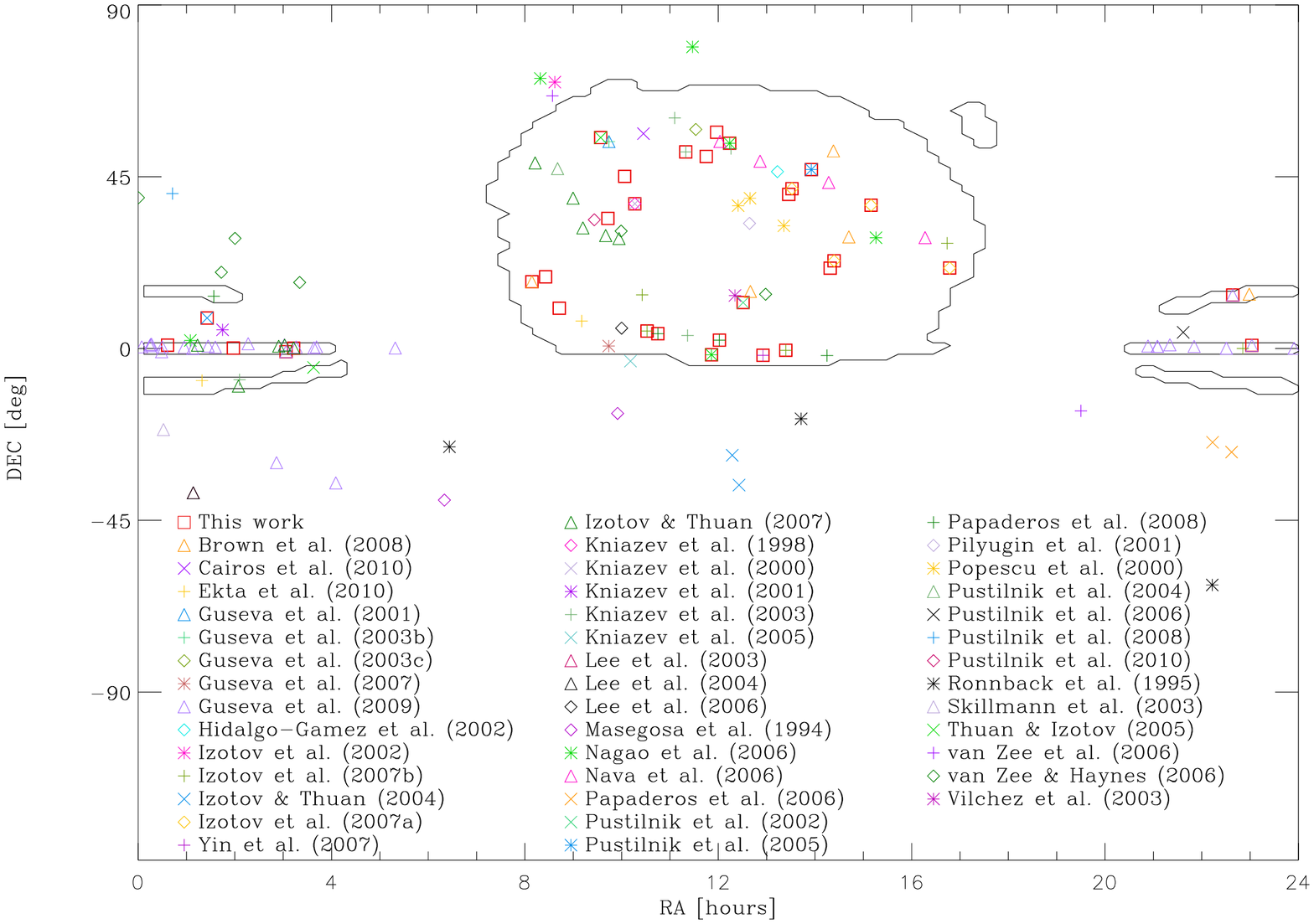}
\caption{Spatial distribution of XMP galaxies, 
both of the 
\modified{
129
}
found in the literature, plus the 32 targets obtained in our 
systematic search through SDSS/DR7.
Twenty one of them coincide.
Different symbols and colors
represent different sources as coded in the inset. In particular,
the square symbols correspond to the galaxies found in our search.  
The contours represent the sky coverage of the spectroscopic sample
of SDSS/DR7. 
Note that one of our targets is outside the main 
spectroscopic sample of SDSS/DR7 (i.e., it lies outside the contours). 
As usual, RA and DEC stand for right ascension and 
declination.
}
\label{spatial}
\end{figure*}

Figure~\ref{spatial} shows the spatial distribution of the low metallicity 
targets, i.e., the 
\modified{
129
}
galaxies obtained from the literature plus the targets 
found in this work.
Note that most works have focused the search in very localized 
areas on the sky. For example, the targets by \citet{2009A&A...505...63G} are clustered 
around the equator, even though the search  is based on SDSS/DR6 which has
a much broader scope. Our systematic search, however,
has selected targets spread throughout the SDSS/DR7 field of view,
that covers a significant part of the sky ($\sim 20$\%).

\begin{figure*}
\includegraphics[width=0.7\textwidth,angle=90]{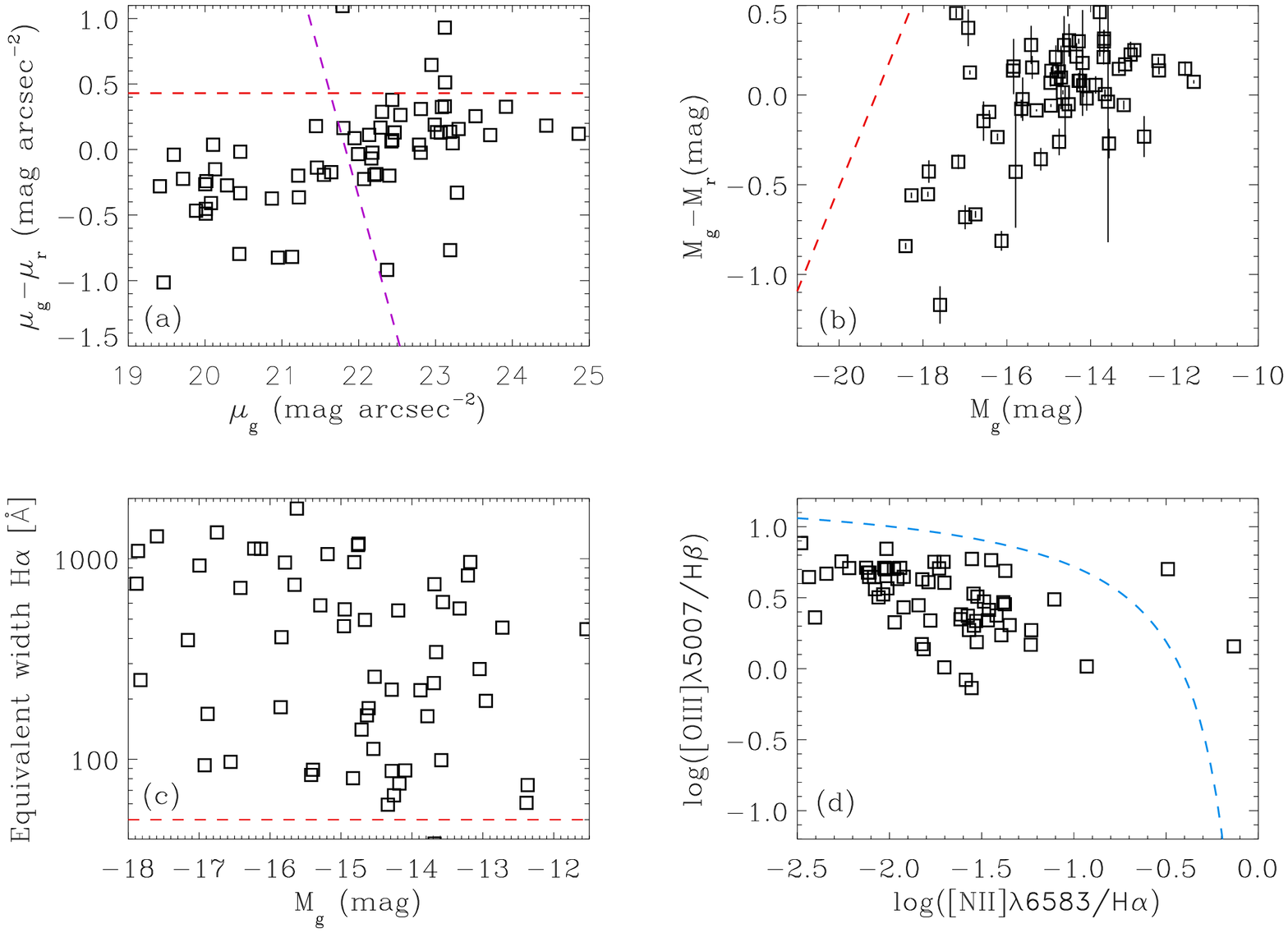}
\caption{
Similar to Fig.~\ref{diag_bcd}, but including previously
known XMP galaxies with spectra 
in SDSS/DR7.  
(a) Color $\mu_g-\mu_r$ vs surface brightness in the $g$ filter
$\mu_g$. Note that these targets have systematically lower 
surface brightness than the ones we find, and many of them  
cannot be considered BCDs according
to the criteria used in \S~\ref{properties}.
(b) Color $M_g-M_r$ vs absolute magnitude $M_g$ in the 
$g$ filter. 
\modified{
The error bars have been taken directly from SDSS.
}
(c) H$\alpha$ equivalent width vs absolute magnitudes.
(d) BPT diagram showing most of
the targets to be starforming galaxies.
The outlier comes from the sample of
special targets with broad emission lines 
selected by \citet{2007ApJ...671.1297I}, 
where the integrated spectra are known to have a
significant SNa and/or AGN contamination. 
In general,
the ratio between ${\rm[NII]}\lambda$6583 and H$\alpha$
is not as small as it is in our selection.
}
\label{diag_xmps}
\end{figure*}

Figure~\ref{diag_xmps} shows diagnostic plots like 
Fig.~\ref{diag_bcd} but for the galaxies in the literature 
with spectra in SDSS. 
Assuming this subset to be representative of the full family,
it shows several systematic differences with respect to the targets 
we have selected. The main one has to do with the ratio between
${\rm[NII]}\lambda$6583 and H$\alpha$, which is smaller
than those characterizing our targets (cf. Fig.~\ref{diag_bcd}d,
and Fig.~\ref{diag_xmps}d). Because of this difference, many of the
XMP found in the literature would have escaped our search, which pick out 
classes with extreme contrast between ${\rm[NII]}\lambda$6583 and H$\alpha$. 
We do not have a final explanation for the difference, but we can offer
two conjectures.
Our search may yield galaxies in the low metallicity
end of the XMP family, i.e., even more extreme than those existing in the 
literature and, consequently, even more metal poor than our estimates based 
on the semi-empirical calibrations by \citet{pet04}. Alternatively, 
our search may be missing a fraction of the XMP galaxies,
where the difference between the [NII] lines and H$\alpha$ is not
so extreme. The two possibilities explain other significant differences 
between our set and the targets 
from the literature. Our galaxies have larger 
surface brightness than those of the XMP
galaxies in the literature (cf. Fig.~\ref{diag_bcd}a and 
Fig.~\ref{diag_xmps}a). 
They have larger H$\alpha$ equivalent widths as well.
The surface brightness difference is attributable to the trend for the most metal poor 
galaxies to be BCDs (see \S~\ref{properties}). Alternatively, the physical 
conditions in the HII regions of the BCDs may differ systematically from other 
galaxies so that the same low metallicities render a particularly small
ratio between ${\rm[NII]}\lambda$6583 and H$\alpha$ in BCDs. In order to sort out
the two possibilities one would have to determine the metallicities 
of our targets in a self-consistent way measuring electron temperatures
and excitations \citep[as in, e.g.,][]{sta04,izo06}. 
Such a detailed analysis clearly goes beyond the scope of the paper, 
and it is planned for following up work.  Unfortunately, the
twenty one candidates 
that were already known to be XMP do not clarify the situation. 
Some have particularly low metallicities, but not all of them
(see the names with asterisks in Table~\ref{revision}).

%

\section{Number density of XMP galaxies in the local universe}\label{number_density}

One of the advantages of carrying out a systematic search in SDSS/DR7 is
starting from a magnitude limited sample. The relatively simple bias produced
by this condition can be corrected for and, thus,  the selected XMP galaxies 
can be used to estimate the volume  number density of these unusual 
objects in the  local universe. Such estimate is described in the present section.

Assume the metallicity $X$ of all galaxies in SDSS/DR7 
to be known. Then the number of galaxies 
per unit volume and metallicity is just
\begin{equation}
n(X)={1\over{\Delta X}}\sum_i{1\over {V_i}}\,\Pi\big({{X_i-X}\over{\Delta X}}\big),
\label{pi_eq}
\end{equation}
where the sum over $i$ includes all the galaxies in the sample, 
$\Delta X$ represents the bin size of the metallicity
histogram, the symbol $\Pi$ stands for the rectangle function,
\begin{equation}
\Pi(x)=\cases{
1&$|x|<1/2$,\cr
0&elsewhere,
}
\label{pi_eq6}
\end{equation}
and $V_i$ represents the maximum volume in which the $i$-th galaxy
of the sample could be observed. Equation~(\ref{pi_eq}) represents
the so-called $V_{max}$ approximation by \citet{sch68}  
used to determine luminosity function of galaxies \citep[e.g.,][]{tak00},
which has been transformed to derive the number density of any 
other physical property of the galaxies \citep[e.g.,][\S~4]{san08}.  
Assume the sample to be magnitude limited, 
so that all galaxies brighter than the apparent magnitude $m_{lim}$
are included. (This is the way the spectroscopic sample of SDSS 
was defined and therefore it must be good approximation.)
Then,
\begin{equation}
V_i={{d_i^3}\over{3}}\Omega,
\label{pi_eq7}
\end{equation}
with $\Omega$ the solid angle covered by the survey and 
$d_i$ the maximum distance at which the $i-$th galaxy can
be observed,
\begin{equation}
\log(d_{i}/{\rm 1 Mpc})= ({m_{lim}-M_i})/5-5,
\label{pi_eq8}
\end{equation}
which only depends on its absolute magnitude $M_i$. 
A few comments and caveats are in order before applying 
equation~(\ref{pi_eq}) to the dataset. 
First, SDSS is not truly magnitude limited. Some bright galaxies are 
not observed because of problems to pack the spectrograph fibers in 
crowded fields \citep[see][]{sto02} and, as usual,  very low surface brightness galaxies 
tend to be missed even if they are luminous \citep[e.g.,][]{bla05}. None of these 
two problems seems to be of relevance for the 
XMP galaxies, which are both 
isolated and high surface brightness (\S~\ref{properties}).  
SDSS/DR7 sets $m_{lim}=17.8$ in $r~$, but our selection also imposes 
a cut at redshift $\le 0.25$ which, in principle, modifies the magnitude-limited 
character of the original sample. This question is of no concern, though. 
Our targets are dwarf galaxies, never reaching a luminosity 
sufficient to be detected beyond the redshift threshold. 
Finally, computing the galaxy metallicities using detailed chemical
abundance analyses \citep[as in, e.g.,][]{sta04,izo06} goes beyond the 
scope of the paper. We estimate the metallicities using the strong line 
calibration of by \citet{pet04} described in \S~\ref{kmeans}.
They are given in Table~\ref{list1}.
Specifically, we employ the calibration in equation~(\ref{pet2}), which seems to be more 
appropriate for XMP galaxies where  N2$\sim -2$ (N2 is the logarithm
of ratio between the  equivalent widths of the lines,
as defined in equation~[\ref{n2def}]).
The use of this approximation for the estimate of the metallicity 
implies that our estimates are only indicative.

The 32 XMP galaxies selected in \S~\ref{kmeans}, together with 
equations~(\ref{pet2}), (\ref{pi_eq}), (\ref{pi_eq6}), (\ref{pi_eq7}), and (\ref{pi_eq8}),
render the local density of galaxies with (oxygen) metallicity 
less than,
approximately, one tenth of the solar value,
$12+\log({\rm O/H})~\leq~7.65$,
\begin{equation}
\int_{-\infty}^{7.65}n(X)dX\,\simeq (1.32\pm 0.23)\cdot10^{-4}\,{\rm Mpc}^{-3}.
\label{numberd1}
\end{equation}
The error bar in the previous expression assumes the 32 targets 
to be drawn from a Poisson distribution. In terms of the total number of 
galaxies, XMP galaxies are just
\begin{equation}
0.10\pm0.02\% ,
\label{per_xmp}
\end{equation}
where the total number of galaxies in the local universe ($\simeq 0.13\,{\rm Mpc}^{-3}$)
has been taken from the normalization of the luminosity functions by \citet{bla03b}. 
These luminosity functions are also based on SDSS, and we use 
a Hubble constant of 70~km\,s$^{-1}$\,Mpc$^{-1}$ to revert their normalization.
The error bar in equation~(\ref{per_xmp}) considers the dispersion 
quoted in equation~(\ref{numberd1}) together with the spread of values
among the different luminosity functions corresponding to the
five SDSS color filter bandpasses.

\begin{figure}
\includegraphics[width=0.45\textwidth]{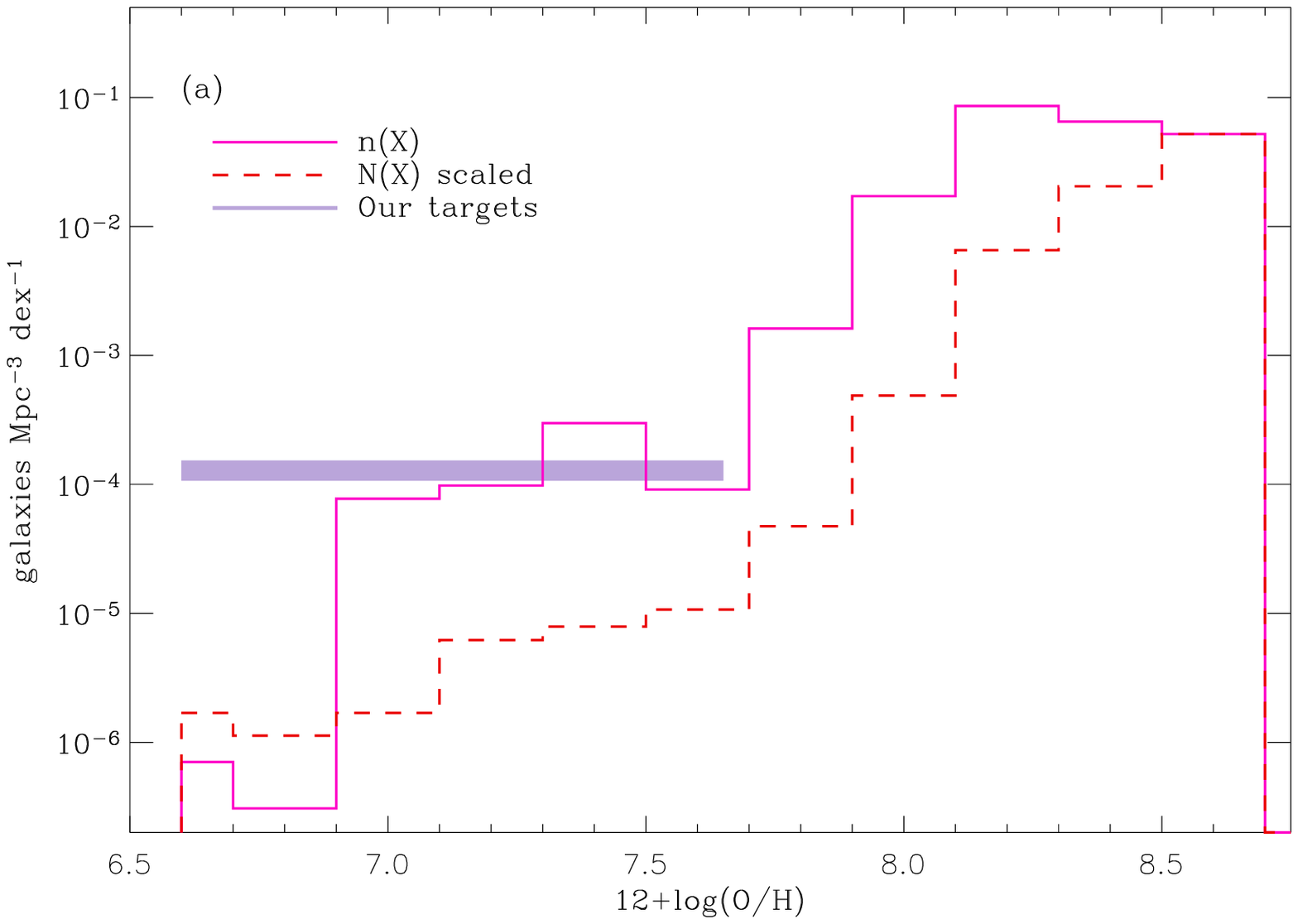}\\
\includegraphics[width=0.45\textwidth]{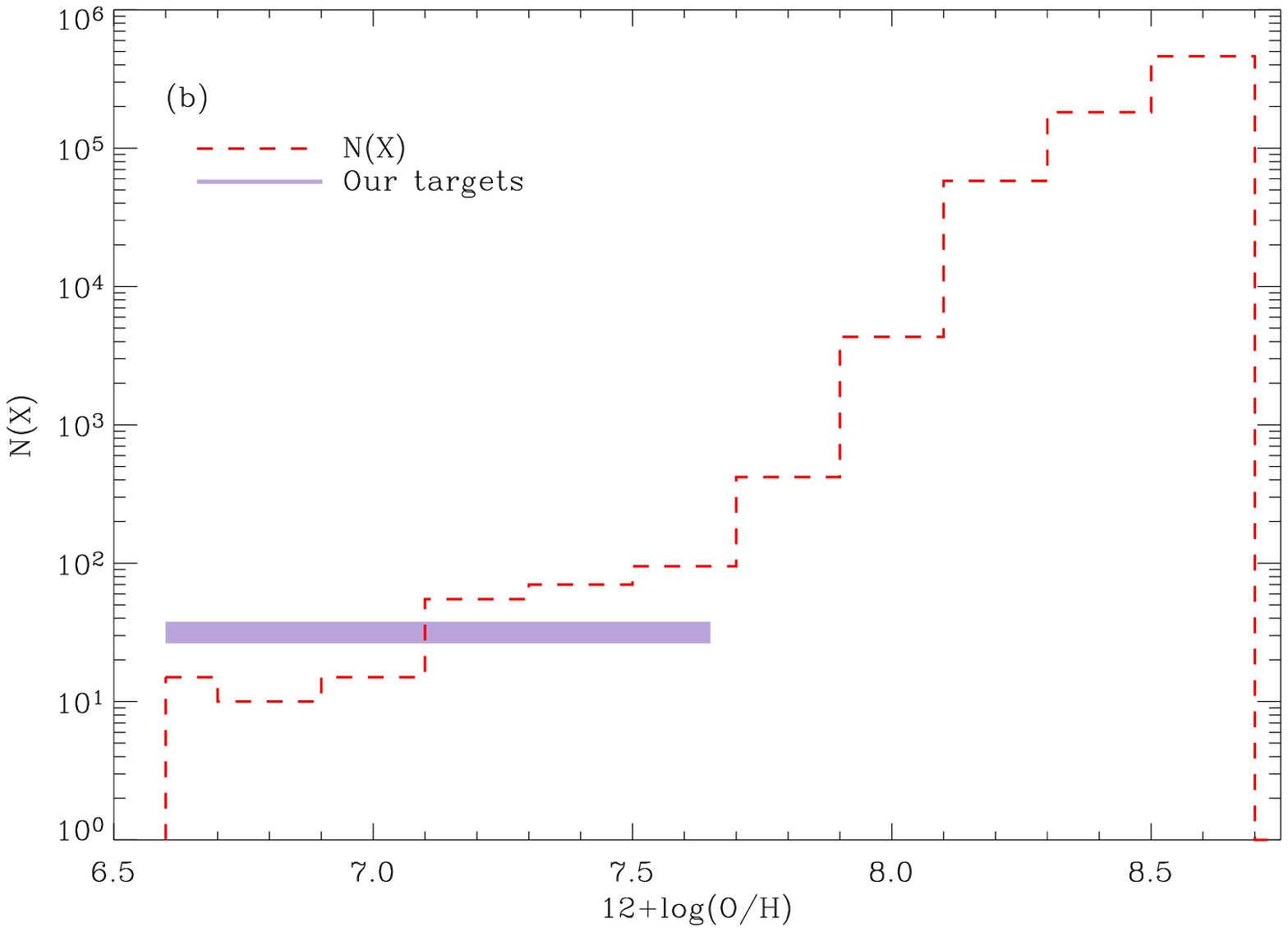}
\caption{
(a) Volume number density of galaxies with a given metallicity
as inferred from the SDSS/DR7 galaxies with emission lines 
($n[X]$, the solid line). The bar represents the average
value inferred from the 32 XMP galaxies identified in this 
study.
(b) Histogram of galaxies with a given metallicity as 
directly observed in SDSS/DR7 (the dashed line). A scaled
up version of this histogram is also shown as the dashed line in
(a).  Note how the already scarce XMP galaxies, represented by the
solid line, are 
hindered even further from observation because they are dwarf, and so, 
only observable in our  neighborhood.
}
\label{histograms}
\end{figure}

Going a step further, we have considered all the galaxies with emission 
lines in SDSS/DR7 to estimate $n(X)$ in a broader range of metallicities.
The SDSS/DR7 data reduction pipeline provides the equivalent widths of 
[NII]$\lambda$6583 and H$\alpha$, 
which we use to estimate oxygen abundances.  
The distribution function $n(X)$ inferred from all these galaxies 
is shown in Fig.~\ref{histograms}a, the solid line.
It has been computed using equation~(\ref{pi_eq}) with $\Delta X=0.2$,
and considering only N2 $\le -0.3$ ($\equiv\,12+\log({\rm O/H})< 8.7$), 
with this upper limit forced by the validity range of the
calibration used to estimate abundances \citep[see][]{pet04}. The
horizontal
bar in Fig.~\ref{histograms}a corresponds to the density in 
equation~(\ref{numberd1}), assuming the 32 XMP galaxies to spread 
one dex in metallicity. It is consistent with this other
independent estimate inferred from the full distribution of 
SDSS/DR7 galaxies -- if one integrates the histogram $n(X)$ in  
Fig.~\ref{histograms}a for galaxies with metallicities smaller than
one-tenth solar ($12+\log({\rm O/H}) \le 7.65$),
the volume number density of XMP galaxies turns out to be, 
$\int_{-\infty}^{7.65}n(X)dX\,\simeq 1.0\cdot10^{-4}\,{\rm Mpc}^{-3}$,
which is close to the figure in equation~(\ref{numberd1}).
%
The integral of $n(X)$ in Fig.~\ref{histograms}a
for all metallicities turns out to be 
$\simeq 5.2\cdot10^{-2}\,{\rm Mpc}^{-3}$, which is a factor
two smaller than,  but consistent
with, the number density of local galaxies from \citet{bla03b} used 
in equation~(\ref{per_xmp}) for normalization.
One have to keep in mind
that the distribution in Fig.~\ref{histograms}a do not consider 
neither galaxies without emission lines nor galaxies with
super-solar metallicities, which all together can easily account for 
the factor two difference. 

XMP galaxies are dwarf and therefore they tend 
to be underrepresented in surveys. In order to illustrate the 
importance of such Malmquist 
bias, Fig.~\ref{histograms}b contains the actual histogram of 
observed galaxies $N(X)$ from which $n(X)$ was derived.
In the parlance of equation~(\ref{pi_eq}), it is given by 
\begin{equation}
N(X)=\sum_i\Pi\big({{X_i-X}\over{\Delta X}}\big).
\label{other_phi}
\end{equation}
Note how the low metallicity tail of $N(X)$ is depressed with respect to
$n(X)$. This difference can be better appreciated in 
Fig.~\ref{histograms}a, which also shows a scaled version of 
$N(X)$ forced to agree with $n(X)$ at the 
bin of highest metallicity. The figure shows how XMP galaxies are 
clearly lacking in the observation  
(the dashed line) as compared to their actual number 
densities (the solid line). We find than only 0.01\%
of the observed galaxies with emission lines are XMP, whereas they
represent 0.2\% considering galaxies in a fixed volume.

Obviously, the above estimates assume the XMP galaxy set to be complete.
However, we cannot discard the existence of missing targets 
(\S~\ref{biblio_search}) and, if so,
the figures we provide have to be scaled up accordingly.
Although it is difficult to tell how much, we will try to make 
an educated guess. About sixty known XMP galaxies
in the SDSS field of view have not been included in our selection 
(see Fig.~\ref{spatial}). They may have been excluded
for several reasons -- 
because they are too faint, because noise in the spectra 
artificially enhances [NII]$\lambda$6583 with respect to H$\alpha$,
because [NII]$\lambda$6583 exceeds the limits of our selection,
and possibly others. Since some of them are proper reasons
for exclusion, the number of sixty undetected galaxies
may be taken as an upper limit for those missing. 
Consequently, the number densities we estimate may 
need to be increased, but only up to a factor of three.
%
%
\section{Conclusions}\label{conclusions}
We have carried out a systematic search for extremely metal poor 
(XMP) galaxies among the spectroscopic sample of SDSS/DR7. 
These objects are rare, and have a clear cosmological interest
as unevolved galaxies probably tracing
physical conditions in
an early phase of the universe 
(see \S~\ref{intro}).  The search is based on the classification 
of a narrow spectral region around H$\alpha$, known to be particularly 
sensitive to the (HII gas) metallicity. Obviously, the almost one million spectra
in the database cannot be inspected individually, and we 
resort to the use of a standard automatic method of classification:
k-means. After two nested runs of the procedure, and a
subsequent cleaning up for artifacts created by the SDSS 
pipeline, we end up with 32 targets (\S~\ref{kmeans}, Table~\ref{list1}).
They represent only 0.01\% of the observed galaxies with 
emission lines.
The final metallicity estimate remains pending, however,
strong-line empirical calibrations by \citet{pet04}
imply their oxygen metallicity to be of the order of
one-tenth solar.
Obviously, our candidates must be studied in detail through imaging 
and spectroscopy in following up work.

In order to put the work into context, we carried out
a bibliographic search for galaxies with metallicity smaller than
one-tenth the solar value (\S~\ref{biblio_search}). 
We find only
129
(Table~\ref{revision}), and only 21
of them overlap with our sample which, consequently, 
provides 11 new XMP 
candidates. The oxygen metallicity of the
21 known targets  
turns out to be $12+\log({\rm O/H)}\simeq 7.61\pm 0.19$. 
These metallicities
are based on electron temperatures and, thus, they are more precise than 
the empirical calibrations we have been employing. Assuming this subset to be
representative of the full sample, it confirms the XMP character of 
our targets.

Our procedure is systematic, therefore, in principle, it should have 
identified all the spectra in SDSS/DR7 where 
H$\alpha\gg {\rm[NII]}\lambda$6583. 
This assumption, together with the fact that 
the SDSS spectroscopic sample is limited in apparent magnitude,
allows us to estimate the volume number density of XMP galaxies.
Using the $V_{\rm max}$ approximation, we estimate it 
to be  $(1.32\pm 0.23)\cdot 10^{-4}~{\rm Mpc}^{-3}$. 
So far as we are aware of, 
it provides the first estimate of 
this number density.
The XMP galaxies represent 0.1\%\ of the galaxies in the local 
volume,  or $\sim$ 0.2\%\ considering only emission line galaxies.

We analyze some of the physical properties of the candidates
in \S~\ref{properties}.  All but four of our XMP candidates turn 
out to be blue compact dwarfs (BCDs). 
Note that this association is by no means trivial. We have selected
our sample according to the shape of their spectra in an narrow 
spectral window around H$\alpha$, and this narrow bit
of spectrum turns out to determine many global properties of the 
galaxy such as color, compactness, and star formation rate. 
We ignore what causes the association between XMPs and 
BCDs. The fact that most metal poor galaxies tend to be BCDs
is known. \citet{kun00} point it out, but warn against a trivial
interpretation since it may reflect an observational bias that
makes it easier to detect high surface brightness objects such as BCDs. 
Actually, there are XMP galaxies whose surface brightness does not 
suffice to call them BCDs (see \S~\ref{biblio_search}). However, 
the fact that our XMP candidates are also BCDs is revealing. 
The surface brightness where SDSS starts having problems of completeness 
is about  $\mu_g > 23.5$ \citep{bla05}, i.e., one magnitude 
fainter than the faintest XMP candidate. Consequently, if we do not find 
low surface brightness galaxies with 
${\rm[NII]}\lambda 6583 \ll {\rm H}\alpha$ is because they do not exist 
in SDSS/DR7.

Among the 32 XMP candidates, 24 of them have either 
cometary shape or are formed by chains of knots.
\citet{2008A&A...491..113P} already noticed the trend for XMP BCDs to
reveal a cometary morphology due to the presence of 
intense star formation at one edge that gradually 
decreases. This shape is not unique to XMP galaxies, but it comprises 
only 10\% of all BCDs \citep{loo85}, whereas it is dominant 
when they are metal poor \citep{2008A&A...491..113P}. 
The origin of the XMP shapes is also unclear. There are arguments
for gravitational triggering due to mergers with low-mass stellar or
gaseous companions,  or for the self propagation of star formation
activity within a pre-existing gas rich galaxy \citep{2008A&A...491..113P}. Recently 
\citet{ekt10} and \citet{ekt08} have 
found that all XMP galaxies have distorted HI morphologies, which may indicate 
the infall of external unenriched gas feeding the starburst and dropping the 
metallicity \citep[e.g.,][]{kew06}. It may also be the signature of gas 
stripping forced by the interaction with an external medium 
\citep[e.g.,][]{gav01,elm10}.

A concluding remark 
is in order. The methods for mining massive 
astronomical databases are still under development. 
The bases are simply too large for the traditional techniques to be 
efficient. In this sense, the current paper presents a new approach that may be of 
interest beyond our particular application. The standard method to find 
XMP galaxies (or galaxies with any other property) would have been to set, 
beforehand, the observational criteria the targets should fulfill. Then
those targets complying with the criteria would have been selected. 
Unfortunately, the criteria are often crude and have large uncertainties, 
which propagate into the selection as false objects sneaking in or 
true objects leaking out.  Here the criteria have not been stipulated 
in advance.  We classify the whole database, and only afterward 
we select those galaxies belonging the classes that have the intended 
property. The search is comprehensive, and the results are robust 
against uncertainties in the selection criteria.

%
%
%
%
%
%
%
%

%
%
\begin{acknowledgements}
Thanks are due to B. Elmegreen for fruitful discussions on 
the origin of the cometary shapes in XMP galaxies.
Thanks are also due to an anonymous referee for 
helping us completing the bibliographic search, as well as for 
insightful comments that prompted an improvement of both
the argumentation and the presentation.
%
%
%
This work has been partly funded by the Spanish MICINN, projects
AYA~2007-67965-03-01,           
AYA~2007-67752-C03-01, and      
AYA~2010-21887-C04-04.          
%
JSA, ALA and CMT are members of the Consolider-Ingenio 2010 Program, grant 
MICINN CSD2006-00070: First Science with GTC.
Funding for the SDSS and SDSS-II has been provided by the Alfred P. Sloan
Foundation, the Participating Institutions, the National Science Foundation, the
U.S. Department of Energy, the National Aeronautics and Space Administration,
the Japanese Monbukagakusho, the Max Planck Society, and the Higher Education
Funding Council for England. The SDSS is managed by the Astrophysical Research 
Consortium for the Participating Institutions (for details,
see the SDSS web site at http://www.sdss.org/).

{\it Facilities:} \facility{Sloan (DR7, spectra)
}
\end{acknowledgements}

%
%

\begin{thebibliography}{103}
\expandafter\ifx\csname natexlab\endcsname\relax\def\natexlab#1{#1}\fi

\bibitem[{{Abazajian} {et~al.}(2009){Abazajian}, {Adelman-McCarthy},
  {Ag{\"u}eros}, {Allam}, {Prieto}, {An}, {Anderson}, {Anderson}, {Annis},
  {Bahcall}, {Bailer-Jones}, {Barentine}, {Bassett}, {Becker}, {Beers}, {Bell},
  {Belokurov}, {Berlind}, {Berman}, {Bernardi}, {Bickerton}, {Bizyaev},
  {Blakeslee}, {Blanton}, {Bochanski}, {Boroski}, {Brewington}, {Brinchmann},
  {Brinkmann}, {Brunner}, {Budav{\'a}ri}, {Carey}, {Carliles}, {Carr},
  {Castander}, {Cinabro}, {Connolly}, {Csabai}, {Cunha}, {Czarapata},
  {Davenport}, {de Haas}, {Dilday}, {Doi}, {Eisenstein}, {Evans}, {Evans},
  {Fan}, {Friedman}, {Frieman}, {Fukugita}, {G{\"a}nsicke}, {Gates},
  {Gillespie}, {Gilmore}, {Gonzalez}, {Gonzalez}, {Grebel}, {Gunn},
  {Gy{\"o}ry}, {Hall}, {Harding}, {Harris}, {Harvanek}, {Hawley}, {Hayes},
  {Heckman}, {Hendry}, {Hennessy}, {Hindsley}, {Hoblitt}, {Hogan}, {Hogg},
  {Holtzman}, {Hyde}, {Ichikawa}, {Ichikawa}, {Im}, {Ivezi{\'c}}, {Jester},
  {Jiang}, {Johnson}, {Jorgensen}, {Juri{\'c}}, {Kent}, {Kessler}, {Kleinman},
  {Knapp}, {Konishi}, {Kron}, {Krzesinski}, {Kuropatkin}, {Lampeitl},
  {Lebedeva}, {Lee}, {Lee}, {Leger}, {L{\'e}pine}, {Li}, {Lima}, {Lin}, {Long},
  {Loomis}, {Loveday}, {Lupton}, {Magnier}, {Malanushenko}, {Malanushenko},
  {Mandelbaum}, {Margon}, {Marriner}, {Mart{\'{\i}}nez-Delgado}, {Matsubara},
  {McGehee}, {McKay}, {Meiksin}, {Morrison}, {Mullally}, {Munn}, {Murphy},
  {Nash}, {Nebot}, {Neilsen}, {Newberg}, {Newman}, {Nichol}, {Nicinski},
  {Nieto-Santisteban}, {Nitta}, {Okamura}, {Oravetz}, {Ostriker}, {Owen},
  {Padmanabhan}, {Pan}, {Park}, {Pauls}, {Peoples}, {Percival}, {Pier}, {Pope},
  {Pourbaix}, {Price}, {Purger}, {Quinn}, {Raddick}, {Fiorentin}, {Richards},
  {Richmond}, {Riess}, {Rix}, {Rockosi}, {Sako}, {Schlegel}, {Schneider},
  {Scholz}, {Schreiber}, {Schwope}, {Seljak}, {Sesar}, {Sheldon}, {Shimasaku},
  {Sibley}, {Simmons}, {Sivarani}, {Smith}, {Smith}, {Smol{\v c}i{\'c}},
  {Snedden}, {Stebbins}, {Steinmetz}, {Stoughton}, {Strauss}, {Subba Rao},
  {Suto}, {Szalay}, {Szapudi}, {Szkody}, {Tanaka}, {Tegmark}, {Teodoro},
  {Thakar}, {Tremonti}, {Tucker}, {Uomoto}, {Vanden Berk}, {Vandenberg},
  {Vidrih}, {Vogeley}, {Voges}, {Vogt}, {Wadadekar}, {Watters}, {Weinberg},
  {West}, {White}, {Wilhite}, {Wonders}, {Yanny}, {Yocum}, {York}, {Zehavi},
  {Zibetti}, \& {Zucker}}]{aba09}
{Abazajian}, K.~N., {Adelman-McCarthy}, J.~K., {Ag{\"u}eros}, M.~A., {et~al.}
  2009, \apjs, 182, 543

\bibitem[{{Amor{\'{\i}}n} {et~al.}(2009){Amor{\'{\i}}n}, {Aguerri},
  {Mu{\~n}oz-Tu{\~n}{\'o}n}, \& {Cair{\'o}s}}]{amo09}
{Amor{\'{\i}}n}, R., {Aguerri}, J.~A.~L., {Mu{\~n}oz-Tu{\~n}{\'o}n}, C., \&
  {Cair{\'o}s}, L.~M. 2009, \aap, 501, 75

\bibitem[{{Amor{\'{\i}}n} {et~al.}(2007){Amor{\'{\i}}n},
  {Mu{\~n}oz-Tu{\~n}{\'o}n}, {Aguerri}, {Cair{\'o}s}, \& {Caon}}]{amo07}
{Amor{\'{\i}}n}, R.~O., {Mu{\~n}oz-Tu{\~n}{\'o}n}, C., {Aguerri}, J.~A.~L.,
  {Cair{\'o}s}, L.~M., \& {Caon}, N. 2007, \aap, 467, 541

\bibitem[{{Asplund}(2005)}]{asp05b}
{Asplund}, M. 2005, \araa, 43, 481

\bibitem[{{Baldwin} {et~al.}(1981){Baldwin}, {Phillips}, \&
  {Terlevich}}]{bal81}
{Baldwin}, J.~A., {Phillips}, M.~M., \& {Terlevich}, R. 1981, \pasp, 93, 5

\bibitem[{{Balogh} {et~al.}(2001){Balogh}, {Pearce}, {Bower}, \&
  {Kay}}]{2001MNRAS.326.1228B}
{Balogh}, M.~L., {Pearce}, F.~R., {Bower}, R.~G., \& {Kay}, S.~T. 2001, \mnras,
  326, 1228

\bibitem[{{Bishop}(2006)}]{bis06}
{Bishop}, C.~M. 2006, Pattern Recognition and Machine Learning (NY: Springer)

\bibitem[{{Blanton} {et~al.}(2003){Blanton}, {Hogg}, {Bahcall}, {Brinkmann},
  {Britton}, {Connolly}, {Csabai}, {Fukugita}, {Loveday}, {Meiksin}, {Munn},
  {Nichol}, {Okamura}, {Quinn}, {Schneider}, {Shimasaku}, {Strauss}, {Tegmark},
  {Vogeley}, \& {Weinberg}}]{bla03b}
{Blanton}, M.~R., {Hogg}, D.~W., {Bahcall}, N.~A., {et~al.} 2003, \apj, 592,
  819

\bibitem[{{Blanton} {et~al.}(2005){Blanton}, {Lupton}, {Schlegel}, {Strauss},
  {Brinkmann}, {Fukugita}, \& {Loveday}}]{bla05}
{Blanton}, M.~R., {Lupton}, R.~H., {Schlegel}, D.~J., {et~al.} 2005, \apj, 631,
  208

\bibitem[{{Blanton} \& {Moustakas}(2009)}]{2009ARA&A..47..159B}
{Blanton}, M.~R. \& {Moustakas}, J. 2009, \araa, 47, 159

\bibitem[{{Bromm} \& {Larson}(2004)}]{bro04}
{Bromm}, V. \& {Larson}, R.~B. 2004, \araa, 42, 79

\bibitem[{{Brown} {et~al.}(2008){Brown}, {Kewley}, \&
  {Geller}}]{2008AJ....135...92B}
{Brown}, W.~R., {Kewley}, L.~J., \& {Geller}, M.~J. 2008, \aj, 135, 92

\bibitem[{{Cair{\'o}s} {et~al.}(2001{\natexlab{a}}){Cair{\'o}s}, {Caon},
  {V{\'{\i}}lchez}, {Gonz{\'a}lez-P{\'e}rez}, \&
  {Mu{\~n}oz-Tu{\~n}{\'o}n}}]{cai01c}
{Cair{\'o}s}, L.~M., {Caon}, N., {V{\'{\i}}lchez}, J.~M.,
  {Gonz{\'a}lez-P{\'e}rez}, J.~N., \& {Mu{\~n}oz-Tu{\~n}{\'o}n}, C.
  2001{\natexlab{a}}, \apjs, 136, 393

\bibitem[{{Cair{\'o}s} {et~al.}(2010){Cair{\'o}s}, {Caon}, {Zurita}, {Kehrig},
  {Roth}, \& {Weilbacher}}]{2010A&A...520A..90C}
{Cair{\'o}s}, L.~M., {Caon}, N., {Zurita}, C., {et~al.} 2010, \aap, 520, A90+

\bibitem[{{Cair{\'o}s} {et~al.}(2001{\natexlab{b}}){Cair{\'o}s},
  {V{\'{\i}}lchez}, {Gonz{\'a}lez P{\'e}rez}, {Iglesias-P{\'a}ramo}, \&
  {Caon}}]{cai01}
{Cair{\'o}s}, L.~M., {V{\'{\i}}lchez}, J.~M., {Gonz{\'a}lez P{\'e}rez}, J.~N.,
  {Iglesias-P{\'a}ramo}, J., \& {Caon}, N. 2001{\natexlab{b}}, \apjs, 133, 321

\bibitem[{{Denicol{\'o}} {et~al.}(2002){Denicol{\'o}}, {Terlevich}, \&
  {Terlevich}}]{den02}
{Denicol{\'o}}, G., {Terlevich}, R., \& {Terlevich}, E. 2002, \mnras, 330, 69

\bibitem[{{Diemand} {et~al.}(2007){Diemand}, {Kuhlen}, \& {Madau}}]{die07}
{Diemand}, J., {Kuhlen}, M., \& {Madau}, P. 2007, \apj, 657, 262

\bibitem[{{Ekta} {et~al.}(2008){Ekta}, {Chengalur}, \& {Pustilnik}}]{ekt08}
{Ekta}, {Chengalur}, J.~N., \& {Pustilnik}, S.~A. 2008, \mnras, 391, 881

\bibitem[{{Ekta} \& {Chengalur}(2010{\natexlab{a}})}]{ekt10}
{Ekta}, B. \& {Chengalur}, J.~N. 2010{\natexlab{a}}, \mnras, 403, 295

\bibitem[{{Ekta} \& {Chengalur}(2010{\natexlab{b}})}]{2010MNRAS.406.1238E}
{Ekta}, B. \& {Chengalur}, J.~N. 2010{\natexlab{b}}, \mnras, 406, 1238

\bibitem[{{Ellison} {et~al.}(2000){Ellison}, {Songaila}, {Schaye}, \&
  {Pettini}}]{ell00}
{Ellison}, S.~L., {Songaila}, A., {Schaye}, J., \& {Pettini}, M. 2000, \aj,
  120, 1175

\bibitem[{{Elmegreen} \& {Elmegreen}(2010)}]{elm10}
{Elmegreen}, B.~G. \& {Elmegreen}, D.~M. 2010, \apj, 722, 1895

\bibitem[{{Everitt}(1995)}]{eve95}
{Everitt}, B.~S. 1995, Cluster Analysis (London: Arnold)

\bibitem[{{Gavazzi} {et~al.}(2001){Gavazzi}, {Boselli}, {Mayer},
  {Iglesias-Paramo}, {V{\'{\i}}lchez}, \& {Carrasco}}]{gav01}
{Gavazzi}, G., {Boselli}, A., {Mayer}, L., {et~al.} 2001, \apjl, 563, L23

\bibitem[{{Gil de Paz} {et~al.}(2003){Gil de Paz}, {Madore}, \&
  {Pevunova}}]{gil03}
{Gil de Paz}, A., {Madore}, B.~F., \& {Pevunova}, O. 2003, \apjs, 147, 29

\bibitem[{{Griffith} {et~al.}(2011){Griffith}, {Tsai}, {Stern}, {Blain},
  {Eisenhardt}, {Harrison}, {Jarrett}, {Madsen}, {Stanford}, {Wright}, {Wu},
  {Wu}, \& {Yan}}]{2011arXiv1106.4844G}
{Griffith}, R.~L., {Tsai}, C.-W., {Stern}, D., {et~al.} 2011, ArXiv e-prints

\bibitem[{{Guseva} {et~al.}(2001){Guseva}, {Izotov}, {Papaderos}, {Chaffee},
  {Foltz}, {Green}, {Thuan}, {Fricke}, \& {Noeske}}]{2001A&A...378..756G}
{Guseva}, N.~G., {Izotov}, Y.~I., {Papaderos}, P., {et~al.} 2001, \aap, 378,
  756

\bibitem[{{Guseva} {et~al.}(2007){Guseva}, {Izotov}, {Papaderos}, \&
  {Fricke}}]{2007A&A...464..885G}
{Guseva}, N.~G., {Izotov}, Y.~I., {Papaderos}, P., \& {Fricke}, K.~J. 2007,
  \aap, 464, 885

\bibitem[{{Guseva} {et~al.}(2003{\natexlab{a}}){Guseva}, {Papaderos}, {Izotov},
  {Green}, {Fricke}, {Thuan}, \& {Noeske}}]{gus03c}
{Guseva}, N.~G., {Papaderos}, P., {Izotov}, Y.~I., {et~al.} 2003{\natexlab{a}},
  \aap, 407, 75

\bibitem[{{Guseva} {et~al.}(2003{\natexlab{b}}){Guseva}, {Papaderos}, {Izotov},
  {Green}, {Fricke}, {Thuan}, \& {Noeske}}]{2003A&A...407...91G}
{Guseva}, N.~G., {Papaderos}, P., {Izotov}, Y.~I., {et~al.} 2003{\natexlab{b}},
  \aap, 407, 91

\bibitem[{{Guseva} {et~al.}(2003{\natexlab{c}}){Guseva}, {Papaderos}, {Izotov},
  {Green}, {Fricke}, {Thuan}, \& {Noeske}}]{2003A&A...407..105G}
{Guseva}, N.~G., {Papaderos}, P., {Izotov}, Y.~I., {et~al.} 2003{\natexlab{c}},
  \aap, 407, 105

\bibitem[{{Guseva} {et~al.}(2009){Guseva}, {Papaderos}, {Meyer}, {Izotov}, \&
  {Fricke}}]{2009A&A...505...63G}
{Guseva}, N.~G., {Papaderos}, P., {Meyer}, H.~T., {Izotov}, Y.~I., \& {Fricke},
  K.~J. 2009, \aap, 505, 63

\bibitem[{{Hidalgo-G{\'a}mez} \& {Olofsson}(2002)}]{2002A&A...389..836H}
{Hidalgo-G{\'a}mez}, A.~M. \& {Olofsson}, K. 2002, \aap, 389, 836

\bibitem[{{Izotov} {et~al.}(2009){Izotov}, {Guseva}, {Fricke}, \&
  {Papaderos}}]{izo09}
{Izotov}, Y.~I., {Guseva}, N.~G., {Fricke}, K.~J., \& {Papaderos}, P. 2009,
  \aap, 503, 61

\bibitem[{{Izotov} {et~al.}(2011){Izotov}, {Guseva}, \&
  {Thuan}}]{2011ApJ...728..161I}
{Izotov}, Y.~I., {Guseva}, N.~G., \& {Thuan}, T.~X. 2011, \apj, 728, 161

\bibitem[{{Izotov} {et~al.}(2004){Izotov}, {Papaderos}, {Guseva}, {Fricke}, \&
  {Thuan}}]{2004A&A...421..539I}
{Izotov}, Y.~I., {Papaderos}, P., {Guseva}, N.~G., {Fricke}, K.~J., \& {Thuan},
  T.~X. 2004, \aap, 421, 539

\bibitem[{{Izotov} {et~al.}(2006{\natexlab{a}}){Izotov}, {Papaderos}, {Guseva},
  {Fricke}, \& {Thuan}}]{2006A&A...454..137I}
{Izotov}, Y.~I., {Papaderos}, P., {Guseva}, N.~G., {Fricke}, K.~J., \& {Thuan},
  T.~X. 2006{\natexlab{a}}, \aap, 454, 137

\bibitem[{{Izotov} {et~al.}(2006{\natexlab{b}}){Izotov}, {Stasi{\'n}ska},
  {Meynet}, {Guseva}, \& {Thuan}}]{izo06}
{Izotov}, Y.~I., {Stasi{\'n}ska}, G., {Meynet}, G., {Guseva}, N.~G., \&
  {Thuan}, T.~X. 2006{\natexlab{b}}, \aap, 448, 955

\bibitem[{{Izotov} \& {Thuan}(2002)}]{2002ApJ...567..875I}
{Izotov}, Y.~I. \& {Thuan}, T.~X. 2002, \apj, 567, 875

\bibitem[{{Izotov} \& {Thuan}(2004{\natexlab{a}})}]{2004ApJ...616..768I}
{Izotov}, Y.~I. \& {Thuan}, T.~X. 2004{\natexlab{a}}, \apj, 616, 768

\bibitem[{{Izotov} \& {Thuan}(2004{\natexlab{b}})}]{2004ApJ...602..200I}
{Izotov}, Y.~I. \& {Thuan}, T.~X. 2004{\natexlab{b}}, \apj, 602, 200

\bibitem[{{Izotov} \& {Thuan}(2007)}]{2007ApJ...665.1115I}
{Izotov}, Y.~I. \& {Thuan}, T.~X. 2007, \apj, 665, 1115

\bibitem[{{Izotov} {et~al.}(2005){Izotov}, {Thuan}, \& {Guseva}}]{izo05}
{Izotov}, Y.~I., {Thuan}, T.~X., \& {Guseva}, N.~G. 2005, \apj, 632, 210

\bibitem[{{Izotov} {et~al.}(2007{\natexlab{a}}){Izotov}, {Thuan}, \&
  {Guseva}}]{2007ApJ...671.1297I}
{Izotov}, Y.~I., {Thuan}, T.~X., \& {Guseva}, N.~G. 2007{\natexlab{a}}, \apj,
  671, 1297

\bibitem[{{Izotov} {et~al.}(2007{\natexlab{b}}){Izotov}, {Thuan}, \&
  {Stasi{\'n}ska}}]{2007ApJ...662...15I}
{Izotov}, Y.~I., {Thuan}, T.~X., \& {Stasi{\'n}ska}, G. 2007{\natexlab{b}},
  \apj, 662, 15

\bibitem[{{Jones} {et~al.}(2004){Jones}, {Saunders}, {Colless}, {Read},
  {Parker}, {Watson}, {Campbell}, {Burkey}, {Mauch}, {Moore}, {Hartley},
  {Cass}, {James}, {Russell}, {Fiegert}, {Dawe}, {Huchra}, {Jarrett}, {Lahav},
  {Lucey}, {Mamon}, {Proust}, {Sadler}, \& {Wakamatsu}}]{jon04}
{Jones}, D.~H., {Saunders}, W., {Colless}, M., {et~al.} 2004, \mnras, 355, 747

\bibitem[{{Kauffmann} {et~al.}(2003){Kauffmann}, {Heckman}, {Tremonti},
  {Brinchmann}, {Charlot}, {White}, {Ridgway}, {Brinkmann}, {Fukugita}, {Hall},
  {Ivezi{\'c}}, {Richards}, \& {Schneider}}]{kau03}
{Kauffmann}, G., {Heckman}, T.~M., {Tremonti}, C., {et~al.} 2003, \mnras, 346,
  1055

\bibitem[{{Kewley} {et~al.}(2006){Kewley}, {Geller}, \& {Barton}}]{kew06}
{Kewley}, L.~J., {Geller}, M.~J., \& {Barton}, E.~J. 2006, \aj, 131, 2004

\bibitem[{{Klypin} {et~al.}(1999){Klypin}, {Kravtsov}, {Valenzuela}, \&
  {Prada}}]{1999ApJ...522...82K}
{Klypin}, A., {Kravtsov}, A.~V., {Valenzuela}, O., \& {Prada}, F. 1999, \apj,
  522, 82

\bibitem[{{Kniazev} {et~al.}(2003){Kniazev}, {Grebel}, {Hao}, {Strauss},
  {Brinkmann}, \& {Fukugita}}]{2003ApJ...593L..73K}
{Kniazev}, A.~Y., {Grebel}, E.~K., {Hao}, L., {et~al.} 2003, \apjl, 593, L73

\bibitem[{{Kniazev} {et~al.}(2005){Kniazev}, {Grebel}, {Pustilnik}, {Pramskij},
  \& {Zucker}}]{2005AJ....130.1558K}
{Kniazev}, A.~Y., {Grebel}, E.~K., {Pustilnik}, S.~A., {Pramskij}, A.~G., \&
  {Zucker}, D.~B. 2005, \aj, 130, 1558

\bibitem[{{Kniazev} \& {Pustil'Nik}(1998)}]{1998BSAO...46...23K}
{Kniazev}, A.~Y. \& {Pustil'Nik}, S.~A. 1998, Bull.~Special Astrophys.~Obs.,
  46, 23

\bibitem[{{Kniazev} {et~al.}(2000){Kniazev}, {Pustilnik}, {Masegosa},
  {M{\'a}rquez}, {Ugryumov}, {Martin}, {Izotov}, {Engels}, {Brosch}, {Hopp},
  {Merlino}, \& {Lipovetsky}}]{2000A&A...357..101K}
{Kniazev}, A.~Y., {Pustilnik}, S.~A., {Masegosa}, J., {et~al.} 2000, \aap, 357,
  101

\bibitem[{{Kniazev} {et~al.}(2001){Kniazev}, {Pustilnik}, {Ugryumov}, \&
  {Pramsky}}]{2001A&A...371..404K}
{Kniazev}, A.~Y., {Pustilnik}, S.~A., {Ugryumov}, A.~V., \& {Pramsky}, A.~G.
  2001, \aap, 371, 404

\bibitem[{{Kunth} \& {{\"O}stlin}(2000)}]{kun00}
{Kunth}, D. \& {{\"O}stlin}, G. 2000, \aapr, 10, 1

\bibitem[{{Kunth} \& {Sargent}(1986)}]{kun86}
{Kunth}, D. \& {Sargent}, W.~L.~W. 1986, \apj, 300, 496

\bibitem[{{Lee} {et~al.}(2003){Lee}, {Grebel}, \&
  {Hodge}}]{2003A&A...401..141L}
{Lee}, H., {Grebel}, E.~K., \& {Hodge}, P.~W. 2003, \aap, 401, 141

\bibitem[{{Lee} {et~al.}(2006){Lee}, {Skillman}, {Cannon}, {Jackson}, {Gehrz},
  {Polomski}, \& {Woodward}}]{2006ApJ...647..970L}
{Lee}, H., {Skillman}, E.~D., {Cannon}, J.~M., {et~al.} 2006, \apj, 647, 970

\bibitem[{{Lee} {et~al.}(2004){Lee}, {Salzer}, \&
  {Melbourne}}]{2004ApJ...616..752L}
{Lee}, J.~C., {Salzer}, J.~J., \& {Melbourne}, J. 2004, \apj, 616, 752

\bibitem[{{Lequeux} {et~al.}(1979){Lequeux}, {Peimbert}, {Rayo}, {Serrano}, \&
  {Torres-Peimbert}}]{leq79}
{Lequeux}, J., {Peimbert}, M., {Rayo}, J.~F., {Serrano}, A., \&
  {Torres-Peimbert}, S. 1979, \aap, 80, 155

\bibitem[{{Lilly} {et~al.}(2003){Lilly}, {Carollo}, \&
  {Stockton}}]{2003ApJ...597..730L}
{Lilly}, S.~J., {Carollo}, C.~M., \& {Stockton}, A.~N. 2003, \apj, 597, 730

\bibitem[{{Loose} \& {Thuan}(1985)}]{loo85}
{Loose}, H. \& {Thuan}, T.~X. 1985, in Star-forming Dwarf Galaxies and Related
  Objects, ed. {D.~Kunth, T.~X.~Thuan, \& J.~Tran Thanh Van} (Gif sur Yvette:
  Editions Fronti\`eres), 73

\bibitem[{{Malmberg}(2005)}]{mal05}
{Malmberg}, D. 2005, Master's thesis, Uppsala University, Uppsala

\bibitem[{{Masegosa} {et~al.}(1994){Masegosa}, {Moles}, \&
  {Campos-Aguilar}}]{1994ApJ...420..576M}
{Masegosa}, J., {Moles}, M., \& {Campos-Aguilar}, A. 1994, \apj, 420, 576

\bibitem[{{Nagao} {et~al.}(2006){Nagao}, {Maiolino}, \&
  {Marconi}}]{2006A&A...459...85N}
{Nagao}, T., {Maiolino}, R., \& {Marconi}, A. 2006, \aap, 459, 85

\bibitem[{{Nava} {et~al.}(2006){Nava}, {Casebeer}, {Henry}, \&
  {Jevremovic}}]{2006ApJ...645.1076N}
{Nava}, A., {Casebeer}, D., {Henry}, R.~B.~C., \& {Jevremovic}, D. 2006, \apj,
  645, 1076

\bibitem[{{Pagel} {et~al.}(1992){Pagel}, {Simonson}, {Terlevich}, \&
  {Edmunds}}]{pag92}
{Pagel}, B.~E.~J., {Simonson}, E.~A., {Terlevich}, R.~J., \& {Edmunds}, M.~G.
  1992, \mnras, 255, 325

\bibitem[{{Papaderos} {et~al.}(2008){Papaderos}, {Guseva}, {Izotov}, \&
  {Fricke}}]{2008A&A...491..113P}
{Papaderos}, P., {Guseva}, N.~G., {Izotov}, Y.~I., \& {Fricke}, K.~J. 2008,
  \aap, 491, 113

\bibitem[{{Papaderos} {et~al.}(2006){Papaderos}, {Guseva}, {Izotov}, {Noeske},
  {Thuan}, \& {Fricke}}]{2006A&A...457...45P}
{Papaderos}, P., {Guseva}, N.~G., {Izotov}, Y.~I., {et~al.} 2006, \aap, 457, 45

\bibitem[{{Peimbert} \& {Torres-Peimbert}(1974)}]{pei74}
{Peimbert}, M. \& {Torres-Peimbert}, S. 1974, \apj, 193, 327

\bibitem[{{P{\'e}rez-Montero} \& {Contini}(2009)}]{2009MNRAS.398..949P}
{P{\'e}rez-Montero}, E. \& {Contini}, T. 2009, \mnras, 398, 949

\bibitem[{{Pettini} \& {Pagel}(2004)}]{pet04}
{Pettini}, M. \& {Pagel}, B.~E.~J. 2004, \mnras, 348, L59

\bibitem[{{Pilyugin}(2001)}]{2001A&A...374..412P}
{Pilyugin}, L.~S. 2001, \aap, 374, 412

\bibitem[{{Popescu} \& {Hopp}(2000)}]{2000A&AS..142..247P}
{Popescu}, C.~C. \& {Hopp}, U. 2000, \aaps, 142, 247

\bibitem[{{Pustilnik} {et~al.}(2004){Pustilnik}, {Kniazev}, {Pramskij},
  {Izotov}, {Foltz}, {Brosch}, {Martin}, \& {Ugryumov}}]{2004A&A...419..469P}
{Pustilnik}, S., {Kniazev}, A., {Pramskij}, A., {et~al.} 2004, \aap, 419, 469

\bibitem[{{Pustilnik} {et~al.}(2006){Pustilnik}, {Engels}, {Kniazev},
  {Pramskij}, {Ugryumov}, \& {Hagen}}]{2006AstL...32..228P}
{Pustilnik}, S.~A., {Engels}, D., {Kniazev}, A.~Y., {et~al.} 2006, Astronomy
  Letters, 32, 228

\bibitem[{{Pustilnik} {et~al.}(2005){Pustilnik}, {Engels}, {Lipovetsky},
  {Kniazev}, {Pramskij}, {Ugryumov}, {Masegosa}, {Izotov}, {Chaffee},
  {M{\'a}rquez}, {Teplyakova}, {Hopp}, {Brosch}, {Hagen}, \&
  {Martin}}]{2005A&A...442..109P}
{Pustilnik}, S.~A., {Engels}, D., {Lipovetsky}, V.~A., {et~al.} 2005, \aap,
  442, 109

\bibitem[{{Pustilnik} \& {Martin}(2007)}]{2007A&A...464..859P}
{Pustilnik}, S.~A. \& {Martin}, J.-M. 2007, \aap, 464, 859

\bibitem[{{Pustilnik} {et~al.}(2002){Pustilnik}, {Martin}, {Huchtmeier},
  {Brosch}, {Lipovetsky}, \& {Richter}}]{2002A&A...389..405P}
{Pustilnik}, S.~A., {Martin}, J.-M., {Huchtmeier}, W.~K., {et~al.} 2002, \aap,
  389, 405

\bibitem[{{Pustilnik} {et~al.}(2008){Pustilnik}, {Tepliakova}, {Kniazev}, \&
  {Burenkov}}]{2008AstBu..63..102P}
{Pustilnik}, S.~A., {Tepliakova}, A.~L., {Kniazev}, A.~Y., \& {Burenkov}, A.~N.
  2008, Astrophysical Bulletin, 63, 102

\bibitem[{{Pustilnik} {et~al.}(2010){Pustilnik}, {Tepliakova}, {Kniazev},
  {Martin}, \& {Burenkov}}]{2010MNRAS.401..333P}
{Pustilnik}, S.~A., {Tepliakova}, A.~L., {Kniazev}, A.~Y., {Martin}, J.-M., \&
  {Burenkov}, A.~N. 2010, \mnras, 401, 333

\bibitem[{{Roennback} \& {Bergvall}(1995)}]{1995A&A...302..353R}
{Roennback}, J. \& {Bergvall}, N. 1995, \aap, 302, 353

\bibitem[{{S\'anchez Almeida} {et~al.}(2010){S\'anchez Almeida}, {Aguerri},
  {Mu\~noz-Tu\~n\'on}, \& {de Vicente}}]{san10}
{S\'anchez Almeida}, J., {Aguerri}, J.~A., {Mu\~noz-Tu\~n\'on}, C., \& {de
  Vicente}, A. 2010, \apj, 714, 487

\bibitem[{{S\'anchez Almeida} {et~al.}(2009){S\'anchez Almeida}, {Aguerri},
  {Mu\~noz-Tu\~n\'on}, \& {Vazdekis}}]{san09}
{S\'anchez Almeida}, J., {Aguerri}, J.~A., {Mu\~noz-Tu\~n\'on}, C., \&
  {Vazdekis}, A. 2009, \apj, 698, 1497

\bibitem[{{S\'anchez Almeida} \& {Lites}(2000)}]{san00}
{S\'anchez Almeida}, J. \& {Lites}, B.~W. 2000, \apj, 532, 1215

\bibitem[{{S\'anchez Almeida} {et~al.}(2008){S\'anchez Almeida},
  {Mu\~noz-Tu\~n\'on}, {Amor\'\i n}, {Aguerri}, {S\' anchez-Janssen}, \&
  {Tenorio-Tagle}}]{san08}
{S\'anchez Almeida}, J., {Mu\~noz-Tu\~n\'on}, C., {Amor\'\i n}, R., {et~al.}
  2008, \apj, 685, 194

\bibitem[{{Sargent} \& {Searle}(1970)}]{sar70}
{Sargent}, W.~L.~W. \& {Searle}, L. 1970, \apjl, 162, L155

\bibitem[{{Schmidt}(1968)}]{sch68}
{Schmidt}, M. 1968, \apj, 151, 393

\bibitem[{{Shi} {et~al.}(2005){Shi}, {Kong}, {Li}, \&
  {Cheng}}]{2005A&A...437..849S}
{Shi}, F., {Kong}, X., {Li}, C., \& {Cheng}, F.~Z. 2005, \aap, 437, 849

\bibitem[{{Skillman} {et~al.}(2003){Skillman}, {C{\^o}t{\'e}}, \&
  {Miller}}]{2003AJ....125..610S}
{Skillman}, E.~D., {C{\^o}t{\'e}}, S., \& {Miller}, B.~W. 2003, \aj, 125, 610

\bibitem[{{Skillman} {et~al.}(1989){Skillman}, {Kennicutt}, \& {Hodge}}]{ski89}
{Skillman}, E.~D., {Kennicutt}, R.~C., \& {Hodge}, P.~W. 1989, \apj, 347, 875

\bibitem[{{Spite} \& {Spite}(1982)}]{spi82}
{Spite}, M. \& {Spite}, F. 1982, \nat, 297, 483

\bibitem[{{Stasi{\'n}ska}(2004)}]{sta04}
{Stasi{\'n}ska}, G. 2004, in Cosmochemistry. The melting pot of the elements,
  ed. C.~{Esteban}, R.~{Garc{\'{\i}}a L{\'o}pez}, A.~{Herrero}, \&
  F.~{S{\'a}nchez} (Cambridge: CUP), 115

\bibitem[{{Stoughton} {et~al.}(2002{\natexlab{a}}){Stoughton}, {Lupton},
  {Bernardi}, {Blanton}, {Burles}, {Castander}, {Connolly}, {Eisenstein},
  {Frieman}, {Hennessy}, {Hindsley}, {Ivezi{\'c}}, {Kent}, {Kunszt}, {Lee},
  {Meiksin}, {Munn}, {Newberg}, {Nichol}, {Nicinski}, {Pier}, {Richards},
  {Richmond}, {Schlegel}, {Smith}, {Strauss}, {SubbaRao}, {Szalay}, {Thakar},
  {Tucker}, {Vanden Berk}, {Yanny}, {Adelman}, {Anderson}, {Anderson}, {Annis},
  {Bahcall}, {Bakken}, {Bartelmann}, {Bastian}, {Bauer}, {Berman},
  {B{\"o}hringer}, {Boroski}, {Bracker}, {Briegel}, {Briggs}, {Brinkmann},
  {Brunner}, {Carey}, {Carr}, {Chen}, {Christian}, {Colestock}, {Crocker},
  {Csabai}, {Czarapata}, {Dalcanton}, {Davidsen}, {Davis}, {Dehnen},
  {Dodelson}, {Doi}, {Dombeck}, {Donahue}, {Ellman}, {Elms}, {Evans}, {Eyer},
  {Fan}, {Federwitz}, {Friedman}, {Fukugita}, {Gal}, {Gillespie}, {Glazebrook},
  {Gray}, {Grebel}, {Greenawalt}, {Greene}, {Gunn}, {de Haas}, {Haiman},
  {Haldeman}, {Hall}, {Hamabe}, {Hansen}, {Harris}, {Harris}, {Harvanek},
  {Hawley}, {Hayes}, {Heckman}, {Helmi}, {Henden}, {Hogan}, {Hogg}, {Holmgren},
  {Holtzman}, {Huang}, {Hull}, {Ichikawa}, {Ichikawa}, {Johnston}, {Kauffmann},
  {Kim}, {Kimball}, {Kinney}, {Klaene}, {Kleinman}, {Klypin}, {Knapp},
  {Korienek}, {Krolik}, {Kron}, {Krzesi{\'n}ski}, {Lamb}, {Leger},
  {Limmongkol}, {Lindenmeyer}, {Long}, {Loomis}, {Loveday}, {MacKinnon},
  {Mannery}, {Mantsch}, {Margon}, {McGehee}, {McKay}, {McLean}, {Menou},
  {Merelli}, {Mo}, {Monet}, {Nakamura}, {Narayanan}, {Nash}, {Neilsen},
  {Newman}, {Nitta}, {Odenkirchen}, {Okada}, {Okamura}, {Ostriker}, {Owen},
  {Pauls}, {Peoples}, {Peterson}, {Petravick}, {Pope}, {Pordes}, {Postman},
  {Prosapio}, {Quinn}, {Rechenmacher}, {Rivetta}, {Rix}, {Rockosi}, {Rosner},
  {Ruthmansdorfer}, {Sandford}, {Schneider}, {Scranton}, {Sekiguchi}, {Sergey},
  {Sheth}, {Shimasaku}, {Smee}, {Snedden}, {Stebbins}, {Stubbs}, {Szapudi},
  {Szkody}, {Szokoly}, {Tabachnik}, {Tsvetanov}, {Uomoto}, {Vogeley}, {Voges},
  {Waddell}, {Walterbos}, {Wang}, {Watanabe}, {Weinberg}, {White}, {White},
  {Wilhite}, {Wolfe}, {Yasuda}, {York}, {Zehavi}, \& {Zheng}}]{sto02}
{Stoughton}, C., {Lupton}, R.~H., {Bernardi}, M., {et~al.} 2002{\natexlab{a}},
  \aj, 123, 485

\bibitem[{{Stoughton} {et~al.}(2002{\natexlab{b}}){Stoughton}, {Lupton},
  {Bernardi}, {Blanton}, {Burles}, {Castander}, {Connolly}, {Eisenstein},
  {Frieman}, {Hennessy}, {Hindsley}, {Ivezi{\'c}}, {Kent}, {Kunszt}, {Lee},
  {Meiksin}, {Munn}, {Newberg}, {Nichol}, {Nicinski}, {Pier}, {Richards},
  {Richmond}, {Schlegel}, {Smith}, {Strauss}, {SubbaRao}, {Szalay}, {Thakar},
  {Tucker}, {Vanden Berk}, {Yanny}, {Adelman}, {Anderson}, {Anderson}, {Annis},
  {Bahcall}, {Bakken}, {Bartelmann}, {Bastian}, {Bauer}, {Berman},
  {B{\"o}hringer}, {Boroski}, {Bracker}, {Briegel}, {Briggs}, {Brinkmann},
  {Brunner}, {Carey}, {Carr}, {Chen}, {Christian}, {Colestock}, {Crocker},
  {Csabai}, {Czarapata}, {Dalcanton}, {Davidsen}, {Davis}, {Dehnen},
  {Dodelson}, {Doi}, {Dombeck}, {Donahue}, {Ellman}, {Elms}, {Evans}, {Eyer},
  {Fan}, {Federwitz}, {Friedman}, {Fukugita}, {Gal}, {Gillespie}, {Glazebrook},
  {Gray}, {Grebel}, {Greenawalt}, {Greene}, {Gunn}, {de Haas}, {Haiman},
  {Haldeman}, {Hall}, {Hamabe}, {Hansen}, {Harris}, {Harris}, {Harvanek},
  {Hawley}, {Hayes}, {Heckman}, {Helmi}, {Henden}, {Hogan}, {Hogg}, {Holmgren},
  {Holtzman}, {Huang}, {Hull}, {Ichikawa}, {Ichikawa}, {Johnston}, {Kauffmann},
  {Kim}, {Kimball}, {Kinney}, {Klaene}, {Kleinman}, {Klypin}, {Knapp},
  {Korienek}, {Krolik}, {Kron}, {Krzesi{\'n}ski}, {Lamb}, {Leger},
  {Limmongkol}, {Lindenmeyer}, {Long}, {Loomis}, {Loveday}, {MacKinnon},
  {Mannery}, {Mantsch}, {Margon}, {McGehee}, {McKay}, {McLean}, {Menou},
  {Merelli}, {Mo}, {Monet}, {Nakamura}, {Narayanan}, {Nash}, {Neilsen},
  {Newman}, {Nitta}, {Odenkirchen}, {Okada}, {Okamura}, {Ostriker}, {Owen},
  {Pauls}, {Peoples}, {Peterson}, {Petravick}, {Pope}, {Pordes}, {Postman},
  {Prosapio}, {Quinn}, {Rechenmacher}, {Rivetta}, {Rix}, {Rockosi}, {Rosner},
  {Ruthmansdorfer}, {Sandford}, {Schneider}, {Scranton}, {Sekiguchi}, {Sergey},
  {Sheth}, {Shimasaku}, {Smee}, {Snedden}, {Stebbins}, {Stubbs}, {Szapudi},
  {Szkody}, {Szokoly}, {Tabachnik}, {Tsvetanov}, {Uomoto}, {Vogeley}, {Voges},
  {Waddell}, {Walterbos}, {Wang}, {Watanabe}, {Weinberg}, {White}, {White},
  {Wilhite}, {Wolfe}, {Yasuda}, {York}, {Zehavi}, \&
  {Zheng}}]{2002AJ....123..485S}
{Stoughton}, C., {Lupton}, R.~H., {Bernardi}, M., {et~al.} 2002{\natexlab{b}},
  \aj, 123, 485

\bibitem[{{Takeuchi} {et~al.}(2000){Takeuchi}, {Yoshikawa}, \& {Ishii}}]{tak00}
{Takeuchi}, T.~T., {Yoshikawa}, K., \& {Ishii}, T.~T. 2000, \apjs, 129, 1

\bibitem[{{Thuan} \& {Izotov}(2005)}]{2005ApJS..161..240T}
{Thuan}, T.~X. \& {Izotov}, Y.~I. 2005, \apjs, 161, 240

\bibitem[{{Tremonti} {et~al.}(2004){Tremonti}, {Heckman}, {Kauffmann},
  {Brinchmann}, {Charlot}, {White}, {Seibert}, {Peng}, {Schlegel}, {Uomoto},
  {Fukugita}, \& {Brinkmann}}]{tre04}
{Tremonti}, C.~A., {Heckman}, T.~M., {Kauffmann}, G., {et~al.} 2004, \apj, 613,
  898

\bibitem[{{van Zee} \& {Haynes}(2006)}]{2006ApJ...636..214V}
{van Zee}, L. \& {Haynes}, M.~P. 2006, \apj, 636, 214

\bibitem[{{van Zee} {et~al.}(2006){van Zee}, {Skillman}, \&
  {Haynes}}]{2006ApJ...637..269V}
{van Zee}, L., {Skillman}, E.~D., \& {Haynes}, M.~P. 2006, \apj, 637, 269

\bibitem[{{V{\'{\i}}lchez} \&
  {Iglesias-P{\'a}ramo}(2003)}]{2003ApJS..145..225V}
{V{\'{\i}}lchez}, J.~M. \& {Iglesias-P{\'a}ramo}, J. 2003, \apjs, 145, 225

\bibitem[{{Viticchi{\'e}} \& {S{\'a}nchez Almeida}(2011)}]{2011A&A...530A..14V}
{Viticchi{\'e}}, B. \& {S{\'a}nchez Almeida}, J. 2011, \aap, 530, A14+

\bibitem[{{Yin} {et~al.}(2007){Yin}, {Liang}, {Hammer}, {Brinchmann}, {Zhang},
  {Deng}, \& {Flores}}]{2007A&A...462..535Y}
{Yin}, S.~Y., {Liang}, Y.~C., {Hammer}, F., {et~al.} 2007, \aap, 462, 535

\end{thebibliography}
%

\end{document}